\documentclass[aps,pra,reprint,twocolumn,superscriptaddress,floatfix]{revtex4-2}
\usepackage{amsmath,amssymb,amsfonts}
\usepackage{graphicx}
\usepackage[caption=false]{subfig} % APS-friendly subfigures
\usepackage{bbold}
\usepackage{xspace}
\usepackage{url}
\usepackage[hidelinks]{hyperref}

\graphicspath{{./figs/}}

\newcommand{\ket}[1]{\left| #1 \right\rangle}
\newcommand{\bra}[1]{\left\langle #1 \right|}	
\newcommand{\exb}{\mathbf{E} \times \mathbf{B}}

\newcommand{\co}{NoMoCou\xspace}
\newcommand{\MS}{M{\o}lmer--S{\o}rensen\xspace}
\newcommand{\MSGATE}{M{\o}lmer--S{\o}rensen gate\xspace}

\begin{document}

\title{Nonlinear Coupling between Motional Modes in Trapped Ion Quantum Processors} 

\author{Wes Johnson}
\affiliation{University of Colorado, Boulder, 80309, USA}
\affiliation{Sandia National Laboratories, Albuquerque, New Mexico, 87185, USA}

\author{Brandon Ruzic}
\affiliation{Sandia National Laboratories, Albuquerque, New Mexico, 87185, USA}

\begin{abstract}
Trapped-ion crystals are a leading platform for quantum information science, but achieving the high-fidelity entangling gates required for fault-tolerant quantum computing becomes harder as system size increases.
As systems scale and adopt new geometries, spectral crowding makes low-order nonlinear resonances between collective motional modes increasingly common and can limit gate performance, especially in monolithic or global-mode architectures. 
We develop a general model to identify and simulate nonlinear motional-mode coupling (\co) arising from third-order Coulomb terms and quantify its impact on the \MSGATE{} across linear chains and 2D crystals in rf and Penning traps.
We delineate the regimes where \co{} dominates the error budget and provide design rules: detune operating points from low-order resonances, tune trap anisotropy to reshape spectra, and shape gate waveforms.
\end{abstract}

\maketitle

\section{Introduction} 
\label{sec:intro}
Fault-tolerant quantum computation requires extremely high-fidelity gates. 
In architectures based on surface codes, this comes at the cost of encoding each logical qubit in thousands of physical qubits \cite{Fowler2012}.
Even with optimized protocols, executing useful quantum algorithms will require millions of physical qubits and gate fidelities that are just now approaching fault tolerant limits \cite{OGorman2017,Loschnauer2024}.
This places growing importance on identifying all physical error sources that could limit gate performance in large-scale systems.

Despite extensive theoretical and experimental studies, the impact of nonlinear motional mode coupling (\co) on entangling gate fidelity remains largely uncharacterized—particularly in global-mode trapped-ion processors, where large ion crystals share collective motional modes. 
In these systems, all-to-all connectivity between qubits is achieved through the collective motion of the ion crystal, whose mode spectrum becomes increasingly dense as the number of ions grows—opening new resonant pathways for unwanted nonlinear interactions. 
Because entangling gates rely on the controlled evolution of these shared modes \cite{Cirac1995,Srensen2000}, preserving their linearity is critical to gate performance.
In contrast to modular or QCCD-based architectures \cite{Kielpinski2002}---which localize entangling operations to small subsets of ions---global-mode systems must contend with the full complexity of their collective dynamics. 
Thus, understanding and mitigating the effects of \co is vital for scaling up such architectures.

Previous work has explored motional nonlinearities in trapped-ion systems, including frequency shifts and mode distortions arising from anharmonic trap potentials and Coulomb interactions~\cite{Wu2013,McAneny2014,Marquet2003,Roos2008,Nie2009,Home2011}. 
In the context of two-dimensional Coulomb crystals, Porras \& Cirac showed that anharmonic couplings between axial and in-plane modes in an ideal triangular lattice can impose temperature-dependent decoherence limits on ``pushing'' entangling gates~\cite{Porras2006}. 
Related studies have examined nonlinearity in the magnetic gradient induced coupling (MAGIC) architecture~\cite{Nagies2025} and more generally in linear chains~\cite{Marquet2003,Roos2008,Nie2009,Home2011}. 
Motional nonlinearity has also been leveraged as a resource for quantum simulation~\cite{Ding2018,Ding2017co}, phonon counting~\cite{Ding2017}, nonlinearity-enabled robust entangling gates~\cite{Le2025}, and quantum thermodynamic experiments~\cite{Maslennikov2019,Nimmrichter2017}, yet its impact on entangling operations in trapped-ion quantum processors remains largely unexplored.
In this work, we systematically study the effects of \co{} on entangling-gate fidelity in large global-mode trapped-ion systems. 
To our knowledge, this is the first systematic, architecture-spanning computation of third-order Coulomb (triad) couplings built from exact finite-crystal equilibria (including Penning and radiofrequency (rf) Paul trap spectra), enabling quantitative triad statistics and gate-fidelity maps. 
To this end, we develop a general technique to identify \co{} in trapped-ion systems and study its effect on the \MSGATE{} fidelity via numerical simulation, applying this framework to multiple architectures, including two-dimensional (2D) ion crystals in Penning traps and rf Paul traps confining both linear ion chains and 2D crystals.

In Section~\ref{sec:background}, we introduce the physical foundations of \co in trapped ion systems. 
We illustrate its dynamics in a simple two-ion crystal and develop a two-level (TL) model to characterize coupling strengths uniformly across trapped ion systems. 
We also define the dimensionless quantum expansion parameter, $\epsilon_0$, which sets the relative strength of normal mode and third order coupling dynamics. 
In Section~\ref{sec:results}, we use this framework to simulate MS gates under realistic conditions, beginning with the two-ion system and extending to large ion crystals. 
We examine how \co scales with system size, assess its presence in experimental 2D Penning and rf Paul trap crystals, and explore its impact on entangling gate fidelities under representative conditions.
We conclude by discussing implications for quantum processor design and strategies to mitigate the effects of \co{} as trapped-ion systems are scaled up.

\section{Background} 
\label{sec:background}

\begin{figure*}[t]
  \centering
  \includegraphics[width=.9\linewidth]{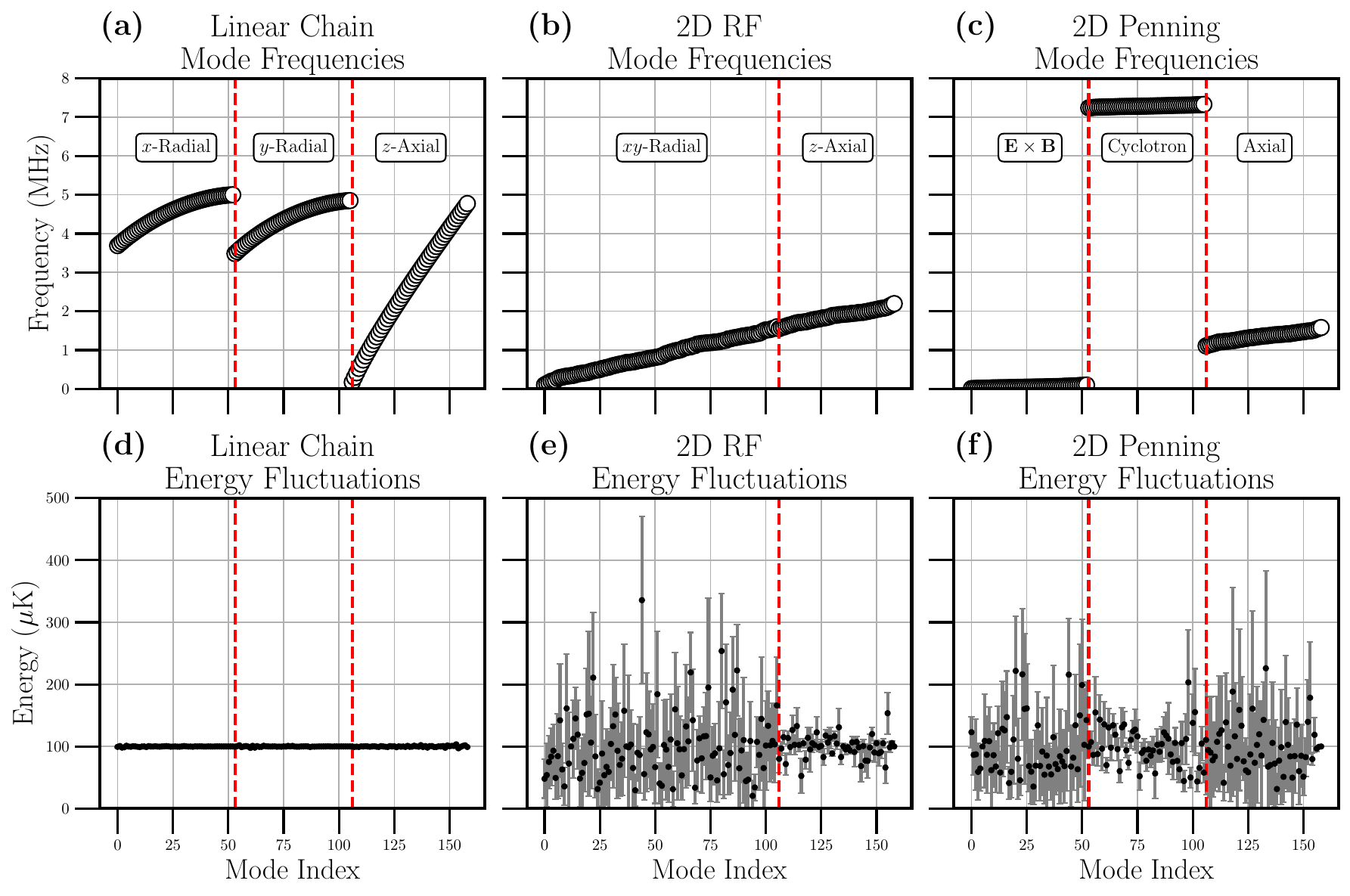}
  \caption{Two-dimensional ion crystals exhibit larger motional-energy fluctuations than linear chains, with Penning-trap 2D arrays showing enhanced axial (drumhead) variability while their highest-frequency cyclotron modes remain the most stable. 
  Top row: normal-mode spectra. Bottom row: per-mode mean energy and standard deviation over a $T_\mathrm{sim}=10~\mathrm{ms}$ classical MD evolution (all $3N$ modes initialized at $100~\mu\mathrm{K}$ with random phases; $N=53$). 
  (a,d) Linear chain in an rf trap: small fluctuations about initial energies. 
  (b,e) 2D rf array: increased fluctuations, especially among radial modes. 
  (c,f) 2D Penning array: larger axial fluctuations than (b,e) while cyclotron modes remain comparatively stable. 
  These trends indicate stronger nonlinear energy exchange (\co) in 2D geometries---most prominently in Penning traps---motivating a quantitative assessment of their impact on gate performance later in the paper.}
  \label{fig:crystal_energy_fluctuation_comparison}
\end{figure*}

Trapped-ion platforms span multiple architectures and ion species \cite{Bruzewicz2019}. 
Here we focus on two widely used systems: Penning traps with $^9\text{Be}^+$ and radio-frequency (rf) Paul traps with $^{171}\text{Yb}^+$ and $^{40}\text{Ca}^+$. 
Penning traps employ a strong magnetic field to realize rigidly rotating two-dimensional (2D) crystals with hundreds of ions for quantum sensing \cite{Gilmore2021} and quantum simulation \cite{Bohnet2016}. 
The collective rotation complicates site-resolved qubit control, though demonstrated strategies mitigate this challenge \cite{Polloreno2022,McMahon2024}. 
In contrast, rf Paul traps confine ions at static equilibrium positions, supporting both linear chains \cite{James1998} and 2D arrays \cite{Kiesenhofer2023}, and are widely used for quantum computing experiments with tens of qubits \cite{Chen2024}. 
Both architectures have enabled quantum simulation and other NISQ-era applications \cite{Britton2012,Bohnet2016,Zhang2017,Pagano2018,Monroe2021}.
There are also proposals for 2D crystals confined with static electric and optical potentials \cite{Sun2024}, hybrid Penning surface-electrode traps \cite{Jain2020,Jain2024}, and 3D arrays in Penning traps \cite{Hawaldar2024,Zaris2025}, which are not explicitly treated here but can be analyzed with the same framework. 

We use molecular dynamics (MD) simulations to compare the nonlinear motional dynamics of these architectures (Section~\ref{subsec:comparative_dynamics_across_different_architectures}). 
Figure~\ref{fig:crystal_energy_fluctuation_comparison} shows that 2D arrays exhibit stronger nonlinear energy exchange than linear chains in simulations based on real experimental setups, with Penning 2D crystals displaying especially large axial (drumhead) mode energy fluctuations. 
These observations motivate a detailed analysis of their impact on quantum-gate performance in later sections.

\subsection{Normal Modes}
\label{subsec:normal_modes}

In both trap architectures, small displacements about equilibrium decompose into collective vibrational normal modes~\cite{James1998,Wang2013,Shankar2020,Dubin2020}. 
These modes mediate interactions between the ions' internal states~\cite{Cirac1995} and underpin entangling operations such as the \MSGATE~\cite{Srensen2000}. 
We adopt a Hamiltonian formulation of normal-mode analysis~\cite{Dubin2020}, allowing us to treat linear and nonlinear dynamics across architectures within the same framework.

\emph{Classical non-dimensionalization.} We normalize lengths by $l_0$, time by $\omega_0^{-1}$, masses by a reference $m_0$, and energies by 
$E_0 = m_0 \omega_0^2 l_0^2$ (definitions in Appendix~\ref{app:quantization}). 
With this choice, mode coordinates are dimensionless:
$Q_n$ and $P_n$ are the canonical coordinates obtained from the linearized dynamics, and the mode frequencies $\omega_n$ are expressed in units of $\omega_0$.

Within this framework, each mode behaves as an independent harmonic oscillator 
\begin{equation}
    \mathcal{H}^{(2)} = \sum_{n=1}^{3N}\frac{\omega_n}{2}\!\left(Q_n^2 + P_n^2\right),
    \label{eq:mode_Hamiltonian}
\end{equation}
where $\omega_n$ is the (dimensionless) angular frequency of the $n^{\text{th}}$ mode and $Q_n,P_n$ are the associated canonical coordinates. 
For a system of $N$ ions there are $3N$ normal modes. 
Additional details and the construction from the linearized equations of motion are provided in Appendix~\ref{app:Hamiltonian}.

\subsection{Nonlinear Coupling}
\label{subsec:nonlinear_coupling}

Motional nonlinearity in trapped-ion systems arises from higher-than-quadratic terms in the total potential energy. 
These anharmonic contributions originate both from the intrinsically nonlinear Coulomb interaction and from deviations of the trapping potential from a perfect harmonic form. 
The latter can be pronounced in surface-electrode rf traps, where ions reside close to electrodes and multi-species operation results in significant nonlinear effects~\cite{Home2011}. 

Depending on the relationship between mode frequencies, nonlinear coupling can produce two qualitatively distinct effects.
On resonance, it enables energy exchange between normal modes, resulting in mode-entanglement~\cite{Marquet2003}. 
Off resonance, it produces amplitude-dependent shifts of mode frequencies due to the excitations in spectator modes~\cite{Roos2008,Nie2009,Home2011,McAneny2014}. 
Either effect can disturb the motional state required for high-fidelity operations such as the \MSGATE{}~\cite{Srensen2000}, by either directly entangling the qubits with spectator modes (resonant \co) or by shifting the bus mode frequency and thus the gate detuning (off-resonant effect) (see Figures~\ref{fig:n2_motional_detuning_vs_gate_time} \& \ref{fig:n2_motional_detuning_vs_gate_time_thermal}). 

We capture these nonlinear effects by extending the mode Hamiltonian to third order,
\begin{equation}
  \mathcal{H}^{(3)} = \frac{1}{6}\sum_{n,m,p=1}^{3N}\;\sum_{X,Y,Z\in\{Q,P\}}
  T^{XYZ}_{nmp}\, X_n Y_m Z_p,
  \label{eq:mode_Hamiltonian_nonlinear}
\end{equation}
where $X_n\in\{Q_n,P_n\}$ are the canonical coordinates of mode $n$ and $T^{XYZ}_{nmp}$ are the (dimensionless) classical third-order coupling coefficients in this scaling. 
The factor $1/6$ avoids overcounting permutations, and $T^{XYZ}_{nmp}$ inherits index symmetries from the underlying potential (e.g., symmetry under permutations of the triplets $(n,X)$, $(m,Y)$, $(p,Z)$). 
Terms with repeated mode indices describe self- and cross-anharmonicities (e.g., $Q_n^3$ or $Q_n P_n^2$), while mixed-mode index terms enable two-mode and three-mode interactions that become efficient near classical sum/difference conditions such as $\omega_p\approx2\omega_n$ or $\omega_p\approx\omega_n+\omega_m$.
The explicit construction of $T^{XYZ}_{nmp}$ from third derivatives of the potential in normal-mode coordinates is provided in Appendix~\ref{app:tressian}. 
In the next subsection we quantize this model, identify the near-resonant processes that survive under the rotating-wave approximation, and determine the relative strength of nonlinear and linear dynamics in terms of the quantum expansion parameter $\epsilon_0$.

\subsection{Quantization \& Quantum Expansion Parameter}
\label{subsec:quantization}

\emph{Quantum non-dimensionalization.} 
We nondimensionalize using a reference angular frequency $\omega_0$ and the quantum energy unit $\hbar\omega_0$, choosing
$E_\mathrm{unit}=\hbar\omega_0=\epsilon_0^2 E_0$ and $T_\mathrm{unit}=\omega_0^{-1}$, so that in these units $\hbar=\epsilon_0^2$ (definitions in Appendix~\ref{app:quantization}). 
The dimensionless quantum expansion parameter
\begin{equation}
  \epsilon_0=\frac{z_0}{l_0},
  \label{eq:quantumExpansionParameter_back}
\end{equation}
compares the single-oscillator length $z_0$ to a characteristic inter-ion length $l_0$ and sets the relative strength of cubic (and higher) terms versus harmonic dynamics. 
The quantum expansion parameter has a very weak dependence on mass and trap frequency, $\epsilon_0\propto m^{-1/6}\,\omega_0^{1/6}$. 
For $^9\mathrm{Be}^+$, $^{40}\mathrm{Ca}^+$, and $^{171}\mathrm{Yb}^+$ at $\omega_0=2\pi\times 2~\mathrm{MHz}$ we obtain $\epsilon_0\approx 2\times 10^{-3}$ to $4\times 10^{-3}$. 

Quantizing Eqs.~\eqref{eq:mode_Hamiltonian} and \eqref{eq:mode_Hamiltonian_nonlinear}, we use $[Q_n,P_m]=i\epsilon_0^2\delta_{nm}$ and define ladder operators
$a_n=(Q_n+iP_n)/(\sqrt{2}\epsilon_0)$ and $a_n^\dagger=(Q_n-iP_n)/(\sqrt{2}\epsilon_0)$, which satisfy $[a_n,a_m^\dagger]=\delta_{nm}$. 
Expressed in the quantum energy unit $\hbar\omega_0$, the harmonic Hamiltonian for mode $n$ is
\begin{equation}
  \mathcal{H}_n^{(2)}
  = \frac{\omega_n}{2}\big(Q_n^2+P_n^2\big)
  = \omega_n \big(a_n^\dagger a_n + \tfrac{1}{2}\big),
  \label{eq:harmonic_quantized_eps}
\end{equation}
where $\omega_n$ is the dimensionless mode frequency in units of $\omega_0$. 

Under the rotating-wave approximation (RWA) and in the interaction picture (with respect to $\sum_n \omega_n a_n^\dagger a_n$), a representative near-resonant three-mode coupling term takes the form  
\begin{equation}
  \mathcal{H}^{\mathrm{RWA}}_{nmp}
  = \epsilon_0\!\left(C^{\mathrm{RWA}}_{nmp}\; a_n a_m a_p^\dagger\, e^{+i\Delta_{nmp} t}
  + \mathrm{H.c.}\right),
  \label{eq:quantum_nonlinear_hamiltonian_back}
\end{equation}
where $\Delta_{nmp}=\omega_p-\omega_n-\omega_m$ is the \emph{nonlinear detuning} (all angular frequencies) and $t$ is in units of $\omega_0^{-1}$. 
The complex coupling $C^{\mathrm{RWA}}_{nmp}$ is determined by the classical third-order coefficients $T^{XYZ}_{nmp}$ (Appendix~\ref{app:quantization}) and inherits their index symmetries. 
We reserve $\delta_\text{gate}$ for the \emph{gate detuning} used in the \MS\ gate, distinct from the nonlinear detuning $\Delta_{nmp}$ introduced here.

\subsection{Comparative Dynamics Across Different Architectures}
\label{subsec:comparative_dynamics_across_different_architectures}

To compare dynamics across architectures, we perform classical molecular dynamics (MD) simulations for three representative systems: a linear chain in an rf trap~\cite{Zhang2017}, a 2D crystal in an rf trap~\cite{Kiesenhofer2023}, and a 2D crystal in a Penning trap~\cite{Gilmore2021}. 
Simulation details are provided in Appendix~\ref{app:simulations}. 
For each system, we simulate $N=53$ ions initialized at $T=100~\mu\mathrm{K}$. 
Initialization is performed by assigning each of the $3N$ normal modes a random phase and an amplitude corresponding to the desired temperature, so that the initial mode energy satisfies $E_n(0)=k_B T$, where $k_B$ is Boltzmann's constant. 
We then evolve the system for a total duration of $T_\mathrm{sim}=10~\mathrm{ms}$ and compute the per-mode energy $E_n(t)$ throughout.

Figure~\ref{fig:crystal_energy_fluctuation_comparison} summarizes these results: the top row displays the normal-mode spectra for each geometry, and the bottom row shows, for each mode, the mean and standard deviation of $E_n(t)$ over the evolution. 
In the absence of nonlinearities, each mode's energy would remain constant under the linearized dynamics; observed fluctuations therefore reflect energy exchange mediated by nonlinear couplings~\cite{Johnson2024}.

The results reveal a clear contrast between the linear chain and 2D configurations. 
In the linear chain, mode energies remain close to their initial values with only small fluctuations. 
In both 2D geometries, fluctuations are significantly larger---especially in the Penning-trap case---indicating stronger nonlinear coupling and more extensive inter-mode energy exchange. 
This highlights the central role of crystal geometry in setting the strength and consequences of \co. 
These observations motivate a more detailed study of nonlinear dynamics in 2D systems, where \co\ can limit gate fidelities or drive thermalization-like behavior even at low temperatures \cite{Tang2021,Johnson2024,Johnson2025}

\subsection{\co in a Two-Ion Crystal}
\label{subsec:two_ion_crystal}

\begin{figure}[htbp] 
	\centering
	\includegraphics[width=1\linewidth]{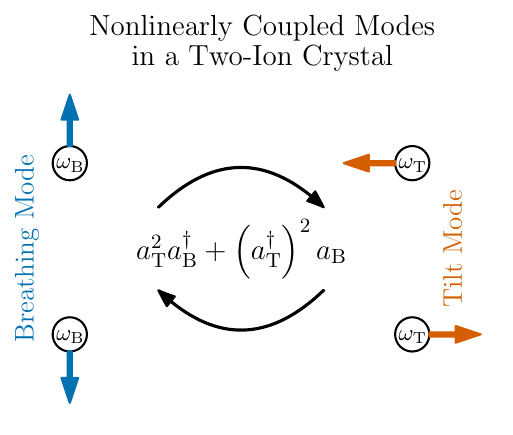}
	\caption{Schematic of the two-ion crystal showing the nonlinear coupling between the tilt mode (red, $\omega_\text{T}$) and the breathing (bus) mode (blue, $\omega_\text{B}$). 
	The interaction involves products of the orthogonal modes' ladder operators and enables energy exchange when the resonance condition $\omega_\text{B} \approx 2\omega_\text{T}$ is satisfied.
	}
	\label{fig:N2_coupling_schematic}
\end{figure}

\begin{figure*}[t]
	\centering
	\includegraphics[width=\textwidth]{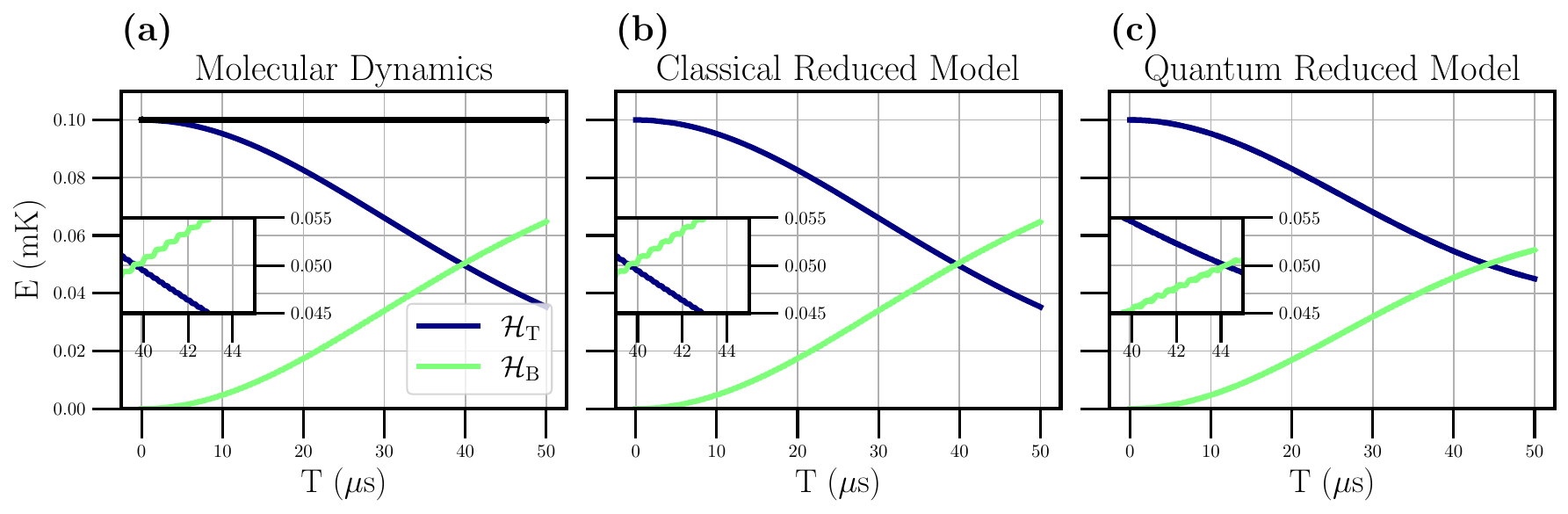}
	\caption{Dynamics of the two-ion crystal showing nonlinear mode coupling (\co) between the tilt (blue) and breathing (green) modes when the resonance condition $\Delta_{\mathrm{TTB}}=\omega_\mathrm{B}-2\omega_\mathrm{T} = 0$ is satisfied.
    \textbf{(a)} Molecular dynamics simulation with the full Coulomb potential. 
    All modes except the breathing mode are initialized to amplitudes corresponding to a temperature of $100~\mu\mathrm{K}$, while the breathing mode is initialized with zero amplitude.
    All other modes are plotted in black.
    \textbf{(b)} Classical reduced model evolving the four canonical variables of the tilt and breathing modes.
    \textbf{(c)} Quantum reduced model with (solid lines) and without (dashed lines) the rotating-wave approximation (RWA). 
    (Note: the dashed RWA lines are indistinguishable from the solid lines on this scale.)
    All simulations are initialized with energy in the tilt mode and the breathing mode in its ground state. 
    The close agreement between (a) and (b) validates the classical reduced model, while (c) shows the quantum oscillations and the accuracy of the RWA.
	}
	\label{fig:two_ion_crystal_coupling_comparison}
\end{figure*}

To illustrate the impact of nonlinear motional mode coupling (\co) described by Eqs.~\eqref{eq:mode_Hamiltonian_nonlinear} and \eqref{eq:quantum_nonlinear_hamiltonian_back}, we analyze a two-ion crystal using both classical and quantum models. 
This is the simplest trapped-ion configuration exhibiting \co\ from the Coulomb interaction alone and provides a clean test bed for the modeling and simulation techniques used throughout this work; related quantum-level observations appear in Refs.~\cite{Roos2008,Ding2017,Ding2017co}.

We take the $z$ direction to be the weakest confinement axis so the ions align axially. 
Each Cartesian direction supports an in-phase (center-of-mass, COM) and an out-of-phase mode. 
Because the Coulomb interaction depends only on relative displacements, it leaves COM frequencies unchanged but shifts and couples out-of-phase modes. 
We focus on the out-of-phase axial \emph{breathing} mode (B) and the out-of-phase radial \emph{tilt} mode (T) along $y$ (the orthogonal $x$ tilt is sufficiently detuned to be neglected). 
Their linear frequencies are
\begin{equation}
  \omega_\mathrm{T}=\sqrt{\omega_y^2-\omega_z^2},\qquad
  \omega_\mathrm{B}=\sqrt{3}\,\omega_z,
  \label{eq:two_ion_frequencies}
\end{equation}
where $\omega_y$ and $\omega_z$ are the single-ion radial and axial trap frequencies, respectively.

\paragraph*{Classical reduced model.}
Restricting the cubic Hamiltonian to the $\{\mathrm{T},\mathrm{B}\}$ subspace yields
\begin{equation}
  \mathcal{H}_\text{2-ion}^{(3)}
  = \xi_\text{class}\, Q_\mathrm{T}^{2} P_\mathrm{B}
  + \chi_\text{class}\, Q_\mathrm{B}^{3},
  \label{eq:2-ion_hamiltonian}
\end{equation}
where $Q_\alpha,P_\alpha$ are the canonical coordinates for mode $\alpha\in\{\mathrm{T},\mathrm{B}\}$, and $\xi_\text{class},\chi_\text{class}$ are dimensionless coefficients obtained from the third-order tensor $T^{XYZ}_{nmp}$ in Eq.~\eqref{eq:mode_Hamiltonian_nonlinear} (see Appendix~\ref{app:simulations} for explicit construction and values). 
The $Q_\mathrm{T}^{2}P_\mathrm{B}$ term drives the near-resonant two-mode process; $Q_\mathrm{B}^{3}$ produces a small self-anharmonic shift of the breathing frequency.

\paragraph*{Quantum interaction and RWA.}
Quantizing with $[Q_n,P_m]=i\epsilon_0^{2}\delta_{nm}$ and $a=(Q+iP)/(\sqrt{2}\epsilon_0)$ (Appendix~\ref{app:quantization}), the leading three-operator term that exchanges energy between the tilt and breathing modes (up-conversion) is, in the interaction picture and under the rotating-wave approximation (RWA),
\begin{equation}
  \mathcal{H}^{\mathrm{RWA}}_\text{2-ion}
  = \epsilon_0\!\left(
    C^{\mathrm{RWA}}_{\mathrm{TTB}}\; a_\mathrm{T}^{2}\, a_\mathrm{B}^\dagger\,
    e^{+i\Delta_{\mathrm{TTB}} t}
    + \mathrm{H.c.}\right),
  \label{eq:2-ion_interaction_hamiltonian}
\end{equation}
where $C^{\mathrm{RWA}}_{\mathrm{TTB}}$ is the time-independent dimensionless coupling coefficient derived from $T^{XYZ}_{nmp}$ in the classical expansion (Appendix~\ref{app:quantization}). 
The operator $a_\mathrm{T}^2 a_\mathrm{B}^\dagger$ annihilates two tilt phonons while creating one breathing phonon; H.c.\ reverses this process. 
The nonlinear detuning is
\begin{equation}
  \Delta_{\mathrm{TTB}}=\omega_\mathrm{B}-2\omega_\mathrm{T}.
  \label{eq:two_ion_detuning}
\end{equation}

\paragraph*{Resonance condition.}
Complete energy exchange occurs near $\Delta_{\mathrm{TTB}}=0$, i.e.,
\begin{equation}
  \Delta_{\mathrm{TTB}}=0
  \quad\Longleftrightarrow\quad
  \frac{\omega_y}{\omega_z}=\frac{\sqrt{7}}{2},
  \label{eq:two_ion_resonance}
\end{equation}
the two-mode ($2\omega_\mathrm{T}\!\approx\!\omega_\mathrm{B}$) coupling resonance discussed in Sec.~\ref{subsec:TLS}.

\paragraph*{Validating reduced models.}
We compare four simulations (Appendix~\ref{app:simulations}): (i) classical molecular dynamics (MD) with the full Coulomb potential; (ii) a classical reduced model (CRM) using Eq.~\eqref{eq:2-ion_hamiltonian}; (iii) a quantum reduced model (QRM) using Eq.~\eqref{eq:2-ion_interaction_hamiltonian} (RWA); and (iv) a QRM without the RWA. 
Figure~\ref{fig:two_ion_crystal_coupling_comparison} shows energy exchange between the modes: (a) MD includes all degrees of freedom; (b) CRM retains only the $\{\mathrm{T},\mathrm{B}\}$ subspace and agrees closely with MD in this regime, validating the third-order coefficients extracted from $T^{XYZ}_{nmp}$. 
Panel (c) compares QRM with and without the RWA: the RWA (dashed) omits small rapid oscillations but reproduces the net energy transfer predicted by the full QRM (solid).

The time when the tilt and breathing modes have equal energy is markedly different between the classical and quantum simulations. 
Our initial conditions correspond to only a few quanta of energy in the tilt mode, so quantum effects are significant.
As $\epsilon_0\!\to\!0$, or equivalently, $|\alpha|^2\!\to\!\infty$, coherent states behave semiclassically and quantum expectation values follow classical trajectories more closely.

These results show that a reduced Hamiltonian captures the essential \co\ physics in the two-ion crystal without full MD, while also highlighting that a quantum description is necessary to faithfully model low-temperature dynamics.	

\subsection{Two-Level reduction for quantifying \co} 
\label{subsec:TLS}	

We characterize two- and three-mode coupling processes near the ground state on equal footing by considering the two-level (TL) Hamiltonian that describes the dynamics of the two lowest-energy Fock states coupled by a given nonlinear process.
In general, each nonlinear interaction Hamiltonian (Eq.~\eqref{eq:quantum_nonlinear_hamiltonian_back}) is block diagonal in the joint Fock basis, with sub-blocks labeled by conserved Manley--Rowe invariants of the resonance~\cite{Dodin2008}.
In this picture, the dynamics are governed by Rabi oscillations whose frequency and amplitude directly measure the interaction strength.
We use the TL Hamiltonian to define a resonance criterion for when a nonlinear interaction is significant, and we apply this criterion to the two-ion crystal example to illustrate the impact of \co\ near the ground state.

At third order in the Hamiltonian, nonlinear interactions fall into two classes \cite{Marquet2003}. 
\emph{Two-mode coupling} corresponds to processes with two identical mode indices (e.g., $a_n^2 a_p^\dagger$), as in the two-ion case of Eq.~\eqref{eq:2-ion_interaction_hamiltonian}.
\emph{Three-mode coupling} corresponds to processes with all mode indices distinct (e.g., $a_n a_m a_p^\dagger$), as in Eq.~\eqref{eq:quantum_nonlinear_hamiltonian_back}.
Figure~\ref{fig:coupling_type_diagram} depicts both cases.
Self-interaction terms (e.g., $a_n^\dagger a_n a_n$) do not survive in the RWA and are not considered here.

For two-mode coupling, the relevant lowest-energy pair is $\ket{2,0}$ and $\ket{0,1}$, denoting two quanta in the lower-frequency mode and one in the higher-frequency mode.
For three-mode coupling, the relevant pair is $\ket{1,1,0}$ and $\ket{0,0,1}$.

The TL Hamiltonian is
\begin{equation}
  \mathcal{H}_{nmp}^{\mathrm{TL}}
  =
  \begin{bmatrix}
    0 & C_{nmp}^{\mathrm{TL}}\,e^{-i\Delta_{nmp} t} \\
    C_{nmp}^{\mathrm{TL}}\,e^{+i\Delta_{nmp} t} & 0
  \end{bmatrix},
  \label{eq:2-level_hamiltonian}
\end{equation}
where $C_{nmp}^{\mathrm{TL}}$ is the matrix element of the nonlinear interaction and $\Delta_{nmp}$ is the nonlinear detuning.
The explicit $C_{nmp}^{\mathrm{TL}}$ depends on whether the process is two-mode or three-mode.

\textbf{Two-mode coupling:}
\begin{equation}
  C_{nnp}^{\mathrm{TL}}=\sqrt{2}\,\epsilon_0\,C_{nnp}^{\mathrm{RWA}},
  \qquad
  \Delta_{nnp}=\omega_p-2\omega_n.
  \label{eq:two_mode_coupling}
\end{equation}

\textbf{Three-mode coupling:}
\begin{equation}
  C_{nmp}^{\mathrm{TL}}=\epsilon_0\,C_{nmp}^{\mathrm{RWA}},
  \qquad
  \Delta_{nmp}=\omega_p-\omega_n-\omega_m,
  \label{eq:three_mode_coupling}
\end{equation}
with the mode ordering chosen such that $\omega_n\le\omega_m<\omega_p$.
The factor of $\sqrt{2}$ in the two-mode case arises from bosonic enhancement of the matrix element when two identical phonons are annihilated.

Assuming the system is initialized in $\ket{0}$ (i.e., $\ket{2,0}$ for two-mode or $\ket{1,1,0}$ for three-mode), the probability to be in $\ket{1}$ (i.e., $\ket{0,1}$ or $\ket{0,0,1}$) at time $t$ is
\begin{align}
  P_{0\to 1}(t)
  &= S_{nmp}^{\mathrm{TL}}\,\sin^2\!\big(\Omega_{nmp}^{\mathrm{TL}}\,t\big), \notag\\
  \Omega_{nmp}^{\mathrm{TL}}
  &= \sqrt{\left(C_{nmp}^{\mathrm{TL}}\right)^2 + \left(\frac{\Delta_{nmp}}{2}\right)^2}, \notag\\
  S_{nmp}^{\mathrm{TL}}
  &= \left(\frac{C_{nmp}^{\mathrm{TL}}}{\Omega_{nmp}^{\mathrm{TL}}}\right)^2.
  \label{eq:2-level_probability}
\end{align}
The oscillation amplitude $S_{nmp}^{\mathrm{TL}}$ is a Lorentzian in $\Delta_{nmp}$, peaking at unity on resonance $\Delta_{nmp}=0$, with width set by $C_{nmp}^{\mathrm{TL}}$.

In our non-dimensional units, $C_{nmp}^{\mathrm{RWA}}=\mathcal{O}(1)$ and $\epsilon_0\sim 10^{-3}$, so the characteristic \emph{frequency} scale of the nonlinear process is $\epsilon_0 C_{nmp}^{\mathrm{RWA}}\omega_0$.
In physical units this corresponds to $\sim 2\pi\times 1$–$10~\mathrm{kHz}$ when $\omega_0\sim 2\pi\times 1$–$3~\mathrm{MHz}$ and $C_{nmp}^{\mathrm{RWA}}$ is order unity.
On resonance, the TL oscillation period $T_{nmp}^{\mathrm{TL}}=2\pi/\Omega_{nmp}^{\mathrm{TL}}$ sets the \co\ timescale near the ground state and is typically $T_{nmp}^{\mathrm{TL}}\sim 0.1$–$1~\mathrm{ms}$, comparable to entangling-gate durations.

The coupling strengthens with occupation number.
For the up-conversion step $\ket{n,m,p}\!\to\!\ket{n-1,m-1,p+1}$ the relevant matrix-element magnitude scales as
\begin{equation}
  \big|\mathcal{M}_{nmp}\big|
  \;=\;
  \epsilon_0\,\big|C^{\mathrm{RWA}}_{nmp}\big|\,\sqrt{n\,m\,(p+1)}
  \;\propto\; E^{3/2},
  \label{eq:matrix_element}
\end{equation}
so \co\ becomes more pronounced at higher energies.

For the two-ion parameters used in Fig.~\ref{fig:two_ion_crystal_coupling_comparison}, the TL model predicts the equal-energy (half-population) time at a quarter period: $t_{1/2}=T_{\mathrm{TL}}/4 \approx 51~\mu\mathrm{s}$, so $T_{\mathrm{TL}}\approx 204~\mu\mathrm{s}$.
In the $100~\mu\mathrm{K}$ simulations of Fig.~\ref{fig:two_ion_crystal_coupling_comparison}, the crossover occurs appreciably faster due to the occupation-enhanced matrix element $\propto \sqrt{n\,m\,(p+1)}$ discussed above.

To decide when a nonlinear interaction is dynamically significant, we define a \textbf{resonance criterion} based on the TL amplitude.
We say the system is in the nonlinear resonance regime when $S_{nmp}^{\mathrm{TL}}\ge 0.1$.
This corresponds to at least $10\%$ population transfer between the lowest-energy coupled Fock states and indicates that nonlinear coupling remains relevant even near the motional ground state.
When no ambiguity arises, we omit the subscript $nmp$ for brevity.

\section{Results}
\label{sec:results}	

This section quantifies the impact of nonlinear motional mode coupling (\co) across representative trapped-ion architectures using the two-level (TL) metric of Sec.~\ref{subsec:TLS}. 
We first validate the TL reduction in a controlled setting: a \MS\ gate mediated by the two-ion breathing mode, where a single dominant two-mode process allows direct comparison among classical MD, classical/quantum reduced models, and the TL predictions (cf.\ Fig.~\ref{fig:two_ion_crystal_coupling_comparison}). 
We then broaden scope from two ions to larger crystals by \emph{enumerating} near-resonant processes: for each configuration we identify triads via a detuning filter $|\Delta_{nmp}|\le\Delta_\mathrm{cut}$, threshold on the cubic tensor magnitude, and apply the TL amplitude test $S^{\mathrm{TL}}\!\ge 0.1$ to label dynamically relevant couplings. 
This pipeline lets us study how the \emph{number and strength of resonances} scale with system size, mode density, and geometry (linear chains, 2D rf arrays, Penning arrays). 
Finally, we connect these counts and TL rates to \emph{gate robustness}: off resonance, nonlinear terms induce AC-Stark/Kerr shifts $\propto g^{2}/\Delta$ that distort \MS\ phase-space loops; on resonance, direct mixing at rate $g=\epsilon_0|C^{\mathrm{RWA}}_{nmp}|$ competes with the \MS\ force. 
We also examine practical mitigations (e.g., multi-loop gates with bus displacement $d=1/\sqrt{2k}$) and compare how architectures differ in their susceptibility due to the density of near-resonant \co. 
Unless otherwise noted, we use the mixed-state fidelity of the pure target spin state and the reduced spin state density matrix at the end of the gate to quantify gate performance \cite{Jozsa1994}. 

\subsection{Why nonlinear coupling matters for \MS\ gates}
\label{subsec:MS_CO_intuition}
Consider an \MS\ gate driven at detuning $\delta_{\mathrm{gate}}$ with first sideband Rabi rate $\eta\,\Omega_r$, together with an off-resonant three-phonon term given in Eq.~\eqref{eq:mode_Hamiltonian_nonlinear} and quantized in Eq.~\eqref{eq:quantum_nonlinear_hamiltonian_back}.
Suppose the gate is mediated by mode $p$ (frequency $\omega_p$) and the three-phonon term couples modes $n$, $m$, and $p$ (frequencies $\omega_n$, $\omega_m$, and $\omega_p$) with nonlinear detuning $\Delta_{nmp}=\omega_p-\omega_n-\omega_m$.
Define
\begin{equation}
  g=\epsilon_0\,\big|C^{\mathrm{RWA}}_{nmp}\big|,
  \qquad
  \Delta=\Delta_{nmp},
\end{equation}
with all frequencies in the same units as Appendix~\ref{app:quantization}.

For $\lvert\Delta\rvert \gg 1/T_{\mathrm{gate}}$ (with $T_{\mathrm{gate}}$ the gate time), time-averaging (Magnus/Schrieffer--Wolff) yields an effective AC--Stark/Kerr shift of the gate mode,
\begin{equation}
  \delta\omega_p \sim \frac{g^2}{\Delta},
\end{equation}
up to constants of order unity from commutators and occupation-number factors from the spectator modes.
Therefore, the effective detuning seen by $p$ during the gate becomes $\delta_\mathrm{eff}=\delta_{\mathrm{gate}}-\delta\omega_p$, resulting in loss of loop closure. 
A significant distortion of the phase-space loop occurs when this nonlinear shift competes with the \MS\ drive.
Using the standard one-loop closure for a maximally entangling \MS\ gate,
\begin{equation}
  \delta_{\mathrm{gate}}\simeq \eta\,\Omega_r,
\end{equation}
the \emph{off-resonant danger criterion} is
\begin{equation}
  \frac{g^2}{\lvert\Delta\rvert} \gtrsim \eta\,\Omega_r,
\end{equation}
again up to $\mathcal{O}(1)$ factors.
In terms of the gate time,
\begin{equation}
  T_{\mathrm{gate}}=\frac{2\pi}{\delta_{\mathrm{gate}}}
  \qquad\Longrightarrow\qquad
  T_{\mathrm{gate}} \gtrsim \frac{\lvert\Delta\rvert}{g^2}.
\end{equation}
Equivalently, dangerous gate times become more prevalent as $\lvert\Delta\rvert$ decreases (closer to a three-phonon resonance) or as $g$ increases (stronger nonlinearity).
This region is indicated in Fig.~\ref{fig:n2_motional_detuning_vs_gate_time}(b), where residual spin--motion entanglement persists at the end of the gate even off resonance.

\emph{On resonance} ($\lvert\Delta\rvert \lesssim 1/T_{\mathrm{gate}}$) the direct mixing at rate $g$ competes with the \MS\ force.
A simple rate comparison applies:
\begin{equation}
  g \gtrsim \eta\,\Omega_r,
\end{equation}
which is restricted to the immediate vicinity of the three-phonon resonance.
This region is indicated in Fig.~\ref{fig:n2_motional_detuning_vs_gate_time}(c), where the spectator mode is excited by the end of the gate.

In summary, off resonance, fidelity impacts occur due to effective shifts in the addressed-mode frequency that scale as $g^2/\Delta$, while on resonance, direct energy exchange occurs at rate $g$---leading to the largest fidelity impacts.
These effects are also enhanced when higher Fock states are populated, as discussed in Sec.~\ref{subsec:TLS}.

\subsection{How many near-resonant couplings should we expect as ion-number grows?}
\label{subsec:scaling_argument} 
To build intuition before presenting numerical results, we give a simplified estimate for how the \emph{number of near-resonant three-mode couplings} scales with the number of ions $N$.

Assume, for heuristic purposes, that normal-mode frequencies are i.i.d.\ uniform on $[0,1]$.
We consider triads $(n,m,p)$ that approximately satisfy the sum rule $\omega_n+\omega_m\approx \omega_p$, i.e.\ whose \emph{nonlinear detuning}
\[
\Delta_{nmp}\equiv \omega_p-\omega_n-\omega_m
\]
lies within a small \emph{detuning window} of width $\Delta_{\mathrm{win}}$:
\[
|\Delta_{nmp}| \le \Delta_{\mathrm{win}}.
\]
(For consistency with Appendix~\ref{app:quantization}, one may take $\Delta_{\mathrm{win}}\equiv \Delta_{\mathrm{cut}}$; we use ``cut'' to emphasize that it is a tunable cutoff parameter). 

Since $\omega_n$ and $\omega_m$ are uniform, their sum $S=\omega_n+\omega_m$ has a triangular pdf on $[0,2]$.
For small $\Delta_{\mathrm{win}}$, the probability that $S$ falls within a width-$2\Delta_{\mathrm{win}}$ window around a given $\omega_p$ scales linearly with the window:
\[
\Pr\big(|S-\omega_p|\le \Delta_{\mathrm{win}}\big)\;\propto\; \Delta_{\mathrm{win}}
\]

There are $3N$ modes in total.
Counting distinct triads with all indices different gives, up to constants,
\[
N_\text{triads} \;\sim\; \binom{3N}{2}\,(3N-2) \;\propto\; N^3.
\]
Multiplying by the window probability yields the expected number of near-resonant three-mode couplings
\begin{equation}
  N_{\mathrm{res}}^{(3)} \;\propto\; N^3\,\Delta_{\mathrm{win}}.
  \label{eq:resonance_scaling}
\end{equation}

\paragraph*{Remarks.}
(i) An analogous estimate for \emph{two-mode} processes (with operator structure $a_n^2 a_p^\dagger$) counts pairs $(n,p)$, giving $N_{\mathrm{res}}^{(2)} \propto N^2\,\Delta_{\mathrm{win}}$.
Thus three-mode opportunities grow parametrically faster with system size.  
(ii) The mapping to physical units is $\Delta_{\mathrm{phys}}=\Delta_{\mathrm{win}}\,\omega_0$ (rad/s) since all frequencies here are normalized by $\omega_0$.  
(iii) This toy model ignores geometry, branch structure, mode participation factors, and tensor selection rules from $T^{XYZ}_{nmp}$, all of which can suppress or enhance specific couplings.
The detailed, geometry-dependent deviations from Eq.~\eqref{eq:resonance_scaling} are quantified in the following sections using the full RWA filter and TL screening.

\subsection{Two-Ion MS gate}
\label{sec:n2_ms_gate}

\begin{figure*}[t]
  \includegraphics[width=\linewidth]{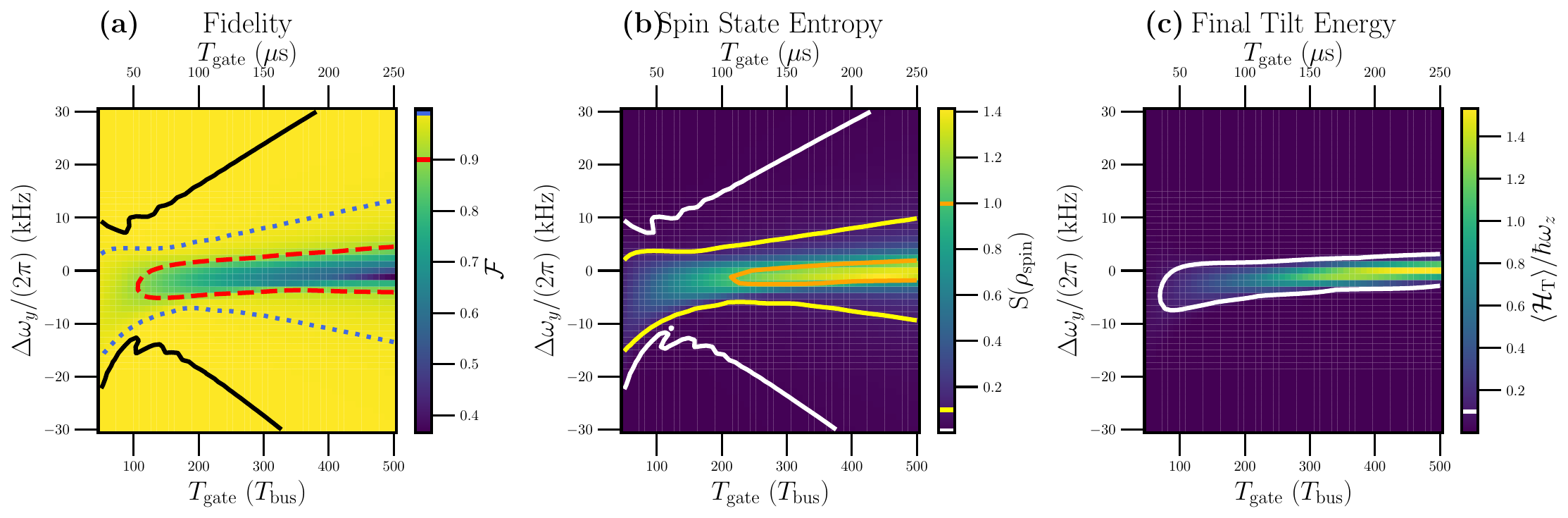}
  \caption{Fidelity of a \MSGATE\ nonlinear motional mode coupling (\co) in a two-ion crystal.
  The breathing mode serves as the \emph{bus}, and the trap is tuned so that the tilt mode is nearly resonant with the bus via the three-phonon (three-wave) interaction ($\omega_\mathrm{B}\approx 2\omega_\mathrm{T}$).
  We plot $\Delta\omega_y\equiv\omega_y-\omega_{y,\mathrm{res}}$, where $\omega_{y,\mathrm{res}}=(\sqrt{7}/2)\,\omega_z$ [Eq.~(\ref{eq:two_ion_resonance})]; this tunes the nonlinear detuning $\Delta_{\mathrm{TTB}}\equiv\omega_\mathrm{B}-2\omega_\mathrm{T}$ through resonance ($\Delta_{\mathrm{TTB}}=0$).
  The \emph{gate detuning} is set by $\delta_\mathrm{gate}=2\pi/T_{\mathrm{gate}}$ for a single-loop gate ($k=1$).
  \textbf{(a)} Bell-state fidelity vs.\ $T_\mathrm{gate}$ and $\Delta\omega_y\equiv\omega_y-\omega_{y,\mathrm{res}}$; contours at $\mathcal{F}=\{0.90,0.99,0.999\}$.
  \textbf{(b)} Spin-subsystem entropy $S_2(\rho_\mathrm{spin})$ for the same scan; contours at $\{0.01,0.1,1\}$.
  Off resonance ($|\Delta_{\mathrm{TTB}}|\gg g$ with $g=\epsilon_0|C^{\mathrm{RWA}}_{\mathrm{TTB}}|$), the dominant effect is a dispersive shift $\propto g^2/|\Delta_{\mathrm{TTB}}|$ that changes the bus-mode frequency and produces gate over- or under-rotations.
  \text{(c)} Final spectator (tilt) energy $\langle \mathcal{H}_T\rangle/\hbar\omega_z$; higher values coincide with the low-fidelity band near $\Delta_{TTB}\approx 0$; contour at $\langle \mathcal{H}_\text{T}\rangle/\hbar\omega_z=.1$ (white solid).
  Here, the \co is resonant, and there is direct competition between the \MS\ drive at longer $T_\text{gate}$ and \co ($g\sim\eta\Omega_r$). 
  }
  \label{fig:n2_motional_detuning_vs_gate_time}
\end{figure*}

\begin{figure*}[t]
  \includegraphics[width=\linewidth]{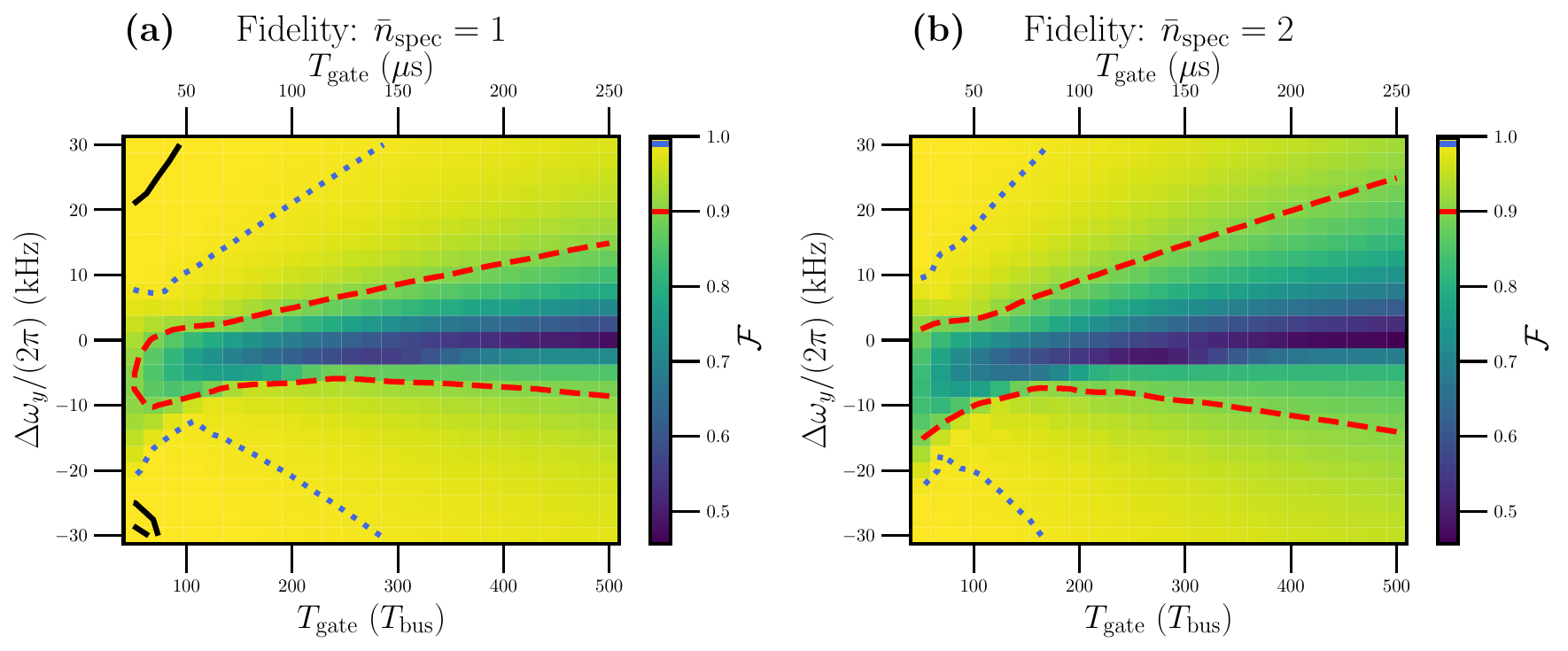}
  \caption{Fidelity of a two-ion \MSGATE\ with nonlinear mode coupling when the spectator (tilt) mode is thermally occupied.
  The bus (breathing) mode is initialized in the ground state and the spins in $\ket{gg}$.
  \textbf{(a)} Spectator initialized in a thermal state with mean occupation $\bar n_{\mathrm{spec}}=1$; the scan over nonlinear detuning and gate time $T_\mathrm{gate}$ matches Fig.~\ref{fig:n2_motional_detuning_vs_gate_time}.
  \textbf{(b)} Same scan with $\bar n_{\mathrm{spec}}=2$.
  Relative to the ground-state case, the parameter region with degraded fidelity is substantially larger even though the bus begins in $\ket{0}$.
  Contours are shown at 0.90 (red dashed), 0.99 (blue dotted), and 0.999 (black solid).}
  \label{fig:n2_motional_detuning_vs_gate_time_thermal}
\end{figure*}

In this section, we simulate the \MSGATE\ in a two-ion system to demonstrate the impact of \co\ on gate fidelity in a physical setting.
We find that \co\ can significantly reduce the fidelity even when the motion is initialized in the ground state, but only in the vicinity of a nonlinear resonance (here, $\Delta_{\mathrm{TTB}}=0$).
Related analytic treatments have studied \MSGATE\ gate errors induced by quartic trap anharmonicity, whereas here we focus on nonlinear interactions between motional modes during entangling gate operations~\cite{Sutherland2022}.

For these simulations the higher-frequency breathing mode mediates the entangling interaction (the \emph{bus}), and the lower-frequency tilt mode is a \emph{spectator}, as depicted in Fig.~\ref{fig:coupling_type_diagram}.
We use \texttt{QuTiP}~\cite{Johansson2013}, with the motional nonlinear Hamiltonian given by Eq.~\eqref{eq:2-ion_interaction_hamiltonian} and the \MS\ gate Hamiltonian of Ref.~\cite{Srensen2000}.
Simulation details are reported in Appendix~\ref{app:simulations}.

In a rotating frame with respect to both spin and motion, the total Hamiltonian is
\begin{equation}
  \mathcal{H}_\text{total}
  \;=\;
  \mathcal{H}_\text{2-ion}^\text{RWA}
  \;+\;
  \mathcal{H}_\text{gate},
  \label{eq:2-ion_ms_hamiltonian}
\end{equation}
where the \MS\ drive is written in quadrature form as
\begin{equation}
  \mathcal{H}_\text{gate}
  \;=\;
  -\sqrt{2}\,\eta\,\Omega_r\,J_y\,
  \Big[\,x\,\cos(\delta_\text{gate} t)\;+\;p\,\sin(\delta_\text{gate} t)\,\Big].
  \label{eq:ms_gate_hamiltonian}
\end{equation}
Here $\eta$ is the Lamb-Dicke parameter, $\Omega_r$ is the per-ion, per-tone carrier Rabi rate, $J_y$ is the collective spin operator, and $\delta_\text{gate}$ is the gate detuning from the motional sidebands $(\omega_0\pm\omega_\mathrm{B})$.  
The bus-mode quadratures and collective spin operator are
\begin{gather}
  x=\frac{1}{\sqrt{2}}\big(a_\mathrm{B}+a_\mathrm{B}^\dagger\big),\qquad
  p=\frac{-i}{\sqrt{2}}\big(a_\mathrm{B}-a_\mathrm{B}^\dagger\big),\notag\\
  J_y=\frac{1}{2}\sum_{i=1}^{N}\sigma_y^{(i)},
\end{gather}
so that $[x,p]=i$ and $a_\mathrm{B} e^{-i\delta_\text{gate} t}+a_\mathrm{B}^\dagger e^{+i\delta_\text{gate} t}
=\sqrt{2}\big[x\cos(\delta_\text{gate} t)+p\sin(\delta_\text{gate} t)\big]$.
This normalization is consistent with Appendix~\ref{app:quantization}, where $x=Q/\epsilon_0$ and $p=P/\epsilon_0$ for the bus mode.
We neglect far-off-resonant terms (including the carrier and counter-rotating components) under the usual Lamb-Dicke and rotating-wave approximations.

The \MSGATE\ was simulated with the same parameters as Sec.~\ref{subsec:two_ion_crystal}, using $^{171}\mathrm{Yb}^+$ as a concrete example.
The system was prepared in the joint motional ground state of the tilt and breathing modes and the ground spin state of the qubits. 
We scanned the gate time $T_\mathrm{gate}$ and the radial trapping frequency $\omega_y$ (the $y$-axis mode).
The target Bell state was $\ket{\psi_\mathrm{target}}=(\ket{gg}-i\ket{ee})/\sqrt{2}$, and $\mathcal{F}$ denotes the fidelity between this target state and the reduced two-spin state at the end of the gate, obtained by tracing out the motional modes~\cite{Jozsa1994}. 

Figure~\ref{fig:n2_motional_detuning_vs_gate_time}\,(a) shows fidelity vs.\ $T_\mathrm{gate}$ and the frequency offset $\Delta\omega_y\equiv \omega_y-\omega_{y,\mathrm{res}}$, where $\omega_{y,\mathrm{res}}=(\sqrt{7}/2)\,\omega_z$ from Eq.~\eqref{eq:two_ion_resonance}.
The secondary $x$-axis reports $T_\mathrm{gate}$ in $\mu$s; the primary axis uses $T_\mathrm{gate}/T_\mathrm{bus}$.
Contours indicate fidelities $0.90$ (red dashed), $0.99$ (blue dotted), and $0.999$ (black solid).
The lowest fidelities occur near the nonlinear resonance ($\Delta_{\mathrm{TTB}}=0$) and for longer gates.
Fidelity improves as $T_\mathrm{gate}$ decreases (less time for nonlinear exchange), and the region with fidelity $<0.99$ \emph{expands} as $T_\mathrm{gate}$ increases, consistent with the off-resonant criterion $T_\mathrm{gate}\gtrsim |\Delta|/g^2$ from Sec.~\ref{subsec:MS_CO_intuition}.
For the scan shown, all $\Delta\omega_y$ points correspond to small but nonzero TL amplitudes $S^{\mathrm{TL}}\!\gtrsim\!10^{-3}$, with the strongest impacts when $S^{\mathrm{TL}}\!\ge\!0.1$—the resonance criterion of Sec.~\ref{subsec:TLS}.

The fidelity map need not be symmetric about $\Delta\omega_y=0$: in the dispersive regime the three-phonon term produces a \emph{signed} AC--Stark/Kerr shift of the bus, $\delta\omega_\mathrm{B} = (g^2/\Delta)(2+4\bar n_\mathrm{spec})$, so the effective detuning is $\delta_\mathrm{eff}=\delta_\mathrm{gate}-\delta\omega_\mathrm{B}$ and the \MSGATE\ loop is under- or over-detuned on opposite sides of the resonance. 
With fixed gate time $T_\mathrm{gate}=2\pi/|\delta_\mathrm{gate}|$, a positive (negative) $\delta\omega_\mathrm{B}$ decreases (increases) $|\delta_\mathrm{eff}|$ and the phase-space loop area, yielding an over- (under-) rotation on opposite sides of the resonance.

Figure~\ref{fig:n2_motional_detuning_vs_gate_time}\,(b) plots the von Neumann entropy $S(\rho_\mathrm{spin})$ of the reduced spin state at the end of the gate, with contours at $1$ (orange), $0.1$ (yellow), and $0.01$ (white).
These contours track the fidelity map in panel (a), indicating strong anti-correlation between gate infidelity and residual spin-motion entanglement.
We use the base two logarithm, so for two qubits the maximum possible spin entropy is $\log_2 4  = 2$ for the maximally mixed state.
The worst-fidelity regions approach this bound near $\Delta_{\mathrm{TTB}}=0$ and longer $T_\mathrm{gate}$.
Away from resonance, the band of reduced but non-negligible fidelity that \emph{fans out} with increasing $T_\mathrm{gate}$ is consistent with AC-Stark/Kerr shifts $\delta\omega_p\!\sim\!g^2/\Delta$ that spoil loop closure (Sec.~\ref{subsec:MS_CO_intuition}).

Figure~\ref{fig:n2_motional_detuning_vs_gate_time}\,(c) shows the final energy of the spectator (tilt) mode in units of $\hbar\omega_z$.
Ideally the spectator remains at its ground energy, but here its final energy correlates with the fidelity map in panel (a).
This is consistent with \co\ between the tilt and breathing modes causing non-ideal \MS\ trajectories: off resonance, dispersive shifts dephase the bus; on resonance, direct exchange at rate $g$ populates the spectator by the end of the gate.
The largest effects occur where $S^{\mathrm{TL}}\!\ge\!0.1$ (Sec.~\ref{subsec:TLS}).

Figure~\ref{fig:n2_motional_detuning_vs_gate_time_thermal}(a) shows the fidelity landscape when the spectator (tilt) mode is initialized thermally with $\bar n_{\mathrm{spec}}=1$.
The same $(\Delta\omega_y,\,T_\mathrm{gate})$ scan as in Fig.~\ref{fig:n2_motional_detuning_vs_gate_time}(a) is used.
Even for short gate times and for $\Delta\omega_y<0$ (below the TTB resonance), regions with fidelity $<0.9$ appear.

Increasing the spectator occupation to $\bar n_{\mathrm{spec}}=2$ [Fig.~\ref{fig:n2_motional_detuning_vs_gate_time_thermal}(b)] expands the degraded-fidelity region markedly.
This trend is consistent with the TL matrix-element scaling from Sec.~\ref{subsec:TLS}, where the effective coupling grows as $\epsilon_0 |C^{\mathrm{RWA}}_{nmp}|\sqrt{n\,m\,(p+1)}$.
In particular, thermal weight at higher $n$ in the spectator enhances both on-resonant exchange (rate $\propto g$) and off-resonant AC-Stark/Kerr shifts ($\propto g^2/|\Delta|$), spoiling phase-space loop closure even when the bus begins in $\ket{0}$.

Taken together with the ground-state results, these data show that \co\ can significantly impact a two-ion \MSGATE\ near a nonlinear resonance and that the impact is exacerbated by spectator thermal occupation.
In larger systems where cooling of modes orthogonal to the bus is less efficient, elevated spectator occupations can therefore widen the “danger region.”
Conversely, for small systems (e.g., two ions) and for the parameters used here, \co\ is unlikely to be limiting if the trap is detuned by a few kHz away from the TTB resonance; if the resonance is encountered, a small frequency retuning avoids it.
\label{sec:two_ion_ms_gate}

\subsection{\MSGATE robustness under three-mode spectator coupling}
\label{subsec:ms_gate_simulations}

\begin{figure}[htpb]
  \includegraphics[width=\linewidth]{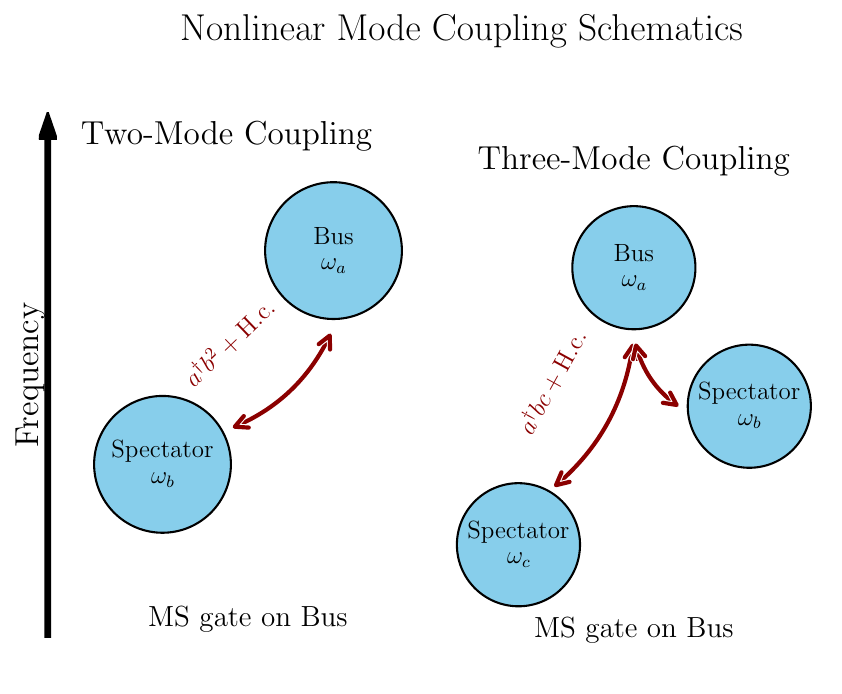}
  \caption{
	Schematics of nonlinear mode coupling: two-mode (left) and three-mode (right).
  	In our simulations the highest-frequency normal mode acts as the \emph{bus} (frequency $\omega_a$) addressed by the gate, while the lower-frequency mode(s) serve as \emph{spectators} (frequencies $\omega_b$, $\omega_c$).
  	Two-mode (parametric) coupling is resonant when the spectator is near half the bus frequency, $2\omega_b \approx \omega_a$; three-mode (sum-frequency) coupling is resonant when two modes nearly sum to a third, $\omega_b + \omega_c \approx \omega_a$.}
  \label{fig:coupling_type_diagram}
\end{figure}

\begin{figure*}[t]
  \includegraphics[width=1.0\linewidth]{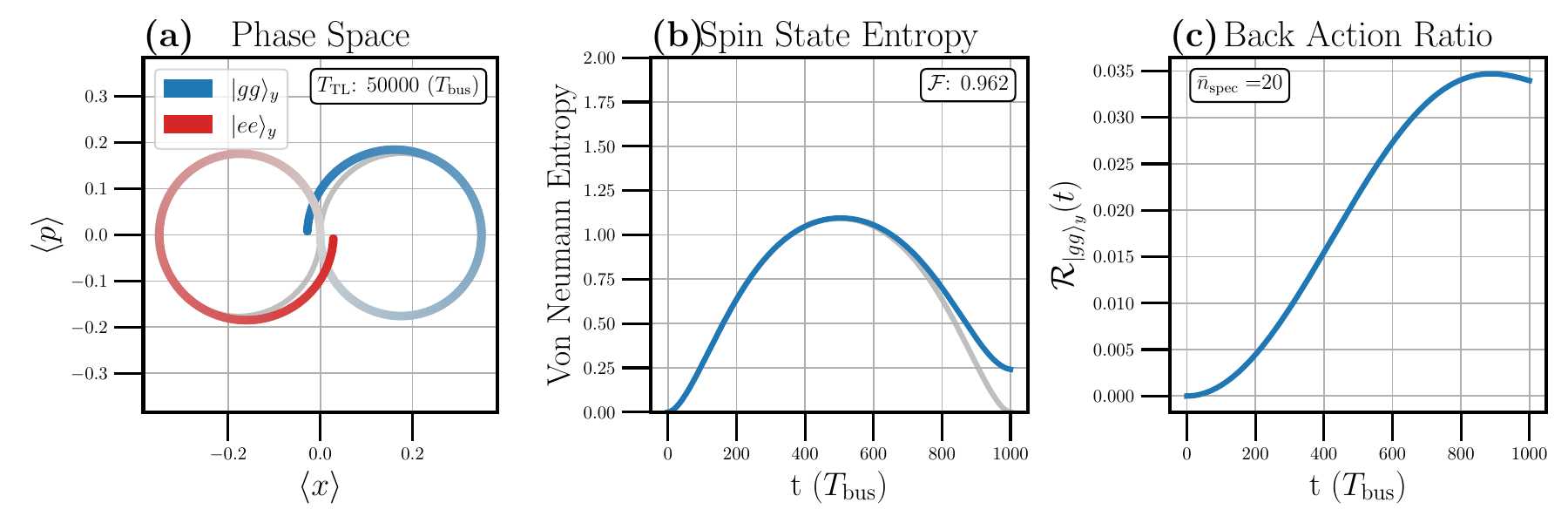}
  \caption{
    Three-mode (sum-frequency) \co{} simulated during an \MS{} gate with mode ratios
    $\omega_a:\omega_b:\omega_c=4:3:1$.
    We normalize by the bus frequency $\omega_a$ so $\omega_b=0.75$ and $\omega_c=0.25$ (with $\omega_b+\omega_c=\omega_a$).
    For $\omega_a/2\pi=2~\mathrm{MHz}$, this corresponds to spectators at $1.5~\mathrm{MHz}$ and $0.5~\mathrm{MHz}$.
    The two-level period is $T_\mathrm{TL}=50{,}000\,T_\mathrm{bus}$ and the gate time is $T_\mathrm{gate}=1{,}000\,T_\mathrm{bus}$, where $T_\mathrm{bus}=2\pi/\omega_a$.
    The spins start in $\ket{gg}$.
    The bus and the intermediate-frequency spectator begin in their motional ground states, while the lowest-frequency spectator is thermal with mean occupation $\bar n_\mathrm{spec}=20$.
    \textbf{(a)} Phase-space trajectories of the $\ket{gg}_y$ and $\ket{ee}_y$ spin branches during the gate.
    Ideal (no \co) trajectories are gray; simulated trajectories are colored from start (light) to end (dark): $\ket{gg}_y$ (blue) and $\ket{ee}_y$ (red).
    Deviations from the ideal circles and lack of loop closure at $T_\mathrm{gate}$ indicate residual spin–motion entanglement.
    \textbf{(b)} Base-2 von Neumann entropy $S_2(\rho_\mathrm{spin})$ of the reduced spin state during the gate (ideal in gray; nonlinear in blue), remaining nonzero at $T_\mathrm{gate}$.
    \textbf{(c)} Nonlinear back-action ratio $\mathcal{R}_{ee_y}(t)$ [Eq.~\eqref{eq:backaction_def}], which compares the instantaneous nonlinear drive on the bus to the \MS{} drive.
    $\mathcal{R}_{ee_y}(t)$ grows during the gate and peaks near the end, indicating increasingly significant nonlinear forcing relative to the \MS{} drive.
  }
  \label{fig:ms_gate_populations_and_energies_3_modes}
\end{figure*}

\begin{figure}[htbp]
  \includegraphics[width=1.0\linewidth]{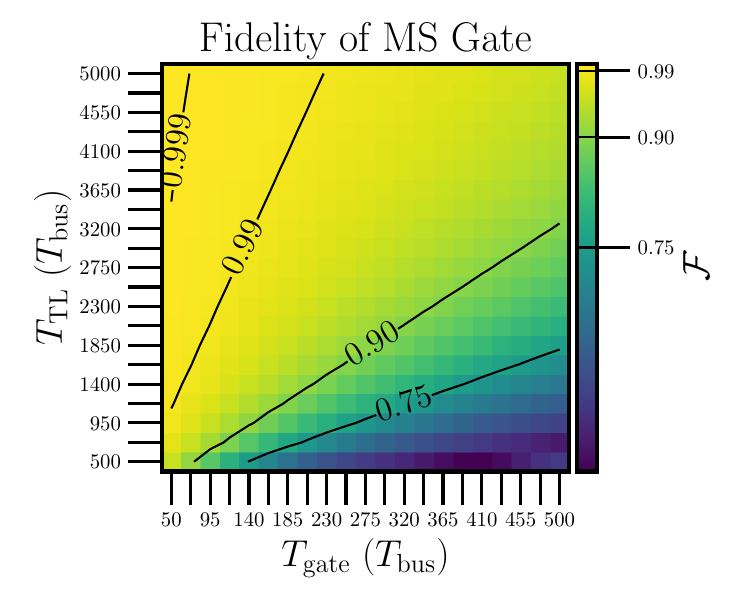}
  \caption{
    An on-resonant three-mode (sum-frequency) \co{} interaction with the same frequency ratios as Fig.~\ref{fig:ms_gate_populations_and_energies_3_modes} is simulated over a grid of two-level (TL) period and gate time.
    We take $\omega_a:\omega_b:\omega_c=4:3:1$ and normalize by the bus frequency ($\omega_a=1$), so $T_\mathrm{bus}=2\pi/\omega_a$.
    The TL period is scanned as $T_\mathrm{TL}\in[500,\,5000]\times T_\mathrm{bus}$ and the gate time as $T_\mathrm{gate}\in[50,\,500]\times T_\mathrm{bus}$.
    Smaller $T_\mathrm{TL}$ corresponds to stronger nonlinear coupling, and in this regime the \MS{}-gate fidelity is significantly reduced, especially for longer $T_\mathrm{gate}$.
  }
  \label{fig:ms_gate_two_level_osc_v_gate_time}
\end{figure}

\begin{figure}[htbp]
  \includegraphics[width=1.0\linewidth]{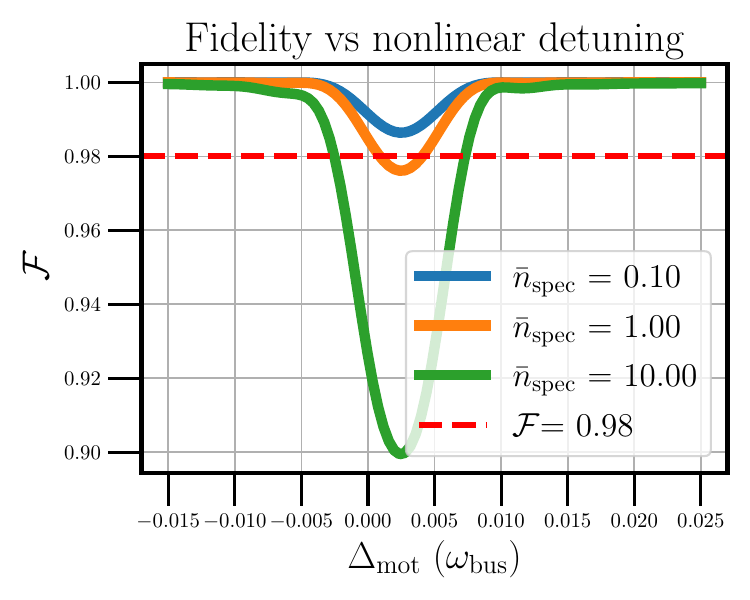}
  \caption{
    MS gate fidelity versus motional detuning \(\Delta_{\mathrm{mot}}\), reported in units of the bus-mode frequency \(\omega_{\mathrm{bus}}\equiv\omega_a\).
    The MS gate is applied to the highest-frequency mode in a three-mode coupling, with the lowest-frequency spectator initialized in a thermal state with \(\bar n_{\mathrm{spec}}=0.1,\,1,\,10\) and the other two modes in their ground states.
    Mode-frequency ratios are chosen to be representative of planar-ion couplings, \(\omega_c:\omega_b:\omega_a=1:3:4\).
    These occupations correspond to Doppler cooling (\(\bar n\sim10\)), sideband cooling (\(\bar n\sim1\)), and near ground-state cooling (\(\bar n\sim0.1\)) for a spectator at \(\omega_{\mathrm{spec}}=2\pi\times1~\mathrm{MHz}\).
    The two-level oscillation period is fixed at \(T_\text{TL}=5000\,T_\text{bus}\), and the gate time at \(T_\text{gate}=200\,T_\text{bus}\).
    The red dashed contour marks 0.98 fidelity, indicating that Doppler-cooled spectator modes can significantly impact gate performance.
  }
  \label{fig:ms_gate_temp_v_resonance_scan}
\end{figure}

\begin{figure}[htbp]
  \includegraphics[width=1.0\linewidth]{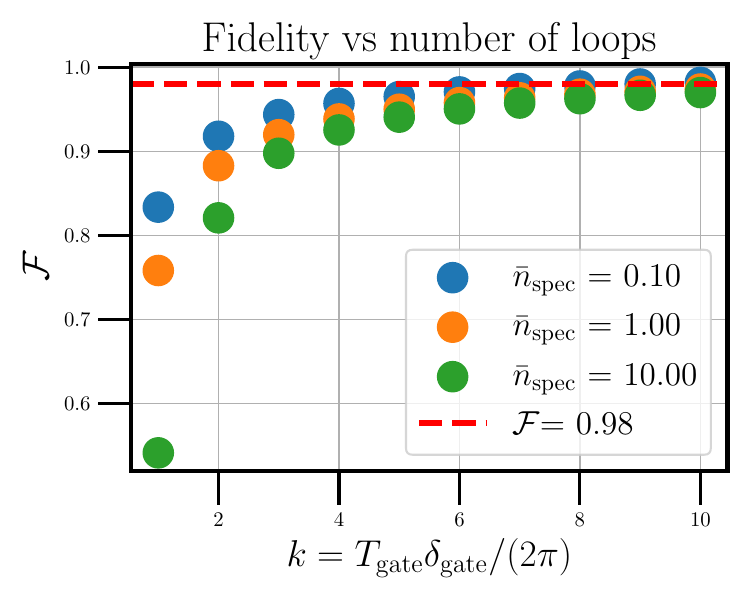}
  \caption{
    Gate fidelity as a function of the number of loops in phase space, $k$, for a fixed gate time $\text{T}_\text{gate} = 500 \times \text{T}_\text{bus}$ and two-level phase period $\text{T}_\text{TL} = 1000 \times T_\text{bus}$.
    Results are shown for three different thermal occupations of the lowest-frequency spectator mode in a three-mode coupling: $\bar{n}_\text{spec}=0.1$, $1$, and $10$.
    The number of loops determines the bus-mode displacement $d = 1/\sqrt{2k}$, with larger $k$ corresponding to smaller displacements and faster phase-space traversal.
    As $k$ increases, requiring higher Rabi frequency $\Omega$ and detuning $\delta_\text{gate}$, the gate becomes more robust to \co, and fidelity improves accordingly.
  }
  \label{fig:ms_gate_fidelity_vs_num_loops}
\end{figure}

In the previous section we studied the \MSGATE\ in a two-ion crystal, where only \emph{two-mode} nonlinear coupling between the radial tilt and axial breathing modes is present.
As system size grows, \emph{three-mode} processes become prevalent (Sec.~\ref{subsec:scaling_argument}), and they are central to the behavior of long chains and 2D ion crystals (Secs.~\ref{subsec:linear_chain_mode_coupling}, \ref{subsec:planar_ion_crystals}).
Here we extend the analysis to include both two- and three-mode \co\ and quantify their impact on \MSGATE\ fidelity in more general settings.

We simulate two- and three-mode interactions in which the highest-frequency mode serves as the \emph{bus} and mediates entanglement between two qubits, while the other one or two modes act as \emph{spectators} (schematics in Fig.~\ref{fig:coupling_type_diagram}).
Parameters are chosen to be representative of 2D crystals rather than tied to a specific device, and are consistent with the ranges identified in Secs.~\ref{subsec:linear_chain_mode_coupling} and \ref{subsec:planar_ion_crystals}.

To illustrate how \co\ affects the gate, we present four results:
\begin{enumerate}
  \item \textbf{Gate trajectories and diagnostics.}
        Phase-space trajectories of the two spin branches, the base-2 von Neumann spin entropy $S_2(\rho_\mathrm{spin})$, and a nonlinear back-action ratio $\mathcal{R}$ (Eq.~\ref{eq:backaction_def}) that compares the instantaneous nonlinear drive on the bus to the \MS\ force (Fig.~\ref{fig:ms_gate_populations_and_energies_3_modes}).
  \item \textbf{Timescale interplay.}
        A scan over the gate time $T_\mathrm{gate}$ and the two-level (TL) oscillation period $T_\mathrm{TL}$ showing where \co\ degrades fidelity (Fig.~\ref{fig:ms_gate_two_level_osc_v_gate_time}).
  \item \textbf{Detuning and temperature.}
        A scan of gate fidelity versus the \emph{nonlinear detuning} $\Delta_\text{mot}$ (Eq.~\ref{eq:three_mode_frequencies}) and spectator thermal occupation $\bar n_\mathrm{spec}$ (Fig.~\ref{fig:ms_gate_temp_v_resonance_scan}).
  \item \textbf{Multi-loop mitigation.}
        A comparison of gate performance as the number of phase-space loops $k$ increases at fixed $T_\mathrm{gate}$ (Fig.~\ref{fig:ms_gate_fidelity_vs_num_loops}); here the maximum bus displacement scales as $d=1/\sqrt{2k}$.
\end{enumerate}

\paragraph*{Gate trajectories and diagnostics.}
In Fig.~\ref{fig:ms_gate_populations_and_energies_3_modes}, we show \MS\ gate performance when a low-frequency spectator is thermally occupied (\(\bar n_\mathrm{spec}=20\)) and resonantly coupled to the bus via an intermediate-frequency spectator (schematic in Fig.~\ref{fig:coupling_type_diagram}(b)).
Details are given in Appendix~\ref{app:simulations}.
Only the lowest-frequency spectator is thermally populated; the bus and the other spectator start in their motional ground states.
This scenario is natural in 2D crystals, where low-frequency radial modes are typically Doppler cooled, while higher-frequency axial modes used for quantum operations can be cooled near the ground state via resolved-sideband or electromagnetically induced transparency (EIT) cooling~\cite{Jordan2019,Guo2024}.
Although two-qubit entangling gates have thus far only been demonstrated in small 2D crystals~\cite{Hou2024_2D}, with most 2D experiments focused on quantum simulation~\cite{Britton2012,Bohnet2016,Kiesenhofer2023,Guo2024}, the results here are relevant for future high-fidelity gates in larger arrays.

We consider mode frequencies in the ratio \(\omega_\mathrm{bus}\!\equiv\!\omega_a:\omega_{\mathrm{spec},1}\!\equiv\!\omega_b:\omega_{\mathrm{spec},2}\!\equiv\!\omega_c=4:3:1\).
For a \(2~\mathrm{MHz}\) bus, this corresponds to spectators at \(1.5~\mathrm{MHz}\) and \(0.5~\mathrm{MHz}\), representative of the couplings in Fig.~\ref{fig:mode_coupling_resonances_planar}(b).
For this demonstration we choose a three-mode coupling with TL period \(T_\mathrm{TL}=50{,}000\,T_\mathrm{bus}\) (with \(T_\mathrm{bus}=2\pi/\omega_\mathrm{bus}\)), typical of the axial–radial couplings summarized in Table~\ref{tab:axial_tl_stats} and Fig.~\ref{fig:mode_coupling_resonances_linear}(a,b).
The gate time is set to \(T_\mathrm{gate}=1{,}000\,T_\mathrm{bus}\), shorter than the multi-millisecond global gates often used in quantum simulations~\cite{Britton2012,Bohnet2016,Kiesenhofer2023,Guo2024}, and longer than two-qubit gates in modest 2D arrays recently demonstrated~\cite{Hou2024_2D}.

When all modes are cooled, the final spin populations closely match the target Bell state and the fidelity reaches \(\sim 0.999\).
With the Doppler-cooled spectator (\(\bar n_\mathrm{spec}=20\)), the fidelity drops to \(\sim 0.962\).
In Fig.~\ref{fig:ms_gate_populations_and_energies_3_modes}(a), the phase-space trajectories of the two spin branches deviate from the ideal circles and do not close at \(T_\mathrm{gate}\), indicating residual spin-motion entanglement (time progression is shown by color darkening; the ideal path is gray).
Figure~\ref{fig:ms_gate_populations_and_energies_3_modes}(b) plots the base-2 von Neumann entropy \(S_2(\rho_\mathrm{spin})\), which peaks mid-gate and remains nonzero at the end, confirming residual spin-motion entanglement.
As a diagnostic, Fig.~\ref{fig:ms_gate_populations_and_energies_3_modes}(c) introduces a nonlinear back-action ratio \(\mathcal{R}\) that compares the instantaneous nonlinear drive on the bus to the \MS\ force:
\begin{equation}
  \mathcal{R}_{ee_y}(t)
  = \frac{|g|}{\eta\,\Omega_r\,\sqrt{2}}\;
    \big|\langle c\,b\rangle_{ee_y}(t)\big|,
  \label{eq:backaction_def}
\end{equation}
where \(g=\epsilon_0|C^{\mathrm{RWA}}_{bca}|\) is the three-mode coupling strength, \(\eta\) is the bus Lamb–Dicke parameter, \(\Omega_r\) is the per-ion per-tone carrier Rabi rate, and \(\langle c\,b\rangle_{ee_y}(t)\) is the expectation value of the spectator-mode operator product conditioned on the \(\ket{ee}_y\) spin branch (see Appendix~\ref{app:simulations} for details).
When \(\mathcal{R}\ll 1\), the nonlinear forcing is negligible compared to the \MS\ drive; when \(\mathcal{R}\gtrsim 1\), the nonlinear forcing is comparable to or larger than the \MS\ drive and significant disruption of the gate is expected.    
The growth of \(\mathcal{R}\) during the gate indicates that the nonlinear forcing becomes increasingly significant relative to the \MS\ drive as the loop evolves.

\paragraph*{TL oscillation period vs.\ gate time.}
Figure~\ref{fig:ms_gate_two_level_osc_v_gate_time} explores how \MS\ fidelity depends on the relative timescales of the gate and the nonlinear mode dynamics by simulating a three-mode interaction with frequency ratios $\omega_a:\omega_b:\omega_c=4:3:1$.
We fix $\epsilon_0=10^{-3}$ and scan the two-level (TL) oscillation period $T_\mathrm{TL}\in[500,\,5000]\;T_\mathrm{bus}$ (by varying the nonlinear coupling strength $g$) and the gate time $T_\mathrm{gate}\in[50,\,500]\;T_\mathrm{bus}$, where $T_\mathrm{bus}=2\pi/\omega_a$.
This range is representative of strong radial-axial couplings identified in long chains near the zig-zag transition (Fig.~\ref{fig:linear_chain_scan_axial_confinement}).
For reference, the two-ion case in Fig.~\ref{fig:n2_motional_detuning_vs_gate_time} corresponds to $T_\mathrm{TL}\!\sim\!400\,T_\mathrm{bus}$, i.e., slightly stronger coupling than the lower edge of the present scan.
All modes are initialized in the joint motional ground state.
Gate operation is most disrupted when $T_\mathrm{TL}\approx T_\mathrm{gate}$; conversely, shorter gates suppress nonlinear effects.
Notably, for $T_\mathrm{gate}=50\,T_\mathrm{bus}$, fidelities $\mathcal{F}\gtrsim 0.99$ are maintained across almost the entire $T_\mathrm{TL}$ range considered.

\paragraph*{Nonlinear detuning and thermal effects.}
To quantify how far off resonance \co{} can impair gate performance, Fig.~\ref{fig:ms_gate_temp_v_resonance_scan} plots the \MS{} gate fidelity as a function of the nonlinear motional detuning
\begin{equation}
\boldsymbol{\omega}=(\omega_c,\omega_b,\omega_a)
=\big(\omega_\mathrm{split}-\Delta_\mathrm{mot},\ \omega_a-\omega_\mathrm{split},\ \omega_a\big),
\label{eq:three_mode_frequencies}
\end{equation}
where $\omega_\text{split}$ sets the frequency splitting between the higher two modes, and $\Delta_\mathrm{mot}$ is varied by changing only the lowest-frequency spectator while holding the bus and intermediate spectator fixed.
In the scan we vary $\omega_c$ by $\Delta_\mathrm{mot}$ while holding $\omega_a$ (bus) and $\omega_b$ (spectator) fixed.
We compare three spectator thermal occupations $\bar n_\mathrm{spec}\in\{0.1,1,10\}$ with parameters representative of long chains: $T_\mathrm{TL}=5{,}000\,T_\mathrm{bus}$ and $T_\mathrm{gate}=200\,T_\mathrm{bus}$.

Near resonance, even mild excitation ($\bar n_\mathrm{spec}=1$) lowers the fidelity below $0.99$.
For Doppler-like occupation ($\bar n_\mathrm{spec}=10$), the fidelity dips to $\sim 0.95$ at resonance and falls below $0.98$ for $\lvert\Delta_\mathrm{mot}\rvert\lesssim 0.01\,\omega_\mathrm{bus}$.
Thus, within a few kHz of nonlinear resonance (for $\omega_\mathrm{bus}/2\pi\sim\mathrm{MHz}$), \co{} can significantly degrade gate performance unless spectator modes are cooled close to the ground state.

These results underscore the importance of cooling low-frequency spectators—particularly in large crystals with many radial modes.
In Penning traps, axial (drumhead) modes are routinely ground-state cooled while radial modes often remain Doppler cooled with low frequencies ($\sim\!10$-$100~\mathrm{kHz}$) and large $\bar n$~\cite{Johnson2024}.
In rf 2D arrays, radial modes are typically only Doppler cooled as well, but their higher frequencies yield smaller thermal occupations~\cite{Kiesenhofer2023}.
See Fig.~\ref{fig:mode_coupling_resonances_planar} for a comparison of radial-mode frequencies in rf and Penning-trap 2D crystals.

\paragraph*{Phase–space loops and displacement control.}
We now examine how increasing the number of phase-space loops during the gate mitigates \co.
Both the laser detuning and the Rabi frequency are scaled with the loop count $k$ so that the total gate time is fixed at $T_\text{gate}=500\,T_\text{bus}$.
We simulate the three-mode sum-frequency scenario above with $T_\text{TL}=1{,}000\,T_\text{bus}$ (a worst-case strong coupling in our scans, see Fig.~\ref{fig:linear_chain_scan_axial_confinement}).
The lowest-frequency spectator starts thermally occupied, $\bar n_\text{spec}\in\{0.1,1,10\}$, while the bus and the other spectator begin in their ground states; Fig.~\ref{fig:ms_gate_fidelity_vs_num_loops} plots the resulting fidelity versus $k$.

For square-pulse \MS\ gates with fixed $T_\text{gate}$, the usual scaling
$\delta_{\mathrm{gate}}\!\propto\!k$ and $\Omega_r\!\propto\!\sqrt{k}$ implies that the bus displacement amplitude scales as
\[
d_\text{max}\;=\;\frac{\eta\,\Omega_r}{\delta_{\mathrm{gate}}}\;=\;\frac{1}{\sqrt{2k}}\;\times(\text{const.}),
\]
so increasing $k$ reduces the maximum excursion in phase space.
Consistent with this, the fidelity improves monotonically with $k$ across all $\bar n_\text{spec}$.
For example, at $\bar n_\text{spec}=10$ the fidelity rises from $\sim 0.55$ to $\sim 0.98$ by $k=5$, at the expense of a $\sqrt{5}\approx 2.24$ increase in $\Omega_r$.

This displacement-based view extends to shaped pulses: the relevant susceptibility to \co{} is governed by the largest coherent displacement of the bus mode during the waveform.
Reducing $d_\text{max}$ suppresses coupling to spectator modes (e.g., $\dot b=-i\,g\,a\,c^\dagger$) and the nonlinear back-action on the bus (captured by the ratio $\mathcal{R}$ defined in Eq~\eqref{eq:backaction_def}), thereby protecting loop closure.
Practically, bounding $d_\text{max}$—either by increasing $k$ in square-pulse gates or by explicitly constraining the waveform's mode-excitation provides a simple knob to keep \co-induced errors negligible even in large systems with many, partially cooled modes.

\paragraph*{Conclusion.}
Nonlinear motional coupling (\co) can disrupt \MS\ gate operation when a coupling is nearly resonant, when the gate duration overlaps the TL exchange timescale, or when spectator modes are thermally occupied.
Our results highlight the value of identifying strong \co\ resonances and improving cooling of low-frequency spectators, which otherwise broaden the detuning window for fidelity loss.
We further show that reducing the maximum bus-mode displacement—e.g., by increasing the number of phase-space loops $k$ or by using amplitude/phase shaping—mitigates \co\ and enables high fidelity even in the presence of nonlinear interactions.
These insights provide practical guidance for designing robust entangling gates in larger trapped-ion systems and for scaling toward high-fidelity quantum processors.

\subsection{\co in experimental linear Chains} 
\label{subsec:linear_chain_mode_coupling}

\begin{figure*}[t]
  \includegraphics[width=1.0\linewidth]{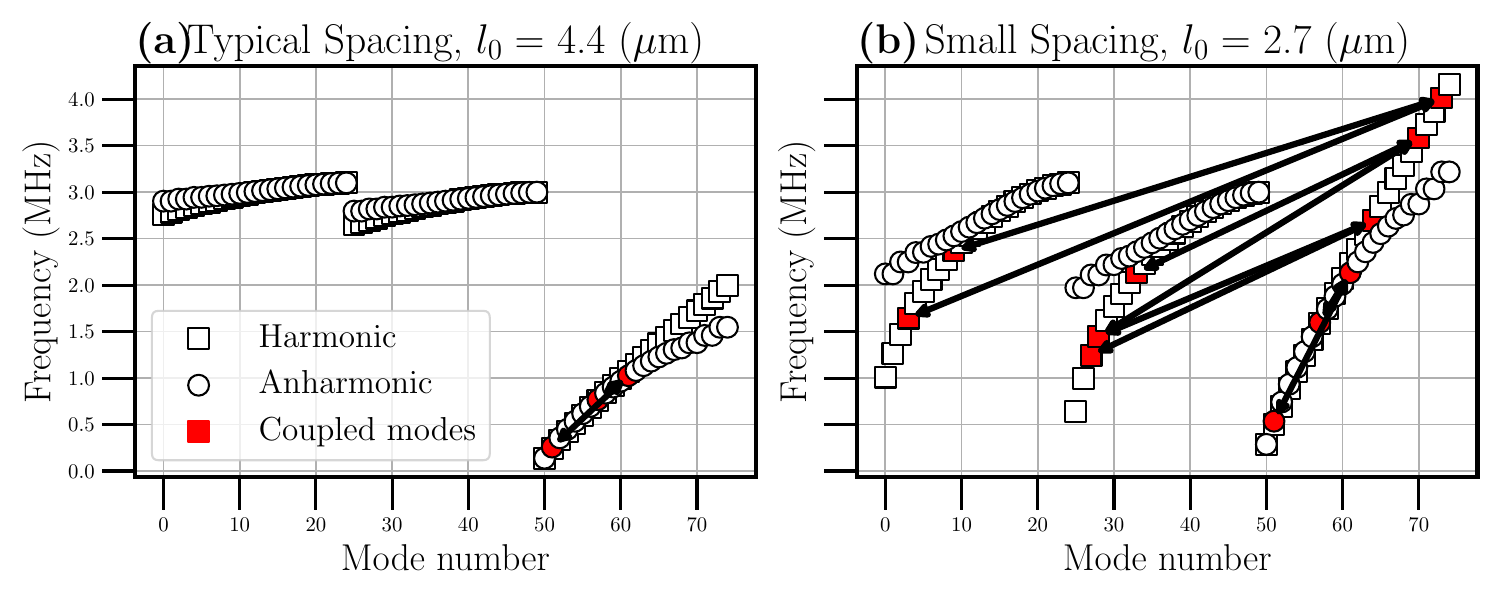}
  \caption{
    Normal-mode spectra for representative linear-ion configurations, with coupled triads shown in red and black arrows connecting the participating modes.
    We compare two $N=25$ cases: one typical of current experiments and one with tighter axial confinement, with $l_0=4.4~\mu\mathrm{m}$ and $l_0=2.7~\mu\mathrm{m}$, respectively; these are \textbf{(a)} and \textbf{(b)}.
    For each $l_0$, we compare (i) an anharmonic axial potential engineered to yield nearly equal spacing at $l_0$ and (ii) a harmonic trap whose axial frequency matches the lowest axial mode of the anharmonic case.
    Radial frequencies are $\omega_x=2\pi\times3.1~\mathrm{MHz}$ and $\omega_y=2\pi\times3.0~\mathrm{MHz}$ unless otherwise noted.
    Red highlights indicate triads $(n,m,p)$ that satisfy the near-resonance condition $\Delta_{nmp}\equiv\omega_p-\omega_m-\omega_n$ with $|\Delta_{nmp}|<0.01\,\omega_0$, exceed the tensor threshold $|T_{nmp}|>10^{-2}$ (natural units), and pass the two-level (TL) criterion described in Sec.~\ref{subsec:TLS}.
  }
  \label{fig:mode_coupling_resonances_linear}
\end{figure*}

\begin{figure*}[t]
  \includegraphics[width=1.0\linewidth]{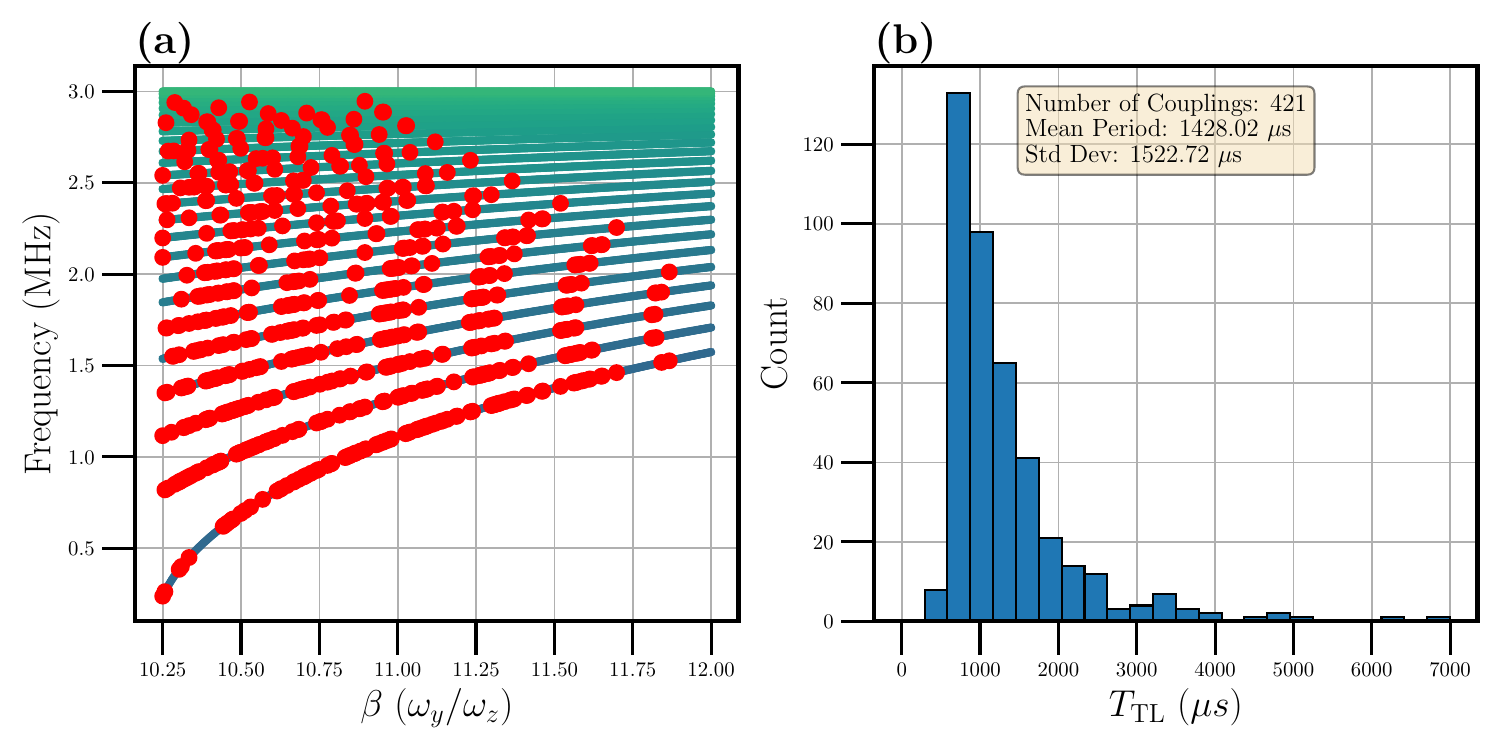}
  \caption{
    Scan of axial confinement near the zigzag regime showing emergence of radial-axial coupling.
    We vary the radial-to-axial ratio $\beta\equiv\omega_y/\omega_z$ with $N=25$, fixing $\omega_y=2\pi\times3.0~\mathrm{MHz}$ and $\omega_x=2\pi\times5.0~\mathrm{MHz}$ (so no coupling to the $x$ branch is expected).
    \textbf{(a)} $y$-branch spectrum versus $\beta$, with coupled triads highlighted in red according to the selection criteria in the text.
    \textbf{(b)} Histogram of two-level oscillation periods $T_\mathrm{TL}$ (in $\mu$s) for all identified radial-axial interactions across the scan.
  }
  \label{fig:linear_chain_scan_axial_confinement}
\end{figure*}

\paragraph*{Results in brief.}
In long linear chains with harmonic axial confinement, tight axial confinement (roughly when $10\,\omega_z \sim \omega_y$) brings radial--axial \co\ resonances onto \MSGATE-relevant timescales ($T_\text{TL}\!\sim\!T_\text{gate}$), whereas adding a modest quartic term that flattens the axial potential reshapes the spectrum to avoid such resonances even at similar inner-ion spacing. 

\paragraph*{Background and motivation.}
Linear ion chains are a well-established platform for quantum processing with trapped ions~\cite{Bruzewicz2019,Monroe2021,Chen2021}.
Chains containing dozens of ions have been used for both quantum simulation~\cite{Pagano2018,Kranzl2022} and quantum computation~\cite{Ringbauer2025,Schwerdt2024,Chen2024,Landsman2019}.

Nonlinear motional mode coupling (\co) in linear chains has been studied theoretically~\cite{Marquet2003}, used in quantum simulation experiments~\cite{Ding2017,Ding2018}, and proposed as a mechanism for unexplained heating in experiments~\cite{Lechner2016}.
Two-dimensional spectroscopy techniques have also been proposed as a diagnostic tool for \co\ in ion crystals~\cite{Lemmer2015}.

A related topic is the frequency shift of motional modes due to anharmonicities in the trapping and Coulomb potentials, examined for mixed-species chains and anharmonic traps~\cite{Home2011} and in the context of quantum operations in small chains~\cite{Roos2008,Nie2009}.
For a broader overview of nonlinear effects in ion chains, see Sec.~7 of Ref.~\cite{Chen2021}.

As demonstrated in the previous section, \co\ can lead to significant errors in quantum gates.
It is therefore important to identify when such interactions arise in experimentally relevant chains and how they can be mitigated.

In this section, we determine the conditions under which \co\ appears in linear chains and quantify the associated timescales.
We compare harmonic versus mildly anharmonic axial trapping potentials and explore confinement settings that approach the zig--zag instability, reporting when radial--axial resonances enter the gate window ($T_\text{TL}\!\sim\!T_\text{gate}$) and how modest axial quartic terms or looser axial confinement push the spectrum back into a safe operating regime.

\paragraph*{Coupling mechanisms and design trade-offs.}
Nonlinear motional mode coupling (\co) in linear ion chains can be classified into two categories: radial-axial coupling and axial-axial coupling.
Due to symmetry consideration, radial-radial \co is not possible in linear ion chains. 

Radial-axial \co is more relevant for quantum processing, as the radial modes are typically used for entangling gates \cite{Zhu2006}. 
This type of coupling can become resonant when either (i) one radial mode is approximately half the frequency of an axial mode (like the two-ion tilt-breathing mode example), or (ii) the sum of two radial mode frequencies matches that of an axial mode \cite{Marquet2003}. 
These interactions are generally stronger than axial-axial couplings. 
See Appendix~\ref{app:tressian} for a discussion based on the symmetry constraints of the nonlinear coupling tensor.

Similarly axial-axial \co occurs when two or three axial modes interact through nonzero third-order coupling coefficients. 
These couplings are usually weak in linear chains, leading to small resonance strengths \(S\) and long TL oscillation periods \(T_\text{TL}\) compared to gate times.

We define the ratio of radial to axial confinement frequencies as \(\beta = \omega_y / \omega_z\), where \(\omega_y\) is the weaker of the two radial confinement frequencies (\(\omega_x > \omega_y\)), and \(\omega_z\) is the axial confinement frequency. 
In typical experiments, \(\beta \gg 1\), meaning the radial confinement is much stronger than the axial confinement. 
This regime improves gate fidelities for operations using radial modes \cite{Zhu2006}.

However, the thermal motion due to soft axial modes can introduce amplitude noise from addressing laser beams, especially in long chains where the resulting error scales as \(N^6\), with \(N\) the number of ions \cite{Cetina2022}. 
This creates a trade-off: stronger axial confinement helps suppress this noise but increases the risk of mode coupling with radial modes.

As the number of ions increases, another challenge emerges—non-uniform ion spacing. 
Tightly packed ions near the center of the chain complicate individual addressing. 
To mitigate this, a quartic term can be added to the axial trapping potential, flattening the potential and promoting more uniform spacing \cite{Lin2009}. 
Such anharmonic potentials also help stabilize long chains by delaying the onset of the zig-zag transition, a second-order structural instability that occurs when the lowest-frequency transverse (zig-zag) mode approaches zero \cite{Fishman2008,Lin2009,Kranzl2022}.

We provide details of the anharmonic potential in Appendix~\ref{app:equally_spaced_ion_chain}, including how the quartic term is tuned to optimize spacing. 
Importantly, modifying the potential also alters the normal mode spectrum, which can change the location and strength of nonlinear mode couplings.

\paragraph*{Spectral comparison: harmonic vs.\ anharmonic (Fig.~\ref{fig:mode_coupling_resonances_linear}).}
To investigate these effects, we compare linear ion chains under two axial potentials—harmonic and anharmonic—and at two inner-ion spacings. 
In the anharmonic case, the quartic term is optimized for uniform spacing. 
In the harmonic case, the axial confinement frequency is chosen to match the lowest axial mode frequency of the anharmonic system, enabling a direct comparison.

By examining the resulting mode spectra, we find that for loosely spaced chains, no radial-axial \co occurs, and the radial modes remain suitable for quantum operations. 
For example, our analysis identifies no radial-axial \co in the $N$ = 53 ion chain simulated in Fig.~\ref{fig:crystal_energy_fluctuation_comparison}. 
In contrast, tightly spaced chains exhibit radial-axial \co in the harmonic potential, while the anharmonic potential suppresses this coupling by maintaining higher radial mode frequencies and lower axial mode frequencies. 
This suggests that anharmonic axial potentials may serve a dual purpose: enabling uniform spacing and preventing unwanted nonlinear interactions.

Figure~\ref{fig:mode_coupling_resonances_linear} compares the mode spectra of linear ion chains confined with either an anharmonic or harmonic axial potential. 
In both cases, \(N = 25\) $^{171}\text{Yb}^+$ ions are considered, with harmonic radial confinement set to \(\omega_x = 2\pi \times 3.1\ \text{MHz}\) and \(\omega_y = 2\pi \times 3.0\ \text{MHz}\). 
The parameters of the system are based on experimental parameters from ref.~\cite{Cetina2022}, with the anharmonic axial potential optimized to yield an inter-ion spacing of approximately \(4.4\ \mu\text{m}\).   
In the harmonic case, the axial confinement strength is adjusted so that the lowest axial mode matches that of the anharmonic case, enabling direct comparison.

In panel (a), mode frequencies are plotted by index and grouped into the \(x\), \(y\), and \(z\) branches. 
Square points represent the harmonic case, while circles denote the anharmonic case. 
Coupled modes are highlighted in red, and black arrows indicate interacting modes. 
Only axial-axial \co is observed, and only in the anharmonic configuration. 
The absence of radial-axial \co is expected, as the radial modes remain higher in frequency than the axial modes, preventing resonance.

Panel (b) shows the same analysis with the ion spacing reduced to \(2.7\ \mu\text{m}\). 
This is achieved by increasing the axial confinement, which pushes the axial mode frequencies higher and lowers the radial mode frequencies. 
As the zig-zag mode softens—approaching zero frequency—it signals proximity to the structural zig-zag transition \cite{Kiethe2021}. 
Near this instability, nonlinear effects such as temperature-dependent frequency shifts become more pronounced.

In the harmonic case, several radial-axial couplings emerge: one involving an \(x\) mode and two involving \(y\) modes. 
In contrast, the anharmonic configuration—with spacing optimized—keeps the radial modes above the axial modes, suppressing these resonances. 
This highlights a key benefit of anharmonic axial potentials: they not only improve ion spacing uniformity but also suppress deleterious nonlinear couplings that could compromise gate fidelity.

Overall, these results suggest that carefully engineering the axial potential can play a critical role in maintaining gate robustness in long ion chains, particularly by avoiding radial-axial \co near structural instabilities.

\paragraph*{Axial-confinement scan and results (Fig.~\ref{fig:linear_chain_scan_axial_confinement}).}
To further investigate the conditions under which radial-axial \co arises in linear ion chains, we perform a scan over the axial confinement frequency approaching the zig-zag transition. 
We simulate a chain of \(N = 25\) \(^{171}\text{Yb}^+\) ions, with radial confinement frequencies fixed at \(\omega_y = 2\pi \times 3.0\ \text{MHz}\) and \(\omega_x = 2\pi \times 5.0\ \text{MHz}\). 
Although the radial confinement frequencies are typically closer together, the stronger confinement along \(x\) ensures that no \co occurs between \(x\)-branch radial modes and axial modes during the scan.

The axial confinement is varied to produce a range of \(\beta = \omega_y / \omega_z\) values from 10.25 to 12, while keeping \(\omega_y\) constant. 
500 instantiations of the system are analyzed for each \(\beta\) value, allowing us to systematically explore the onset of resonant radial-axial \co interactions as the zig-zag transition is approached.
The resonance criteria $S > 0.1$ is used to identify significant \co interactions, where \(S\) is the resonance strength defined in section~\ref{subsec:TLS}.
The results of this scan are shown in Fig.~\ref{fig:linear_chain_scan_axial_confinement}.

Figure~\ref{fig:linear_chain_scan_axial_confinement}(b) presents a histogram of the resonance strengths for the coupled modes identified in panel~(a), quantified via the TL oscillation period $T_\text{TL}$, measured in microseconds. 
The average $T_\text{TL}$ is $\sim 1400~\mu\text{s}$, though the distribution is skewed toward longer times with a large standard deviation of $\sim 1500~\mu\text{s}$. 
The typical (median) resonance is closer to $1000~\mu\text{s}$. 
This is comparable to gate times in current long chains ($N=30$), where MS gates span $550$-$883~\mu\text{s}$ with a median of $672~\mu\text{s}$~\cite{Chen2024}, and longer than the $\sim 200~\mu\text{s}$ two-qubit gates achieved in smaller ($N=5$) chains using robust cardioid MS pulses~\cite{Manovitz2022}.

Coupling is most commonly observed in the lower-frequency \(y\)-modes. 
This is significant because these modes, having shorter wavelengths, are less susceptible to electric field noise and are often used to implement high-fidelity quantum gates \cite{Kalincev2021}. 
However, some modes are symmetry-protected from coupling: the center-of-mass (COM) mode does not participate in \co, and the tilt mode—the second-highest frequency \(y\)-mode with long wavelength—also remains uncoupled throughout the scan. 
Notably, we found no radial-axial \co above \(\beta = 12\), suggesting that in this regime, the radial modes are safely isolated from axial-mode resonances and thus suitable for quantum processing.

This supports the range of \(T_\text{TL}\) values used in the previous section to model the impact of \co on quantum gate performance. 
Notably, a subset of interactions occurs on timescales comparable to those of typical gates (\(\sim 500\ \mu\text{s}\)), indicating that resonant \co near the zig-zag transition can pose a real threat to gate fidelity. 
Among the 500 simulated configurations, roughly 80\% exhibit at least one instance of radial-axial \co, but none are observed above \(\beta = 12\).

\paragraph*{Implications for gates and mitigation.}
This section demonstrates that while linear ion chains are generally robust against \co, they are not immune to it. 
In particular, radial-axial \co can arise near the zig-zag transition, where axial modes become high in frequency and radial modes soften. 
These resonances occur under experimentally relevant conditions and may interfere with quantum gate operations.

However, we find that applying an anharmonic axial trapping potential—designed to produce uniform ion spacing—can prevent such resonances by maintaining a favorable mode frequency separation. 
This same potential also allows for more uniform ion spacing along the chain, and the stabilization of longer ion chains \cite{Lin2009,Kranzl2022}.

Our analysis focuses on near-ground-state behavior, but in systems where axial modes are only Doppler cooled, thermal excitations may further enhance coupling effects. 
Nonetheless, these results suggest that \co is not a fundamental limitation for quantum processing in linear ion chains. 
Instead, careful trap engineering—particularly via anharmonic axial potentials—offers a practical route to suppress unwanted mode couplings and improve the robustness of large-scale trapped-ion quantum processors.

\subsection{\co in experimental 2D crystals}
\label{subsec:planar_ion_crystals}

\begin{figure}[htbp]
	\includegraphics[width=1.0\linewidth]{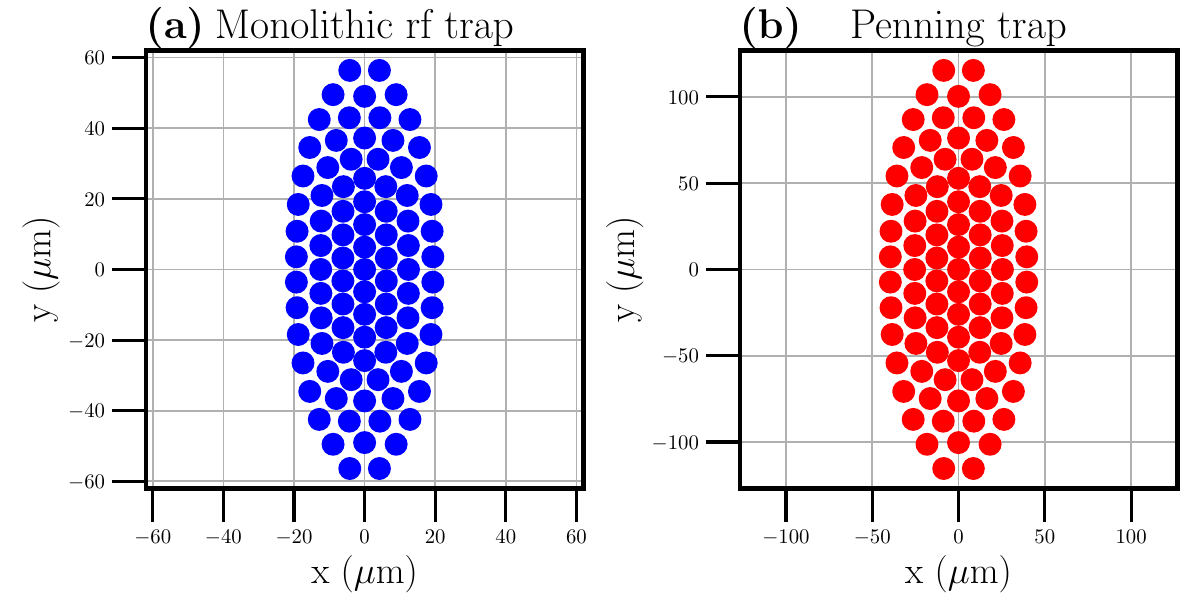}
	\caption{
		The equilibrium configurations of experimental 2D ion crystals. 
		\textbf{(a)} A 2D Penning trap ion crystal with $N=91$ ions, typical of NIST experiments~\cite{Bohnet2016}, with radial anisotropy chosen to match the 2D rf crystal.
		\textbf{(b)} A 2D rf trap ion crystal with $N=91$ ions based on Ref.~\cite{Kiesenhofer2023} parameters. 
		}	
	\label{fig:equilibrium_positions_planar}
\end{figure}

\begin{figure*}[t] 
	\includegraphics[width=1.0\linewidth]{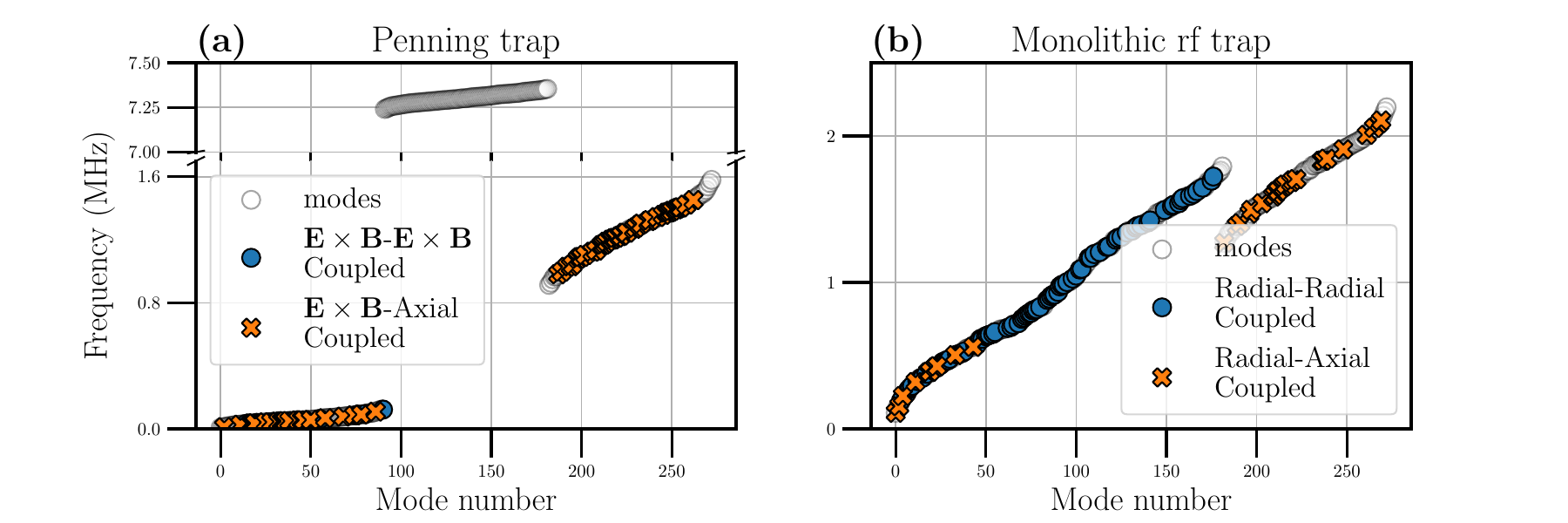}
	\caption{
		The mode spectra, with coupled modes highlighted in red, for experimental 2D ion crystal configurations. 
		Blue circle highlighted modes represent radial-radial couplings, while orange x's represent radial-axial couplings.
		\textbf{(a)} A 2D Penning trap ion crystal shown in [Fig.~\ref{fig:equilibrium_positions_planar}(a)]. 
		The radial modes separate into two distinct branches due to the strong magnetic field, with the highest-frequency cyclotron modes (middle) being the most stable.
		Note the $y$-axis break to show the cyclotron branch at >7 MHZ. 
		Meanwhile, low-frequency $\exb$ modes (left) are mediate inner-branch and axial couplings.
		The last N modes are the axial (sometimes referred to as `drumhead') modes of the crystal, typically used for interacting qubits.
		\text{(b)} A 2D rf trap ion crystal shown in [Fig.~\ref{fig:equilibrium_positions_planar}(b)].
		The radial modes form a single branch.
		Most radial modes are coupled to each other, and several low-frequency radial modes mediate \co with the axial modes.
		}
	\label{fig:mode_coupling_resonances_planar}
\end{figure*}

\paragraph*{Results in brief.}
In large 2D ion crystals, nonlinear motional mode coupling (\co) is \emph{more prevalent} than in linear chains but typically \emph{slower}: the associated two-level (TL) oscillation periods $T_\text{TL}$ are often $10$--$100~\text{ms}$, well above typical two-qubit \MSGATE\ durations yet comparable to millisecond-scale interactions used in quantum simulation. 
Consequently, \co\ primarily pressures \emph{radial-mode cooling} and \emph{gate-times}: improved cooling of radial motion and shorter entangling times reduce its impact on quantum operations, while longer interactions can accumulate coherent effects if not mitigated.
In contrast to linear chains---where radial modes serve as the entangling bus and soft axial modes are typically only Doppler cooled---in 2D crystals the \emph{axial} modes are the bus. 
This is significant because a soft radial spectator together with an intermediate axial mode can couple to a higher-frequency axial mode, opening a failure route absent in linear chains, where two radial modes must sum to an axial mode to satisfy the resonance.

\paragraph*{Scope, systems, and context.}
In this section, we investigate \co\ in large 2D ion crystals confined in either Penning traps or radio-frequency (rf) Paul traps. 
Porras and Cirac showed via perturbation theory on an ideal infinite lattice that this mechanism produces temperature-dependent entangling gate errors in 2D crystals~\cite{Porras2006}. 
Here, we identify the presence of \co triads in finite crystal equilibria calculated from parameters used in experiments, quantify their strength, and evaluate their potential impact on quantum information protocols. 
2D ion crystals provide a promising route toward scaling to hundreds of qubits; their geometry enables compact lattices and has already supported large-crystal quantum simulation~\cite{Gilmore2021,Bohnet2016,McMahon2024,Wolf2024,Kiesenhofer2023,Guo2024}. 
Despite challenges such as single-ion addressing and micromotion in rf traps, recent demonstrations show high-fidelity quantum logic in small 2D arrays and indicate that micromotion errors can be engineered away via coherent gate design~\cite{Wang2015,Hou2024_2D}. 
These results suggest that 2D crystals can support large-scale quantum processing, but may require improved cooling of radial modes and shorter gate times to mitigate \co.

\paragraph*{Mode structure in 2D (Penning vs.\ rf).}
In both Penning and rf Paul traps, the axial (out-of-plane, “drumhead”) modes typically mediate entangling gates. 
In Penning traps, in-plane addressing is complicated by rigid crystal rotation; in rf traps, axial modes are largely immune to micromotion and thus attractive for quantum logic~\cite{Kiesenhofer2023,Wang2015,Wang2020,Qiao2021,OGorman2017,Richerme2016,Kaufmann2012}. 
The in-plane spectra differ strongly between platforms: rf crystals present a dense, lower-frequency radial band (from weaker radial confinement), whereas Penning crystals separate radial motion into high-frequency cyclotron branches and low-frequency $\exb$ branches. 
Cyclotron modes dominate kinetic energy; $\exb$ modes carry most potential energy and can disrupt operations if not sufficiently cooled~\cite{Shankar2020,Johnson2024}.
2D geometries therefore introduce coupling pathways that are absent in linear chains: the thermally occupied, low-frequency radial modes can resonantly couple to the axial \emph{bus} modes, whereas in linear chains the analogous soft axial modes cannot couple to the radial bus until their frequencies approach those of the radial modes.
Consistent with this picture, Figure~\ref{fig:crystal_energy_fluctuation_comparison} shows that 2D configurations exhibit enhanced mode-energy fluctuations---an indicator of \co{}---while the $N=53$ linear chain under typical operating conditions does not exhibit such fluctuations.

\paragraph*{Modeling framework and configurations.}
We model the 2D rf crystal within the \emph{pseudopotential approximation}~\cite{James1998}, which neglects micromotion while accurately capturing secular motion; Penning and rf normal modes and frequencies are obtained via the Hamiltonian formalism of Ref.~\cite{Dubin2020}, detailed in Appendix~\ref{app:Hamiltonian} with numerical parameters in Appendix~\ref{app:simulations}. 
Experimentally, axial modes are typically near ground state via EIT or sideband cooling~\cite{Jordan2019,Shankar2019,Qiao2021,Kiesenhofer2023,Guo2024}, while radial modes are often only Doppler cooled—with Penning $\exb$ modes especially susceptible to residual heating~\cite{Torrisi2016,Johnson2024}. 
For a direct cross-platform comparison, we also consider a Penning configuration engineered to replicate the rf crystal's equilibrium positions up to a global length rescaling [Fig.~\ref{fig:equilibrium_positions_planar}(a,b)], allowing us to isolate dynamical differences from structural ones.  

\paragraph*{Coupling survey and observations (Fig.~\ref{fig:mode_coupling_resonances_planar}).}
We survey \co\ by identifying mode triplets and labeling couplings with the criterion $S>0.1$ (see Eq.~\eqref{eq:2-level_probability} and Appendix~\ref{app:simulations}). 
In Fig.~\ref{fig:mode_coupling_resonances_planar}(a,b) for $N=91$ ions, uncoupled modes are white circles with black edges; radial-axial couplings are orange X's; radial-radial couplings are blue circles. 
Penning crystals show \emph{radial-axial} couplings dominated by low-frequency $\exb$ modes; no cyclotron-involving couplings are observed, suggesting these modes could be gate-capable if made addressable. 
In rf crystals, \emph{radial-radial} couplings are prominent due to the dense radial band, with some radial-axial instances; only the lowest radial modes tend to couple to axial modes. 
This pattern is consistent with the enhanced mode-energy fluctuations seen for 2D crystals relative to chains [Fig.~\ref{fig:crystal_energy_fluctuation_comparison}].

\paragraph*{Timescales, implications, and mitigation.}
Coupling strengths in 2D crystals are relatively weak compared with linear chains: TL oscillation periods $T_\text{TL}$ for axial-involving triads typically lie in the $10$--$100~\text{ms}$ range—well above recently demonstrated two-qubit \MSGATE\ durations in small 2D crystals~\cite{Hou2024_2D} yet comparable to global interactions in quantum simulation~\cite{Guo2024,Bohnet2016}. 
Summary statistics for $T_\text{TL}$ over axial-involving triads from Fig.~\ref{fig:mode_coupling_resonances_planar} appear in Table~\ref{tab:axial_tl_stats}. 
\footnote{Each triad typically involves two axial modes (three-mode coupling), so the number of impacted axial modes is roughly double the number of triads.} 
By contrast, linear chains often exhibit $T_\text{TL}$ in the hundreds of microseconds, overlapping typical gate times and requiring tighter controls. 

Gate-level impact depends strongly on time and temperature. 
With $T_\text{gate}=1000\,T_\text{bus}$ (e.g., $500~\mu\text{s}$ for a $2~\text{MHz}$ bus), near-ground-state operation remains largely unaffected (fidelity $>0.999$ in our three-mode model of a soft radial coupled to two axial modes in Fig.~\ref{fig:ms_gate_populations_and_energies_3_modes}). 
Thermal occupation of the soft radial mode—ubiquitous for Doppler-cooled modes—degrades fidelity appreciably [Fig.~\ref{fig:ms_gate_populations_and_energies_3_modes}(b)], motivating \emph{improved radial cooling} and \emph{shorter gates} in 2D architectures. 
For quantum simulation, where interactions commonly run for milliseconds, \co\ is a coherent error source that demands shaping/shortening of interaction windows and attention to radial temperatures. 
In Penning-based sensing, the axial COM mode remains symmetry-decoupled from $\exb$ modes and experiences no rf-induced heating, providing additional robustness~\cite{Gilmore2021}. 
Overall, \co\ is not a fundamental limitation but an engineering constraint—addressed by cooling the softest radial modes and keeping interaction times well below the relevant $T_\text{TL}$.

\begin{table}[t]
\caption{Axial two-level (TL) period statistics for an $N=91$ ion crystal. 
$N_\text{triads}$ is the number of axial-involving triads; the TL statistics are in milliseconds (ms).}
\label{tab:axial_tl_stats}
\begin{ruledtabular}
\begin{tabular}{l c c c c c c}
System & $N_\text{triads}$ & \multicolumn{5}{c}{TL period (ms)} \\
\cline{3-7}
 &  & Min & Q1 & Median & Q3 & Max \\
Penning         & 25 & 17.90 & 29.02 & 52.39 & 107.72 & 369.92 \\
Monolithic rf   & 14 & 11.00 & 24.02 & 40.73 & 53.17  & 123.05 \\
\end{tabular}
\end{ruledtabular}
\end{table}

\begin{table*}[t]
\caption{Summary of two-level (TL) periods, type of coupling, and mitigations across architectures.
\textbf{Mode keys:} A: axial; R: radial; C: cyclotron; E: $\exb$; ZZ: zig--zag transition. 
\textbf{Coupling keys:} RA: radial--axial; RR: radial--radial; AA: axial--axial; E--A: $\exb$--axial.}
\label{tab:summary_TL_mitigations_wide}
\squeezetable
\begingroup
\footnotesize
\setlength{\tabcolsep}{4pt}
\renewcommand{\arraystretch}{1.05}
\begin{ruledtabular}
\begin{tabular}{l l l l}
System & $T_\text{TL}$ & Couplings & Mitigations / notes \\
\hline
Linear chain (rf)
& 0.5--10 ms (near ZZ)
& RA; AA weak
& Loosen $\omega_z$ or add axial quartic; avoid ZZ \\
2D rf
& 10--100 ms
& RR; RA
& Improve planar radial cooling; shorter $T_\text{gate}$ \\
2D Penning
& 10--100 ms
& E--A; RR; no C observed
& Improve $\exb$ cooling; axial COM protected; consider cyclotron bus \\
\end{tabular}
\end{ruledtabular}
\endgroup
\end{table*}

\section{Conclusion}
\label{sec:conclusion}
Fault-tolerant quantum computation demands extremely high-fidelity entangling gates, placing a premium on identifying and mitigating even subtle, coherent error channels. 
In global-mode trapped-ion processors, the increasing density of collective modes with system size opens additional near-resonant pathways for nonlinear motional mode coupling (\co). Here we developed a general, architecture-agnostic framework to detect, quantify, and simulate \co\ and used it to map out when and how it impacts \MSGATE\ performance across linear chains and 2D ion crystals.

\paragraph*{What we learned.}
Near the ground state, the characteristic two-level (TL) exchange timescale associated with \co\ is typically $T_\mathrm{TL}\!\sim\!0.5$--$10~\mathrm{ms}$, i.e., commensurate with standard \MSGATE\ durations. 
In long \emph{linear chains}, radial--axial resonances become likely as axial confinement tightens toward the zig-zag regime, with measured TL periods clustering around the millisecond scale. 
Crucially, we find these resonances can be engineered away---either by loosening axial confinement or by adding a modest quartic term that flattens the axial potential, which both reduces axial frequencies and lifts radial branches to reopen spectral gaps.

In \emph{2D ion crystals} (rf and Penning), nonlinear coupling is more prevalent but typically weaker at the axial modes used for gates: TL periods for axial-involving triads lie in the $\sim\!10$--$100~\mathrm{ms}$ range. 
The 2D case, however, introduces a distinct failure route: low-frequency, Doppler-cooled radial spectators (radial modes in rf arrays and $\exb$ modes in Penning systems) can couple to near-ground-state axial modes, broadening the window where thermal occupation degrades gate closure. 
This explains the stronger mode-energy fluctuations we observe in 2D configurations relative to the linear chain reference (Fig.~\ref{fig:crystal_energy_fluctuation_comparison}).

\paragraph*{Design guidance.}
Three practical levers emerge. 
(i) \emph{Shape the axial spectrum} in chains (looser confinement or mild anharmonicity) to avoid radial--axial resonances. 
(ii) \emph{Cool the soft spectators}, especially the lowest-frequency radial/$\exb$ modes in 2D crystals; when those modes are cold, axial gates are largely insensitive on operational timescales. 
(iii) \emph{Bound the bus displacement} during \MSGATE\ (e.g., more phase-space loops or displacement-constrained waveforms) to suppress nonlinear back-action and keep loop closure robust even when \co\ is present.

\paragraph{Mitigations against frequency shifts and \co{}.}
Several gate designs suppress sensitivity to static or slowly varying motional-frequency offsets.
Balanced-Gaussian waveforms increase robustness to slowly varying motional mode frequency noise~\cite{Ruzic2024}.
Walsh-modulated \MSGATE{}s use phase flips drawn from Walsh sequences to cancel low-order detuning errors and residual displacements~\cite{Hayes2012}.
Frequency-modulated \MSGATE{}s designed via optimal control further reduce sensitivity to mode-frequency drifts in chains~\cite{Leung2018,Kang2021}.
Additional robustness can be obtained with multi-loop phase-space trajectories (cf. Fig.~\ref{fig:ms_gate_fidelity_vs_num_loops}). 
Finally, alternate entangling gate mechanisms, such as sub-microsecond Rydberg gates may suppress motional decoherence including \co{}~\cite{Zhang2020}.

\paragraph*{Why it matters.}
Today's two-qubit gates in 2D geometries sit around the percent scale, whereas fault tolerance targets $10^{-3}$-$10^{-4}$ per gate depending on code and overhead. 
Our results show that \co\ need not be a fundamental blocker: in chains it can be \emph{designed out} spectrally, and in 2D it can be \emph{made irrelevant} at gate times with better radial cooling and modest increases in speed.
Looking forward, there is a scientific upside: trapped-ion crystals offer strong, tunable, and \emph{coherent} inter-oscillator nonlinearities, making them a uniquely clean arena for studies of quantum thermodynamics, nonequilibrium transport, and chaos at the single-quantum level. 
In this sense, \co\ is a feature as well as a bug—deleterious for computation if ignored, but a powerful knob for interrogating thermalization in engineered quantum matter.

\paragraph*{Outlook: other architectures.}
Our framework extends directly to platforms beyond linear chains and 2D crystals. 
Shuttling of linear chains in QCCD architectures may transiently access near-resonant \co\ conditions~\cite{Murali2022,Valentini2025}. 
Bilayer crystals and proposed 3D trapped-ion arrays introduce denser mode manifolds that may host additional near-resonant \co\ pathways~\cite{Hawaldar2024,Zaris2025}. 
Hybrid Penning--rf surface-electrode arrays and optical static-electrode monolithic traps likewise present distinctive spectral structures where the same diagnostics and mitigations apply~\cite{Jain2020,Jain2024,Sun2024}. 
We anticipate that modest mode frequency tuning, targeted cooling of the softest spectators, and mode-displacement-bounded \MSGATE\ waveforms will remain effective tools for suppressing \co\ in these architectures as well.

\paragraph*{At-a-glance.}
In long linear chains, TL periods associated with near-resonant mode triads involving the radial modes cluster around the $0.5$-$10~\mathrm{ms}$ scale, overlapping typical \MSGATE\ durations; avoiding the zig-zag transition either with weaker axial confinement or with mild quartic anharmonicity can eliminate these resonances. 
In 2D crystals (rf Paul and Penning), axial-involving triads are more common but typically weaker, with TL periods $\sim 10$-$100~\mathrm{ms}$; here, Doppler-cooled soft radial spectators broaden the detuning window where thermal occupation degrades loop closure, so improved radial cooling and modestly shorter gates help.
Table~\ref{tab:summary_TL_mitigations_wide} reports, for each architecture, the two-level (TL) period range, the types of coupling triads observed, and the mitigations that can be applied to suppress \co\ effects on gate performance.

\begin{acknowledgments}
The authors would like to thank Jon Sterk for helpful comments on the manuscript and acknowledge the use of SandiaAI Chat for editing assistance.
Wes Johnson would like to thank the CSRI summer intern program and program coordinators. 
This research was supported by the U.S. Department of Energy, Oﬃce of Science, Oﬃce of Advanced Scientific Computing Research Quantum Testbed Program, and the Laboratory Directed Research and Development program at Sandia National Laboratories.
This work was performed, in part, at the Center for Integrated Nanotechnologies, an Office of Science User Facility operated for the U.S. Department of Energy (DOE) Office of Science. 
Sandia National Laboratories is a multimission laboratory managed and operated by National Technology \& Engineering Solutions of Sandia, LLC, a wholly owned subsidiary of Honeywell International, Inc., for the U.S. DOE's National Nuclear Security Administration under contract DE-NA-0003525. 
The views expressed in the article do not necessarily represent the views of the U.S. DOE or the United States Government.
\end{acknowledgments}

\appendix

\section{Hamiltonian Matrices}
\label{app:Hamiltonian}
In this section, we derive the Hamiltonian for a system of interacting ions. 
We consider a static potential, which could contain contributions from the pseudo-potential produced by the radio-frequency (rf) field in a Paul trap, or the effective potential experienced by ions rotating about the magnetic field axis in a Penning trap. 
This general approach is based on ref.~\cite{Dubin2020}, which provides a systematic way to derive the Hamiltonian matrix and perform the normal-mode analysis for both Penning and rf traps.

The Lagrangian for a system of $N$ ions in a trap can be written as the sum of the kinetic and potential energies of the ions, including contributions from the vector potential due to the magnetic field: 
\begin{equation}
    \mathcal{L} = \sum_{i=1}^N \frac{m_i}{2} \left| \dot{\mathbf{r}}_i \right|^2 - U_i(\mathbf{R}) + q_i \left( \dot{\mathbf{r}}_i \cdot \mathbf{A}_i(\mathbf{r}_i) \right),
\label{eq:Lagrangian_ions}
\end{equation}
where $m_i$ is the mass of the $i$th ion, $\mathbf{r}_i$ is the position of the $i$th ion, $U_i$ is the potential energy of the $i$th ion, $q_i$ is the charge of the $i$th ion, $\mathbf{A}_i$ is the vector potential experienced by the $i$th ion.   
Note, that in general, $U_i$ depends on the positions of all the ions, $\mathbf{R} = (\mathbf{x}, \mathbf{y}, \mathbf{z})$, where $\mathbf{x} = (x_1, x_2, \ldots, x_N)$, and similarly for $\mathbf{y}$ and $\mathbf{z}$.

In the Penning trap, the vector potential corresponds to the magnetic field aligned along the axis of the trap, $\mathbf{B} = \nabla \times \mathbf{A} = B_0 \hat{z}$, where $B_0$ is the magnitude of the magnetic field and $\hat{z}$ is the unit vector in the $z$ direction.
Due to the cylindrical symmetry of the Penning trap, the vector potential is naturally expressed in the symmetric gauge as $\mathbf{A}_i = \frac{1}{2} B_0 \left( x_i \hat{y} - y_i \hat{x} \right)$, where $x_i$ and $y_i$ are the Cartesian coordinates of the $i$th ion.

The equations of motion can be linearized by expanding the potential energy of the system to second order about the equilibrium positions of the ions~\cite{Wang2013,James1998,Shankar2019,Dubin2020}.
This expansion leads to the stiffness matrix $\mathbb{K}$, which describes the linear restoring forces acting on the ions when they are displaced from their equilibrium positions, see Appendix~\ref{app:tressian} for the derivation of the stiffness matrix.    
To connect the Lagrangian formalism used in other works \cite{Wang2013,James1998,Shankar2020} to the Hamiltonian formalism \cite{Dubin2020} used in this work, we begin by writing the second order energy of the system as a function of the ion position displacements and velocities.
We show that the matrix $\mathbb{E}$ associated with the second order energy $E^{(2)}$ can be transformed into the Hamiltonian matrix $\mathbb{H}$. 

The second order energy of the linearized system can be expressed in terms of the stiffness matrix $\mathbb{K}$ and the mass matrix $\mathbb{M}$: 
\begin{equation} 
    \mathbb{E} = \begin{bmatrix} \mathbb{K} & \mathbb{0} \\ \mathbb{0} & \mathbb{M} \end{bmatrix},\quad E^{(2)} = \frac{1}{2} \bra{X} \mathbb{E} \ket{X},
\label{eq:energy_matrix}
\end{equation}
where $\mathbb{M}$ is the mass matrix of the system, with the masses of the ions repeated three times along the diagonal, $\mathbb{0}$ is a square matrix of zeros size $3N\times3N$, and $\ket{X} = (\mathbf{q}, \dot{\mathbf{R}})$ is the vector of ion position displacements and velocity coordinates.   
$\mathbf{q} = \mathbf{R} - \mathbf{R}_0$ is the vector of ion position displacements from their equilibrium positions. 

A transformation of $\mathbb{E}$ to the Hamiltonian matrix, $\mathbb{H}$, can be accomplished by introducing the conjugate momenta, $\mathbf{p} = \frac{\partial \mathcal{L}}{\partial \dot{\mathbf{r}}}$. 
The non-symplectic transformation matrix $\mathbb{T}$ is applied to the position and velocity coordinates, $\ket{X}$, to obtain the canonical coordinates, $\ket{X^\prime} = \mathbb{T} \ket{X}$, where $\ket{X^\prime} = (\mathbf{q}, \mathbf{p})$ is the vector of position and momentum coordinates.
The matrix $\mathbb{T}$ is given by:
\begin{equation}
    \mathbb{T} = \begin{bmatrix} \mathbb{I} & \mathbb{0} \\ \mathbb{B} & \mathbb{M} \end{bmatrix}, 
\label{eq:canonical_transformation_matrix}
\end{equation}
where $\mathbb{I}$ is the identity matrix, and $\mathbb{B}$ is the matrix that depends on the magnetic field. 
In the case of the rf  trap, $\mathbb{B} = \mathbb{0}$, and $\mathbb{T}$ is a block diagonal matrix.
In the case of a Penning trap $\mathbb{B}$ is given by:
\begin{equation}
    \mathbb{B} = \begin{bmatrix} \mathbb{0} & \mathbb{C} & \mathbb{0} \\ -\mathbb{C} & \mathbb{0} & \mathbb{0} \\ \mathbb{0} & \mathbb{0} & \mathbb{0} \end{bmatrix},
\label{eq:canonical_transformation_matrix_B}
\end{equation}
where all matrices are of size $N\times N$, and $\mathbb{C}$ is the matrix with the values $\frac{1}{2} (q_i B_0 - 2 m_i \omega_r)$ for each ion $i$ along the diagonal, and zeros elsewhere.    
$\omega_r$ is the collective frequency at which the ions rotate about the magnetic field axis in the Penning trap.  
The Hamiltonian matrix is then given by $\mathbb{H} = (\mathbb{T}^{-1})^T \mathbb{E} (\mathbb{T}^{-1})$, where $(\mathbb{T}^{-1})^T$ is the transpose of the inverse of the canonical transformation matrix.  

The linearized equations of motion of the system can easily be derived from the Hamiltonian matrix, $\mathbb{H}$, by applying the symplectic matrix, yielding the dynamical matrix $\mathbb{D}$: 
\begin{equation}
    \mathbb{D} = \mathbb{J} \mathbb{H}, \quad \mathbb{J} = \begin{bmatrix} \mathbb{0} & \mathbb{I} \\ -\mathbb{I} & \mathbb{0} \end{bmatrix},   
\label{eq:dynamical_matrix}
\end{equation}
where $\mathbb{I}$ is the identity matrix of size $3N \times 3N$, and $\mathbb{0}$ is the square matrix of zeros of size $3N \times 3N$.   
The eigenvectors and eigenvalues of the dynamical matrix, $\mathbb{D}$, can be used to construct the symplectic transformation, $\mathbb{S}$, to the normal mode coordinates. 
The construction of $\mathbb{S}$ is given in ref.~\cite{Dubin2020}. 
$\mathbb{S}$ diagonalizes $\mathbb{H}$, $\mathbb{H}^{\prime} = \mathbb{S}^T \mathbb{H} \mathbb{S}$, where $\mathbb{H}^{\prime} = \text{diag} \left( \mathbf{\Omega}, \mathbf{\Omega} \right)$, and $\mathbf{\Omega}$ is the vector of normal mode frequencies. 
The linearized energy of the system can now be expressed in terms of the normal mode energies as: 
\begin{equation}
    E = \frac{1}{2} \bra{X^{\prime\prime}} \mathbb{H}^{\prime} \ket{X^{\prime\prime}} = \frac{1}{2} \sum_{n=1}^{3N} \omega_n^2 \left( Q_n^2 + P_n^2 \right),
\label{eq:normal_mode_energy}
\end{equation}
where $\ket{X^{\prime\prime}} = \left( Q_1, Q_2, \ldots, Q_{3N}, P_1, P_2, \ldots, P_{3N} \right)$ is the vector of canonical normal mode coordinates, and $Q_n$ and $P_n$ are the canonical position and momentum normal-mode coordinates of the $n^\text{th}$ normal mode, respectively.

This approach provides a systematic way to derive the normal-mode Hamiltonian matrix via linear transformations of the system coordinates, starting from the energy matrix in position displacement and velocity coordinates in Eq.~(\ref{eq:energy_matrix}). 
Specifically, the transformation sequence $(\mathbf{q}, \dot{\mathbf{R}}) \xrightarrow{\mathbb{T}} (\mathbf{q}, \mathbf{p}) \xrightarrow{\mathbb{S}^{-1}} (\mathbf{Q}, \mathbf{P})$ maps the system into canonical normal mode coordinates. 
In the next section, we derive the third-order anharmonic correction to the Hamiltonian and use these same transformations to express it in terms of the normal modes.

\section{Expansion of Potential}
\label{app:tressian}
In this section, we derive the expansion of the potential energy for a system of trapped ions. 
We explicitly compute the third-order expansion of the Coulomb potential in three dimensions, and express the result in terms of the normal mode coordinates using a linear transformation. 

Greek indices $\alpha, \beta, \text{ and } \gamma \in \{x, y, z\}$ denote Cartesian variable, while $i, j, k \in \{1, 2, \ldots, N\}$ label ions. 
Finally, $a, b, \text{ and } c$ = 1, 2, 3, $\ldots 3N$, where $3N$ is the total number of Cartesian coordinates, denote the Cartesian coordinate index.
For example, $R_a$ is the $a^\text{th}$ component of the vector $\mathbf{R} = (\mathbf{x}, \mathbf{y}, \mathbf{z})$, where $\mathbf{x} = (x_1, x_2, \ldots, x_N)$, and similarly for $\mathbf{y}$ and $\mathbf{z}$.

The total potential energy of the ion crystal, $U(\mathbf{R})$, is a function of all $3N$ Cartesian ion coordinates. 
The equilibrium configuration of the ion crystal is given by $\mathbf{R}_0$, satisfying $\frac{\partial U}{\partial \alpha_i}\big |_{\mathbf{R}_0} = 0$ for all components $\alpha_i$. 
We define displacements from equilibrium as $\mathbf{q} = \mathbf{R} - \mathbf{R}_0$, and expand the potential as: 
\begin{gather}
    U(\mathbf{q}) = \frac{1}{2} \sum_{a,b} K_{ab} q_a q_b + \frac{1}{6} \sum_{a,b,c} T_{abc} q_a q_b q_c + \ldots \\
    K_{ab} = \frac{\partial^2 U}{\partial R_a \partial R_b}\big |_{\mathbf{R}_0} \\
    T_{abc} = \frac{\partial^3 U}{\partial R_a \partial R_b \partial R_c}\big |_{\mathbf{R}_0}
\label{eq:potential_expansion_Cartesian}
\end{gather}
where the constant term is dropped because it does not affect the dynamics of the ion crystal, and the linear term is zero at equilibrium by definition.
$K_{ab}$ and $T_{abc}$ are the second and third derivatives of the potential energy with respect to the Cartesian coordinates evaluated at the equilibrium of the crystal. 
$K_{ab}$ is the stiffness matrix of the ion crystal, and $T_{abc}$ is a tensor of coefficients that describe the lowest order anharmonicity in the potential energy of the ion crystal. 
The normal modes of the ion crystal can be calculated from $\mathbb{K}$, the stiffness matrix, as shown in Appendix~\ref{app:Hamiltonian}, where $K_{ab}$ is just the index notation for the stiffness matrix $\mathbb{K}$. 

The potential energy of the ion crystal can be separated into two parts, $U = U_{\text{trap}} + U_{\text{Coulomb}}$, where $U_{\text{trap}}$ is the trapping potential, and $U_{\text{Coulomb}}$ is the Coulomb repulsion between the ions.
$U_\text{trap}$ is often nearly harmonic, meaning the Coulomb potential is the dominant source of anharmonicity in the potential energy of the ion crystal and will be the focus of this section.
The derivatives of the Coulomb potential energy are derived below. 

The Coulomb potential energy of the ion crystal is given by:
\begin{equation} 
    U_{\text{Coulomb}} = \frac{1}{2} \sum_{i=1}^{N} \sum_{j\neq i}^{N} \frac{k_e q^2}{|\mathbf{r}_i - \mathbf{r}_j|} 
\label{eq:coulomb_potential_dimensionful}
\end{equation}
where $q$ is the charge of the ions (assumed to be the same for all ions), $k_e$ is the Coulomb constant, and $\mathbf{r}_i$ is the position of the $i^\text{th}$ ion. 
The factor of $1/2$ in the sum is to avoid double counting the interactions between pairs of ions.

To simplify the expressions, we introduce characteristic scales for frequency ($\omega_0$), mass ($m_0$), length ($l_0$), and energy ($E_0$) for the ion crystal.
A natural choice is $\omega_0 = \omega_z$, the axial trapping frequency, and $m_0 = \min(m_i)$, the lightest ion mass in the crystal. 
The characteristic energy and length scale of the system are defined as:
\begin{equation}
    E_0 = m_0 \omega_0^2 l_0^2, \quad l_0 = \left( \frac{k_e q^2}{m_0 \omega_0^2} \right)^{1/3}.
\label{eq:characteristic_energy_and_length}
\end{equation}
which simplifies the form of the dimensionless Coulomb potential energy when the energy is rescaled by $E_0/2$ and the length is rescaled by $l_0$:
\begin{equation}
    U_{\text{Coulomb}} = \sum_{i=1}^{N} \sum_{j\neq i}^{N} \frac{1}{r_{ij}}
\label{eq:coulomb_potential_dimensionless}
\end{equation}
where $r_{ij} = \left( x_{ij}^2 + y_{ij}^2 + z_{ij}^2 \right)^{1/2}$, and $x_{ij} = x_i - x_j$, and similarly for $y_{ij}$ and $z_{ij}$.

\subsection*{Jacobian of the Coulomb Potential}  

For legibility, the index of the Cartesian coordinate will be placed as a superscript, while the ion index will be placed as a subscript.   
The first derivative of the Coulomb potential with respect to the Cartesian coordinates of the ions are given by: 
\begin{equation} 
    F_i^\alpha = \frac{\partial}{\partial \alpha_i} U_\text{coulomb} = - \sum_{j\ne i}^{N} \frac{\alpha_{ij}}{r_{ij}^3}, \quad \text{where} \quad \alpha_{ij} = \alpha_i - \alpha_j. 
\label{eq:gradient_coulomb_potential}
\end{equation}
Note, that since the Jacobian of the total potential is zero at equilibrium, the Coulomb derivatives in Eq.~(\ref{eq:gradient_coulomb_potential}) will exactly cancel the derivatives of the trapping potential. 
The derivatives in Eq.~(\ref{eq:gradient_coulomb_potential}) can be organized into a vector of length $3N$. 

\subsection*{Hessian of the Coulomb Potential}

The Hessian of the Coulomb potential is given by the second derivatives of the Coulomb potential with respect to the Cartesian coordinates of the ions: 
\begin{equation}
    \begin{gathered}
    H^{\alpha, \alpha}_{i,i} = \frac{\partial}{\partial \alpha_i} F_i^\alpha = - \sum_{j\ne i}^{N} \left [ \frac{1}{r_{ij}^3} - 3 \frac{\alpha_{ij}^2}{r_{ij}^5}\right ] \\
    H^{\alpha, \alpha}_{i,j} = \frac{\partial}{\partial \alpha_i} F_j^\alpha =  \left [\frac{1}{r_{ij}^3} - 3 \frac{\alpha_{ij}^2}{r_{ij}^5} \right ]  \\
    H^{\alpha, \beta}_{i,i} = \frac{\partial}{\partial \alpha_i} F_i^\beta = 3 \sum_{j\ne i}^{N} \frac{\alpha_{ij} \beta_{ij}}{r_{ij}^5}   \\ 
    H^{\alpha, \beta}_{i,j} = \frac{\partial}{\partial \alpha_i} F_j^\beta = -3 \frac{\alpha_{ij} \beta_{ij}}{r_{ij}^5} \\
    H^{\alpha, \alpha} = H^{\alpha, \alpha}_{i,i} + H^{\alpha, \alpha}_{i,j} \\
    H^{\alpha, \beta} = H^{\alpha, \beta}_{i,i} + H^{\alpha, \beta}_{i,j}   
    \end{gathered}
\label{eq:hessian_coulomb_potential}
\end{equation}
where the derivatives have been organized into two groups: those where the Cartesian coordinates are the same, $H^{\alpha, \alpha}$, and those where the Cartesian coordinates are different, $H^{\alpha, \beta}$. 
Each of these has diagonal $i=j$ (same-ion) and off-diagonal $i \ne j$ (different-ion) terms
These derivatives can be organized into a stiffness matrix below: 
\begin{equation}
    \mathbb{K} = \begin{bmatrix} H_{xx} & H_{xy} & H_{xz} \\ H_{yx} & H_{yy} & H_{yz} \\ H_{zx} & H_{zy} & H_{zz} \end{bmatrix}. 
    \label{eq:hessian_matrix}
\end{equation}
The geometry of the equilibrium configuration of the ion crystal can significantly simplify the form of the stiffness matrix.
For example, in the case of a linear ion crystal, the stiffness matrix will be block-diagonal since $H_{xy} = H_{xz} = H_{yz} = 0$. 
As a consequence, the normal modes of a linear ion crystal separate into three independent branches, one for each Cartesian coordinate.

\subsection*{Tressian of the Coulomb Potential}

The third order derivatives of the potential energy can be organized into a tensor of rank three, which will be referred to as the \textbf{Tressian tensor}. 
The third-order derivatives of the potential energy define a rank-3 tensor, referred to as the \textbf{Tressian}, which encodes the leading-order anharmonic couplings between the modes. 
First we consider derivatives where the Cartesian symbols are the same, $T^{\alpha,\alpha,\alpha}$. 
\begin{equation}
    \begin{gathered}
        T^{\alpha,\alpha,\alpha}_{i,i,i} = \frac{\partial}{\partial \alpha_i} H^{\alpha,\alpha}_{i,i} = - \sum_{j\ne i}^{N} \left [ -9\frac{\alpha_{ij}}{r_{ij}^5} + 15 \frac{\alpha_{ij}^3}{r_{ij}^7} \right ] \\  
        T^{\alpha,\alpha,\alpha}_{i,i,j} = \frac{\partial}{\partial \alpha_i} H^{\alpha,\alpha}_{i,j} =  \left [ -9\frac{\alpha_{ij}}{r_{ij}^5} + 15 \frac{\alpha_{ij}^3}{r_{ij}^7} \right ] \\    
        T^{\alpha,\alpha,\alpha}_{i,j,k} = \frac{\partial}{\partial \alpha_i} H^{\alpha,\alpha}_{j,k} = 0 
    \end{gathered}
\label{eq:tressian_coulomb_potential_1} 
\end{equation}
Next the cases for $T^{\alpha,\alpha,\beta}$, where $\alpha \ne \beta$: 
\begin{equation}
    \begin{gathered}
       T^{\alpha,\alpha,\beta}_{i,i,i} = \frac{\partial}{\partial \alpha_i} H^{\alpha,\beta}_{i,i} = 3 \sum_{j\ne i}^{N} \left [ \frac{\beta_{ij}}{r_{ij}^5} - 5 \frac{\alpha_{ij}^2 \beta_{ij}}{r_{ij}^7} \right ] \\  
       T^{\alpha,\alpha,\beta}_{i,i,j} = \frac{\partial}{\partial \alpha_i} H^{\alpha,\beta}_{i,j} = -3 \left [ \frac{\beta_{ij}}{r_{ij}^5} - 5 \frac{\alpha_{ij}^2 \beta_{ij}}{r_{ij}^7} \right ] \\   
       T^{\alpha,\alpha,\beta}_{i,j,i} = \frac{\partial}{\partial \alpha_i} H^{\alpha,\beta}_{j,i} = T^{\alpha,\alpha,\beta}_{i,i,j} \\ 
       T^{\alpha,\alpha,\beta}_{i,j,k} = \frac{\partial}{\partial \alpha_i} H^{\alpha,\beta}_{j,k} = 0  
    \end{gathered}
\label{eq:tressian_coulomb_potential_2} 
\end{equation}
Finally the cases for $T^{\alpha,\beta,\gamma}$, where $\alpha \ne \beta \ne \gamma$: 
\begin{equation} 
    \begin{gathered}
        T^{\alpha,\beta,\gamma}_{i,i,i} = \frac{\partial}{\partial \alpha_i} H^{\beta,\gamma}_{i,i} = -15 \sum_{j\ne i}^{N} \frac{\alpha_{ij} \beta_{ij} \gamma_{ij}}{r_{ij}^7} \\  
        T^{\alpha,\beta,\gamma}_{i,i,j} = \frac{\partial}{\partial \alpha_i} H^{\beta,\gamma}_{i,j} = 15 \frac{\alpha_{ij} \beta_{ij} \gamma_{ij}}{r_{ij}^7} \\ 
        T^{\alpha,\beta,\gamma}_{i,j,i} = \frac{\partial}{\partial \alpha_i} H^{\beta,\gamma}_{j,i} = T^{\alpha,\beta,\gamma}_{i,i,j} \\  
        T^{\alpha,\beta,\gamma}_{j,i,i} = \frac{\partial}{\partial \alpha_j} H^{\beta,\gamma}_{i,i} = T^{\alpha,\beta,\gamma}_{i,i,j} \\    
        T^{\alpha,\beta,\gamma}_{i,j,k} = \frac{\partial}{\partial \alpha_i} H^{\beta,\gamma}_{j,k} = 0 
    \end{gathered}
\label{eq:tressian_coulomb_potential_3}
\end{equation}
The subblocks of the Tressian tensor are assembled from the components derived in Eqs.~(\ref{eq:tressian_coulomb_potential_1})--(\ref{eq:tressian_coulomb_potential_3}) as follows: 
\begin{equation} 
    \begin{gathered} 
        T^{\alpha,\alpha,\alpha} = T^{\alpha,\alpha,\alpha}_{i,i,i} + T^{\alpha,\alpha,\alpha}_{i,i,j} \\
        T^{\alpha,\alpha,\beta} = T^{\alpha,\alpha,\beta}_{i,i,i} + T^{\alpha,\alpha,\beta}_{i,i,j} + T^{\alpha,\alpha,\beta}_{i,j,i} \\
        T^{\alpha,\beta,\gamma} = T^{\alpha,\beta,\gamma}_{i,i,i} + T^{\alpha,\beta,\gamma}_{i,i,j} + T^{\alpha,\beta,\gamma}_{i,j,i} + T^{\alpha,\beta,\gamma}_{j,i,i}. 
    \end{gathered}
\label{eq:tressian_subblocks}
\end{equation}

Similarly to the Hessian, the Tressian can be organized into a tensor of rank three composed of subblocks: 
\begin{equation} 
        T = \begin{bmatrix}
                \begin{bmatrix}
                    T_{\text{xxx}} & T_{\text{xxy}} & T_{\text{xxz}} \\
                    T_{\text{xyx}} & T_{\text{xyy}} & T_{\text{xyz}} \\
                    T_{\text{xzx}} & T_{\text{xzy}} & T_{\text{xzz}} \\
                \end{bmatrix} \\
                \begin{bmatrix}
                    T_{\text{yxx}} & T_{\text{yxy}} & T_{\text{yxz}} \\
                    T_{\text{yyx}} & T_{\text{yyy}} & T_{\text{yyz}} \\
                    T_{\text{yzx}} & T_{\text{yzy}} & T_{\text{yzz}} \\
                \end{bmatrix} \\
                \begin{bmatrix}
                    T_{\text{zxx}} & T_{\text{zxy}} & T_{\text{zxz}} \\
                    T_{\text{zyx}} & T_{\text{zyy}} & T_{\text{zyz}} \\
                    T_{\text{zzx}} & T_{\text{zzy}} & T_{\text{zzz}} \\
                \end{bmatrix} 
            \end{bmatrix}. 
\label{eq:tressian_tensor}
\end{equation}

\paragraph*{Symmetry of the Tressian tensor.}
Geometric symmetry of the equilibrium ion configuration forces many third-order derivatives of the Coulomb potential (the “Tressian”) to vanish, thereby forbidding entire classes of nonlinear couplings. 
In both linear and 2D crystals, this simplifies the tensor structure and constrains which mode branches can couple at third order. 

\emph{Three distinct Cartesian indices.} 
Terms with three different Cartesian symbols are proportional to products of inter-ion coordinate differences, $T_{\alpha\beta\gamma}\propto \alpha_{ij}\beta_{ij}\gamma_{ij}/r_{ij}^7$ (\ref{eq:tressian_coulomb_potential_3}). 
In a linear chain ($x_i=y_i=0$) or a planar crystal (say $z_i=0$), at least one factor vanishes, so all sub-blocks like $T_{\text{xyz}}$ are identically zero. 
Hence, couplings among \emph{three different} branches are ruled out in 1D and 2D geometries but are allowed in full 3D crystals.

\emph{Two identical + one distinct index.} 
For $T_{\alpha\alpha\beta}$ with $\alpha\neq\beta$ (\ref{eq:tressian_coulomb_potential_2}), the tensor contains an overall factor of $\beta_{ij}$ and terms $\propto \alpha_{ij}^2\beta_{ij}/r_{ij}^7$. 
In a linear chain, $x_{ij}=y_{ij}=0$ but $z_{ij}\neq 0$, so the only nonzero cross-branch sub-blocks are $T_{\text{xxz}}$ and $T_{\text{yyz}}$ (while $T_{\text{xxy}}$, $T_{\text{yyx}}$ vanish). 
This admits radial-axial coupling but forbids $x$-$y$ or other in-plane cross-couplings at third order.

\emph{All three indices identical.} 
For $T_{\alpha\alpha\alpha}$ (\ref{eq:tressian_coulomb_potential_1}), the components scale with $\alpha_{ij}$ and $\alpha_{ij}^3$. 
Thus $T_{\text{zzz}}\neq 0$ in a linear chain (since $z_{ij}\neq 0$), whereas $T_{\text{xxx}}=T_{\text{yyy}}=0$ (because $x_{ij}=y_{ij}=0$). 
Purely radial three-mode couplings are therefore symmetry-forbidden at third order in linear chains; axial-axial couplings remain allowed but are typically weak.

\emph{Consequences for resonances.} 
In linear chains, the only symmetry-allowed third-order pathways that \emph{involve} radial modes are those mediated by an axial mode via $T_{\text{xxz}}$ or $T_{\text{yyz}}$—i.e., (i) a two-mode (2:1) condition $2\omega_{r}\approx \omega_{z}$ or (ii) a three-mode sum condition $\omega_{r_1}+\omega_{r_2}\approx \omega_{z}$. 
Since in typical experiments the axial branch is lowest in frequency, these radial-axial matches are off-resonant, and no observable radial-radial coupling appears at third order—consistent with Fig.~\ref{fig:crystal_energy_fluctuation_comparison}(a).

\subsection*{Transformation to Normal Modes}  

In Appendix~\ref{app:Hamiltonian}, the Hamiltonian matrix $\mathbb{H}$ is diagonalized to obtain the normal modes of the ion crystal; we therefore refer to the map below as the \emph{normal-mode transformation}. Using the previous transformations, the Tressian tensor can be expressed in normal-mode coordinates. This allows us to write the third-order anharmonic coupling in terms of the canonical normal-mode variables, $Q_n$ and $P_n$, where nonlinear couplings between modes can be directly analyzed.

We extend the Tressian to a $6N\times 6N\times 6N$ tensor in position–velocity space by padding with zeros. Let $\ket{X}\in\mathbb{R}^{6N}$ collect Cartesian positions and velocities, and let $\ket{Z}\in\mathbb{R}^{6N}$ collect normal-mode coordinates as $\ket{Z}=(Q_1,\ldots,Q_{3N},P_1,\ldots,P_{3N})^\top$. The transformation to normal-mode coordinates is
\begin{equation}
    \ket{Z} = \mathbb{A}\,\ket{X},\qquad \mathbb{A} = \mathbb{S}^{-1}\mathbb{T},
\label{eq:composite_transformation_matrix}
\end{equation}
where $\mathbb{S}$ and $\mathbb{T}$ are defined in Appendix~\ref{app:Hamiltonian}. In this basis the Hamiltonian matrix is diagonal (normal-mode basis).

The Tressian in the normal-mode basis is obtained by the standard change-of-basis for a rank-3 covariant tensor under $Z=\mathbb{A}X$:
\begin{equation}
    T^{(\mathrm{nm})}_{a b c}
    = \sum_{i,j,k=1}^{6N} T_{i j k}\,(\mathbb{A}^{-1})_{a i}\,(\mathbb{A}^{-1})_{b j}\,(\mathbb{A}^{-1})_{c k},
\label{eq:tressian_normal_modes}
\end{equation}
where $a,b,c$ index normal-mode phase-space coordinates and $i,j,k$ index Cartesian phase-space coordinates.

The Tressian elements in the normal-mode basis couple triples of normal-mode variables, so the third-order Hamiltonian reads
\begin{equation}
\label{eq:thirdOrderCorrectionTerms}
    E^{(3)}=\frac{1}{3!}\sum_{n,m,p=1}^{3N}\ \sum_{X,Y,Z\in\{Q,P\}}
    T^{XYZ}_{nmp}\, X_n Y_m Z_p,
\end{equation}
where $X_n, Y_m, Z_p$ are the normal-mode variables (position or momentum) associated with modes $n,m,p$, respectively.

\section{Quantization, Interaction Picture, and RWA}
\label{app:quantization}
In the previous appendix section, we derived expressions for the second- and third-order contributions to the energy in terms of normal-mode variables,
\begin{equation}
  E \;=\; E^{(2)} + E^{(3)} + \cdots,
  \label{eq:totalEnergy_rescale}
\end{equation}
with classical, non-dimensional Hamiltonians obtained by scaling lengths by $l_0$, time by $\omega_0^{-1}$, masses by $m_0$, and energies by $E_0 \equiv m_0 \omega_0^2 l_0^2$.
Throughout, all angular frequencies are reported in units of $\omega_0$.

\emph{Quantum non-dimensionalization.}
Introducing the quantum scales $\hbar\omega_0$ and $z_0=\sqrt{\hbar/(m_0\omega_0)}$ defines the (dimensionless) quantum expansion parameter
\begin{equation}
  \epsilon_0 \;=\; \left(\frac{\hbar\omega_0}{E_0}\right)^{1/2}
  \;=\; \frac{z_0}{l_0}
  \;=\; \sqrt{\frac{\hbar}{m_0\omega_0 l_0^{2}}}
  \;\ll\; 1,
  \label{eq:quantumExpansionParameter_rescale}
\end{equation}
which sets the scale of anharmonic terms.

\emph{Canonical and ladder operators.}
Promote the normal-mode variables to operators with
\begin{equation}
  [Q_n,P_m] \;=\; i\,\epsilon_0^2\,\delta_{nm},\qquad
  [Q_n,Q_m]=[P_n,P_m]=0,
  \label{eq:canonicalCommutationRelations_rescale}
\end{equation}
and define
\begin{gather}
  Q_n = \frac{\epsilon_0}{\sqrt{2}}\,(a_n+a_n^\dagger),\qquad
  P_n = \frac{\epsilon_0}{i\sqrt{2}}\,(a_n-a_n^\dagger),\notag\\
  [a_n,a_m^\dagger]=\delta_{nm},\quad [a_n,a_m]=[a_n^\dagger,a_m^\dagger]=0.
  \label{eq:raisingLoweringOperators_rescale}
\end{gather}

\emph{Harmonic Hamiltonian.}
For the quadratic piece,
\begin{equation}
  \tilde{\mathcal{H}}^{(2)} \equiv \frac{\mathcal{H}^{(2)}}{E_0}
  = \sum_n \frac{\omega_n}{2}\,(Q_n^2+P_n^2),
\end{equation}
substituting Eq.~\eqref{eq:raisingLoweringOperators_rescale} and dividing by $\epsilon_0^2$ yields the quantum Hamiltonian in units of $\hbar\omega_0$:
\begin{equation}
  \frac{\hat{\mathcal{H}}^{(2)}}{\hbar\omega_0}
  = \sum_n \omega_n \Big(a_n^\dagger a_n + \tfrac{1}{2}\Big),
  \label{eq:secondOrderHamiltonian_rescale}
\end{equation}
where the (dimensionless) $\omega_n$ are already expressed in units of $\omega_0$.
We drop the constant zero-point term when convenient.

\emph{Degree counting and $\epsilon_0$.}
Let $\tilde{\mathcal{H}}^{(d)}$ be a classical term of total degree $d$ in $(Q,P)$.
Each $Q$ or $P$ contributes a factor of $\epsilon_0$ under Eq.~\eqref{eq:raisingLoweringOperators_rescale}, so the operator substitution gives an overall $\epsilon_0^{\,d}$.
Converting from the classical energy unit $E_0$ to $\hbar\omega_0=\epsilon_0^2 E_0$ contributes an additional factor $1/\epsilon_0^2$.
Thus
\begin{equation}
  \frac{\hat{\mathcal{H}}^{(d)}}{\hbar\omega_0}
  \;\sim\; \epsilon_0^{\,d-2}
  \qquad\text{(up to numerical $1/\sqrt{2}$ factors).}
\end{equation}
In particular, cubic terms ($d=3$) carry an overall $\epsilon_0$, while quartic terms ($d=4$) carry $\epsilon_0^{2}$.
For typical trapped-ion parameters $\epsilon_0\sim 2\times 10^{-3}$--$4\times 10^{-3}$, quartic and higher terms are suppressed by $\epsilon_0^2\approx 4\times 10^{-6}$--$1.6\times 10^{-5}$ relative to the harmonic piece.
Accordingly, in this work we neglect $d\ge 4$ terms; however, the fourth-order terms contain number-conserving products such as $a_n^\dagger a_n\,a_m^\dagger a_m$ that survive the RWA and can become important away from strong third-order resonances \cite{Nie2009,Home2011}. 

\emph{Interaction picture and RWA.}
Going to the interaction picture with respect to $\hat{\mathcal{H}}^{(2)}$ sends $a_n \mapsto a_n e^{-i\omega_n t}$.
Retaining only near-resonant contributions of the cubic Hamiltonian (rotating-wave approximation) gives
\begin{equation}
  \frac{\hat{\mathcal{H}}^{(3)}_\mathrm{RWA}}{\hbar\omega_0}
  = \epsilon_0 \sum_{n,m,p} C^{\mathrm{RWA}}_{nmp}\; a_n a_m a_p^\dagger\, e^{+i\Delta_{nmp} t}
  + \mathrm{H.c.},
  \label{eq:thirdOrderHamiltonian_rescale}
\end{equation}
with the \emph{nonlinear detuning}
\begin{equation}
  \Delta_{nmp} \;=\; \omega_p - \omega_n - \omega_m,
  \label{eq:detuning_rescale}
\end{equation}
and $t$ measured in units of $\omega_0^{-1}$.
Here and throughout, $\Delta$ denotes a \emph{nonlinear detuning}, and the \emph{gate} detuning is reserved as $\delta$ in \MS\ contexts.

\emph{Coefficients from the classical tensor.}
The time-independent couplings $C^{\mathrm{RWA}}_{nmp}$ are obtained by writing the classical cubic Hamiltonian
\[
\tilde{\mathcal{H}}^{(3)} \;=\; \frac{1}{6}\sum_{n,m,p}\sum_{X,Y,Z\in\{Q,P\}} T^{XYZ}_{nmp}\, X_n Y_m Z_p,
\]
substituting Eq.~\eqref{eq:raisingLoweringOperators_rescale}, keeping only RWA-allowed operator products, and collecting prefactors.
This algebra is automated in the code via symbolic manipulation (e.g., \texttt{sympy}).
Since $\Delta_{nmp}$ depends only on $\{\omega_k\}$, it can be pre-computed to preselect candidate triads efficiently, we denote the choice of resonance frequency cutoff as $\Delta_\mathrm{cut}$. 
We calculate only terms with $|\Delta_{nmp}| \le \Delta_\mathrm{cut}$ (typically $\Delta_\mathrm{cut}\sim 0.01$--$0.1$) to define “near-resonant.” 

\section{Simulations}
\label{app:simulations}
In this section, we describe the simulation parameters used in the experiments, including the code and packages utilized for running the simulations and performing the analysis.
Our analysis and simulations are conducted using Python. 
For quantum simulations, we use the \texttt{QuTiP} package \cite{Johansson2013}, for symbolic manipulations used to derive the Hamiltonians we use \texttt{SymPy} \cite{Meurer2017}, and for transforming the Tressian tensor to the mode basis we leverage GPU acceleration with \texttt{TensorFlow}.
For ion numbers $N = 100$, the use of GPU acceleration is crucial to identify \co in the ion crystal in a reasonable time frame.

Although \texttt{QuTiP} has built-in support for simulating thermal states, these simulations can become computationally expensive for large thermal occupations of the motional modes. 
Therefore, we implement a weighted sum over pure state evolutions to reconstruct the thermal state. 
Our approach is briefly described here. 

We consider a system of $M$ motional modes, where the Fock basis of each mode is truncated to a maximum occupation number $n_j^{\text{max}}$.
$n_j^{\text{max}}$ is chosen such that with the initial thermal occupation $\bar{n}_j$, the cumulative probability of Fock states above $n_j^{\text{max}}$ is negligible.
To notate the occupation numbers of the modes, we define a vector of Fock numbers $\mathbf{n}$:
\begin{equation}
    \begin{gathered}
    \mathbf{n} = (n_0, n_1, \ldots, n_M), \\ 
    \sum_\mathbf{n} = \sum_{n_1=0}^{n_1^{\text{max}}} \sum_{n_2=0}^{n_2^{\text{max}}} \cdots \sum_{n_M=0}^{n_M^{\text{max}}}, 
    \end{gathered}
    \label{eq:sum_over_M_modes_sims}
\end{equation}
where the sum is over all composite Fock states of the motional modes. 

In our simulations, typically the lowest mode's thermal occupation, $\bar{n}_0$, is non-zero, while the other modes are initialized in their ground states.
In general the initial state of the system is given by a density matrix $\rho(0)$, which is a weighted sum over the pure states of the motional modes and spins, each with a probability $p_\mathbf{n}$:  
\begin{equation} 
    \begin{gathered}
    \rho(0) = \sum_\mathbf{n} p_\mathbf{n} \ket{\psi_\mathbf{n}(0)}\bra{\psi_\mathbf{n}(0)}, \\ 
    \text{where } \ket{\psi_\mathbf{n}(0)} = \ket{\text{spin}_0} \otimes \ket{\mathbf{n}},
    \end{gathered}
    \label{eq:initial_rho_sims}
\end{equation}
where $\ket{\text{spin}_0}$ is the initial state of the spins, which is typically the ground state $\ket{gg}$ in our simulations.

The thermal occupation probabilities for each mode $p_{n_j}^{(j)}$ are renormalized such that the sum over all Fock states up to $n_j^{\text{max}}$ is equal to 1. 
The occupation probability of a composite Fock state $\mathbf{n}$ is given by the product of the probabilities of each mode's occupation number $n_j$:
\begin{equation} 
    p_\mathbf{n} = \prod_{j=0}^{M} p_{n_j}^{(j)}, \text{ where } p_{n_j}^{(j)} = \frac{\bar{n}_j^{n_j}}{(\bar{n}_j + 1)^{n_j+1}}.  
    \label{eq:thermal_occupation_probabilities_sims}
\end{equation}

Because the state $\rho(0)$ is a weighted sum over pure states, the time evolution of the system can be computed by evolving each pure state $\ket{\psi_\mathbf{n}(t)}$ according to the Schrödinger equation:  
\begin{equation}
    i\frac{d}{dt}\ket{\psi_\mathbf{n}(t)} = \mathcal{H}(t) \ket{\psi_\mathbf{n}(t)}, \quad \rho(t) = \sum_\mathbf{n} p_\mathbf{n} \ket{\psi_\mathbf{n}(t)}\bra{\psi_\mathbf{n}(t)}, 
    \label{eq:rho_t_sims}
\end{equation}
where $\mathcal{H}(t)$ is the Hamiltonian of the system. 

To compute the expectation value of an operator $O$ at time $t$, we use the weighted sum over the evolved pure states:
\begin{equation}
    \langle O(t) \rangle = \sum_\mathbf{n} p_\mathbf{n} \bra{\psi_\mathbf{n}(t)} O \ket{\psi_\mathbf{n}(t)}.
    \label{eq:expectation_value_sims}
\end{equation}

This approach allows us to efficiently simulate the dynamics even when the thermal occupations of the motional modes become relatively large. 
Typically, only the lowest frequency mode has a non-zero thermal occupation in our simulations, which significantly reduces the number of pure state evolutions required.
The number of Fock state simulations is chosen so that the cumulative probability of the neglected states is below a specified threshold, typically $10^{-3}$.
For each pure state evolution, the Fock basis of the simulation is taken to be the 15 or $2\times (N_{Fock}+3)$, whichever is larger, where $N_\text{Fock}$ is the current pure state's Fock number in the lowest mode.

\paragraph{Molecular Dynamics Mode initialization.}
We initialize specific collective modes by adding a weighted superposition of normal--mode eigenvectors to the equilibrium state.

\emph{rf--trap simulations.}
The eigenvector matrix $\mathrm{evs}$ is stacked as $[\,\Delta\mathbf{r}\ (3N);\ \Delta\mathbf{v}\ (3N)\,]$.
For each selected mode $m$ with real amplitude $a_m$, we compute
\begin{gather} 
\delta\mathbf{r}_i^{(m)} = a_m\,\mathrm{Re}\!\big[\mathrm{evs}_{1:3N,m}\big]_i\,\ell_0, \notag\\
\delta\mathbf{v}_i^{(m)} = a_m\,\mathrm{Re}\!\big[\mathrm{evs}_{3N+1:6N,m}\big]_i\,v_0, \notag
\end{gather} 
reshape to $(N,3)$, sum over the chosen modes, and add the totals to each particle’s position and velocity.

\emph{Penning--trap simulations.}
Here $\mathrm{evs}$ is stored componentwise as
$[\,\Delta x,\Delta y,\Delta z,\Delta v_x,\Delta v_y,\Delta v_z\,]$ for all $N$ ions.
After forming $\delta x_i,\delta y_i,\delta z_i$ (scaled by $\ell_0$) and $\delta v_{x,i},\delta v_{y,i},\delta v_{z,i}$ (scaled by $v_0$) in the lab frame, we transform velocities into the crystal’s rotating frame at angular frequency $\omega_r$:
\[
\delta\mathbf{v}^{(\mathrm{rot})}
= \delta\mathbf{v}^{(\mathrm{lab})} - \boldsymbol{\omega}_r \times \delta\mathbf{r},
\qquad
\boldsymbol{\omega}_r = \omega_r \hat{\mathbf{z}}.
\]
Componentwise,
\[
\delta v_x \leftarrow \delta v_x + \omega_r\,\delta y,\quad
\delta v_y \leftarrow \delta v_y - \omega_r\,\delta x,\quad
\delta v_z \leftarrow \delta v_z.
\]
We then add the summed $\delta\mathbf{r}$ and rotated $\delta\mathbf{v}$ to the ensemble.
Amplitudes $a_m$ are set per mode with random phases, and $\ell_0$ and $v_0$ are the natural length and velocity scales, $v_0 = \omega_0 l_0$, where $l_0$ is defined in Eq.~\eqref{eq:characteristic_energy_and_length}, and $\omega_0$ is a characteristic frequency scale (e.g., the single-ion axial trapping frequency $\omega_z$ in linear chains). 

\subsection{Simulation Details by Figure}

In this section we provide details on the parameters and methods used to generate each figure in the paper.
We use the mixed-state fidelity to quantify gate performance \cite{Jozsa1994} with the target spin state $\ket{\psi_\mathrm{target}} = 1/\sqrt{2}(\ket{gg}-i\ket{ee})$ and the final reduced spin state $\rho_\mathrm{spin}$:
\begin{equation}
\mathcal{F} = \bra{\psi_\mathrm{target}} \rho_\mathrm{spin} \ket{\psi_\mathrm{target}}.
\end{equation}

\textbf{Figure~\ref{fig:crystal_energy_fluctuation_comparison}}
In this figure MD simulations \cite{Tang2019,Tang2021,Johnson2024} of different ion crystals are compared. 
The cyclotronic integrator \cite{Patacchini2009} used for the Penning trap simulation used is described in refs.~\cite{Tang2019,Zaris2025}, while the rf trap integrator is a verlet integrator \cite{Verlet1967} with the rf potential treated in the pseudopotential approximation using the same code-base.
All ion crystals contain $N = 53$ ions and are initialized with random phases and mode amplitudes corresponding to a temperature of 100 $\mu$K.
All simulations are evolved for 10 ms with a time step of 1 ns, and the trajectories are saved every 100 time steps. 
$N = 53$ is chosen to make comparisons with the linear ion chain described in the experiment of ref.~\cite{Zhang2017}. 
In the rf 2D crystal and linear ion crystal simulations $^{171}\text{Yb}^+$ ions are used, with a mass of $m = 170.9363315$ amu.
In the Penning 2D crystal simulations, $^9\text{Be}^+$ ions are used, with a mass of $m = 9.0121822$ amu.
The trapping potential in this experiment was nearly harmonic. 
We chose trapping parameters based on those used in the experiment, with $\omega_z = 2\pi \times 170$ kHz, $\omega_y = 2\pi \times 4.85$ MHz, and $\omega_x = 2\pi \times 5.0$ MHz.
Although the trapping potentials are different in the 2D ion crystals, we fix the aspect ratios of the trapping potentials to be the same. 
The rf 2D ion crystal frequencies were chosen to match Reference~\cite{Kiesenhofer2023}, however, we note that unlike the experiment, which used $^{40}\text{Ca}^+$ ions, we use $^{171}\text{Yb}^+$ ions in our simulations to make comparisons with the linear chain more direct. 
The trapping parameters for the rf 2D ion crystal are $\omega_z, \omega_y, \omega_x = 2\pi \times 343$ kHz, $2\pi \times 680$ kHz, and $2\pi \times 2196$ kHz, respectively.    
This yields parameters for the Penning trap ion crystal of $\omega_z = 2\pi\times 1.58$ MHz, $\omega_r \approx 2\pi\times 188.74$ kHz, $B = 4.4488$ T, and $\delta \approx 3.574\times 10^{-2}$. 
The simulations are evolved for 10 ms with a time step of 1 ns, and the mode energy and its standard deviation are calculated over the full simulation time.

\textbf{Figure~\ref{fig:two_ion_crystal_coupling_comparison}.}
We compare three simulation techniques for the two-ion coupling problem in a linear crystal.
First, we integrate the full Coulomb dynamics using a molecular-dynamics (MD) code~\cite{Tang2019,Tang2021,Johnson2024} with a timestep of $1~\mathrm{ns}$ and total duration $50~\mu\mathrm{s}$.
Trap frequencies are chosen so that the breathing-mode frequency is approximately twice the radial tilt-mode frequency, $\omega_\mathrm{B}\approx 2\omega_\mathrm{T}$, satisfying the resonance condition in Eq.~\eqref{eq:two_ion_resonance}.
The axial frequency is $\omega_z$.
The weaker radial frequency is $\omega_y$ (chosen near resonance via $\omega_\mathrm{B}\approx 2\omega_\mathrm{T}$).
The orthogonal radial frequency is set to $\omega_x=2\pi\times 10~\mathrm{MHz}$ to keep it spectrally isolated.
All angular frequencies are reported in units of $\omega_0$; here we take $\omega_0\equiv \omega_z$ for the two-ion example (see Appendix~\ref{app:quantization}).

The \emph{classical reduced model} (CRM) evolves only the tilt and breathing canonical variables under the reduced Hamiltonian in Eq.~\eqref{eq:2-ion_hamiltonian}, using a symplectic integrator for inseparable Hamiltonians~\cite{Pihajoki2014}.
The \emph{quantum reduced model} (QRM) uses the interaction Hamiltonian in Eq.~\eqref{eq:2-ion_interaction_hamiltonian} with and without the rotating-wave approximation (RWA).
QRM dynamics are computed in Fock space with \texttt{QuTiP}~\cite{Johansson2013}.
Each mode's Fock basis is truncated so that the cumulative probability outside the truncated space remains below $10^{-3}$ when the system is initialized in coherent states whose amplitudes reproduce the target mean energies (e.g., $0.1~\mathrm{mK}$ for the tilt mode).
Canonical variables, non-dimensionalization, and ladder-operator conventions follow Appendix~\ref{app:quantization}.

\paragraph{Reduced two-mode Hamiltonian and classical coefficients.}
With the parameters used for Fig.~\ref{fig:two_ion_crystal_coupling_comparison}, the dimensionless linear frequencies are
\[
\omega_\mathrm{T}=\frac{\sqrt{3}}{2},\qquad \omega_\mathrm{B}=\sqrt{3}.
\]
Therefore
\begin{align}
\frac{\mathcal{H}_\text{lin}}{E_0}
&=\frac{\omega_\mathrm{T}}{2}\!\left(Q_\mathrm{T}^2+P_\mathrm{T}^2\right)
 +\frac{\omega_\mathrm{B}}{2}\!\left(Q_\mathrm{B}^2+P_\mathrm{B}^2\right)\notag\\
&=\underbrace{\frac{\sqrt{3}}{4}}_{0.4330127}\!\left(Q_\mathrm{T}^2+P_\mathrm{T}^2\right)
 +\underbrace{\frac{\sqrt{3}}{2}}_{0.8660254}\!\left(Q_\mathrm{B}^2+P_\mathrm{B}^2\right).
\end{align}
To leading nonlinear order (with our mode-phase convention),
\[
\frac{\mathcal{H}_\text{nl}}{E_0} 
= \xi_\text{class}\,P_\mathrm{B}\,Q_\mathrm{T}^2 \;-\; \chi\,P_\mathrm{B}^{\,3} \;+\; \cdots,
\]
with 
\[
\xi_\text{class}=\sqrt{\frac{\omega_\mathrm{B}^{3}}{2\,\omega_\mathrm{T}^{2}}}, 
\qquad
\chi=\frac{\sqrt{2\,\omega_\mathrm{B}}}{3}.
\]
Using $\omega_\mathrm{T}=\sqrt{3}/2$ and $\omega_\mathrm{B}=\sqrt{3}$ gives
\[
\xi_\text{class}=\sqrt{2\sqrt{3}},\quad 
\chi=\frac{\sqrt{2\sqrt{3}}}{3}\ldots,
\]
so
\[
\frac{\mathcal{H}_\text{nl}}{E_0}
= \underbrace{1.861209718}_{\xi_\text{class}}\,P_\mathrm{B} Q_\mathrm{T}^2 \;-\;
\underbrace{0.620403239}_{\chi}\,P_\mathrm{B}^3+\cdots,
\]
which matches the coefficients computed from the Tressian-tensor construction (Appendix~\ref{app:tressian}) and used by the code (assertions at relative tolerance $10^{-6}$).

\paragraph{Quantum RWA coupling and identification with Appendix~\ref{app:quantization}.}
For the near-resonant three-wave term $P_\mathrm{B} Q_\mathrm{T}^2$ (total degree $d=3$), the mapping in Appendix~\ref{app:quantization} contributes an overall $\epsilon_0$ in $\hat{\mathcal{H}}/(\hbar\omega_0)$.
Writing $\tilde\omega=\omega/\omega_0$, and letting $\tilde z_0$ denote the dimensionless half-separation (so $l_0=2\tilde z_0$), the interaction in the RWA can be written as
\[
\frac{\mathcal{H}_\text{int}^{(\mathrm{RWA})}}{\hbar\omega_0}
= \xi\,\big(a_\mathrm{T}^{2}a_\mathrm{B}^\dagger + a_\mathrm{T}^{\dagger 2}a_\mathrm{B}\big),
\qquad
\xi=\frac{\epsilon_0}{8\,\tilde z_0}\,\frac{\tilde\omega_\mathrm{B}^{3/2}}{\tilde\omega_\mathrm{T}},
\]
consistent with the general form in Eq.~\eqref{eq:thirdOrderHamiltonian_rescale} and with Ref.~\cite{Ding2017co}.
In our mode-analysis normalization $\tilde z_0=1/2$, so
\[
\xi=\frac{\epsilon_0}{4}\sqrt{\frac{\tilde\omega_\mathrm{B}^{3}}{\tilde\omega_\mathrm{T}^{2}}}
=\frac{\epsilon_0}{2\sqrt{2}}\;\xi_\text{class}.
\]
Equivalently, in the main-text notation of Eq.~\eqref{eq:2-ion_interaction_hamiltonian} one may identify $\xi=\epsilon_0\,C^{\mathrm{RWA}}_{\mathrm{TTB}}$, i.e.,
\[
C^{\mathrm{RWA}}_{\mathrm{TTB}}=\frac{1}{2\sqrt{2}}\;\xi_\text{class},
\]
which matches the tensor-derived value used in the simulations.

\textbf{Figure~\ref{fig:n2_motional_detuning_vs_gate_time} \& \ref{fig:n2_motional_detuning_vs_gate_time_thermal}}
We simulate a two-ion \MSGATE\ using the motional Hamiltonian from mode analysis together with the MS drive of Eq.~(\ref{eq:2-ion_ms_hamiltonian}). 
The trapping parameters (and thus the motional Hamiltonian) match Fig.~\ref{fig:two_ion_crystal_coupling_comparison}; in particular, the tilt and breathing mode frequencies are $\omega_\mathrm{T}=\sqrt{3}/2$ and $\omega_\mathrm{B}=\sqrt{3}$ in our units, and the nonlinear coupling coefficient $\xi$ is as defined in the preceding appendix section.

\emph{Motional Hamiltonian and quantization.}
Given the trap parameters, we compute the normal-mode frequencies and cubic coupling tensor, then form the classical reduced two-mode Hamiltonian $H=H_\mathrm{lin}+H_\mathrm{nl}$. 
We quantize with the expansion parameter $\epsilon_0$ (defined earlier in the appendix) and construct the RWA ladder-form $H_\mathrm{mot}^{(\mathrm{RWA})}$ used for time evolution.

\emph{MS drive and gate parameters.}
The MS drive is applied equally to both ions on the breathing (gate) mode. 
For a desired gate time $T_\mathrm{gate}$ and loop number $k\in\mathbb{N}$, we use
\begin{equation}
\delta_\mathrm{MS}=\frac{2\pi k}{T_\mathrm{gate}},
\qquad
\Omega_r=\frac{\delta_\mathrm{MS}}{2\sqrt{k}\,\eta_\mathrm{B}},
\label{eq:ms_gate_parameters_sims}
\end{equation}
and set $k=1$ unless otherwise stated. 
To connect with experimental parameters we take a Raman wavelength $\lambda=355~\mathrm{nm}$ and a nearly counter-propagating geometry so that $\Delta k\simeq 2(2\pi/\lambda)$. 
The Lamb--Dicke parameter for the breathing mode is
\begin{equation}
\eta_\mathrm{B}=\frac{\eta_0}{\sqrt{2\sqrt{3}}}, 
\qquad
\eta_0=\Delta k\,\sqrt{\frac{\hbar}{2m\omega_z}},
\label{eq:lamb_dicke_parameter_sims}
\end{equation}
with $m$ the mass of $^{171}\mathrm{Yb}^+$. 
This yields $\eta_\mathrm{B}\approx 0.096$, representative of typical experiments.

\emph{Target state and metrics.}
The target Bell state is
\begin{equation}
\ket{\Psi_\mathrm{target}}
=\frac{1}{\sqrt{2}}(\ket{gg}-i\ket{ee}).  
\label{eq:target_bell_state_sims}
\end{equation}
Fidelity is 
$\mathcal{F}=\bra{\Psi_\mathrm{target}}\rho_\mathrm{spin}(T_\mathrm{gate})\ket{\Psi_\mathrm{target}}$, where $\rho_\mathrm{spin}(T_\mathrm{gate})$ is the reduced spin state at the end of the gate.  
We also report the two-qubit von Neumann entropy $S=-\mathrm{Tr}\!\big(\rho_\mathrm{spin}\log_2\rho_\mathrm{spin}\big)$ of the reduced spin state $\rho_\mathrm{spin}$.
The energy of the tilt (spectator) mode is reported at the end of the gate, $\langle \mathcal{H}_\mathrm{T}\rangle/\hbar\omega_z$, across the scan range in Fig.~\ref{fig:n2_motional_detuning_vs_gate_time_thermal}(c).

\emph{Scans and truncation (ground-state panel).}
For Fig.~\ref{fig:n2_motional_detuning_vs_gate_time} we use a $51\times51$ grid over
$N_\mathrm{period}\in[50,500]$ and $\Delta\omega_y/2\pi\in[-30,30]~\mathrm{kHz}$; the top axis converts $N_\mathrm{period}$ to $T_\mathrm{gate}$.
Both modes start in $\ket{0}$ and the spins in $\ket{gg}$. 
We truncate the Fock basis as $[N_\mathrm{T},N_\mathrm{B}]=[22,10]$ for all simulations.  

\emph{Scans and truncation (thermal spectator panel).}
For Fig.~\ref{fig:n2_motional_detuning_vs_gate_time_thermal} we use a $25\times25$ grid over the same detuning range and $N_\mathrm{period}\in[50,500]$. 
The tilt (spectator) mode is initialized in a thermal state with $\bar n_\mathrm{spec}=1$ or $2$ while the breathing mode starts in $\ket{0}$; spins start in $\ket{gg}$. 
We truncate the Fock basis as $[N_\mathrm{T},N_\mathrm{B}]=[22,10]$ for all simulations.  

\textbf{Figure~\ref{fig:ms_gate_populations_and_energies_3_modes}.}
We plot \textbf{(a)} the phase-space trajectories of the $\ket{gg}_y$ and $\ket{ee}_y$ spin branches during the \MS\ gate, \textbf{(b)} the base-2 spin entropy during the gate, and \textbf{(c)} a back-action metric that compares the nonlinear forcing on the bus mode to the \MS\ drive.
Here the $y$-basis states are $\ket{g}_y=\tfrac{1}{\sqrt{2}}(\ket{g}-i\ket{e})$ and $\ket{e}_y=\tfrac{1}{\sqrt{2}}(\ket{e}+i\ket{g})$.

\paragraph*{Initialization and ensemble method.}
The system is initialized in the spin state $\ket{gg}$ in the computational ($z$) basis.
We consider three motional modes ordered by frequency $\omega_c<\omega_b<\omega_a$, with the highest-frequency mode ($a$) used as the \emph{bus}.
For this figure, the bus ($a$) and intermediate spectator ($b$) start in their ground states, while the lowest-frequency spectator ($c$) is prepared in a thermal state with mean occupation $\bar{n}_\text{spec}=20$.
Thermal dynamics are simulated by an ensemble of pure Fock states with thermal weights, and expectation values are reconstructed via the ensemble formula of Eq.~(\ref{eq:expectation_value_sims}).
Unless otherwise stated, we use $N_b=4$ and $N_a=8$ Fock states for the higher-frequency modes; for each ensemble member with spectator occupancy $n_c$, the pure-state evolution of mode $c$ is truncated at $N_c=n_c+3$.
Convergence was verified by increasing each cutoff until all plotted quantities changed by less than $10^{-3}$.

\paragraph*{Spin observables and diagnostics.}
Spin-resolved phase-space trajectories are computed using the projectors $P_{gg,y}=\ket{gg}_y\!\bra{gg}_y$ and $P_{ee,y}=\ket{ee}_y\!\bra{ee}_y$ to form the conditional expectation values
\begin{equation}
  \langle x \rangle_{s}(t)
  = \frac{\mathrm{Tr}\!\left[(P_{s}\!\otimes x)\,\rho(t)\right]}{\mathrm{Tr}\!\left[(P_{s}\!\otimes \mathbb{I})\,\rho(t)\right]},
  \quad s\in\{gg_y,ee_y\},
\end{equation}
where $x=(a+a^\dagger)/\sqrt{2}$, and $\langle p \rangle_{s}(t)$ is defined similarly with $p=-i(a-a^\dagger)/\sqrt{2}$.
The base-2 von Neumann entropy of the reduced spin state, $S_2(\rho_\mathrm{spin})$, is computed as in Fig.~\ref{fig:n2_motional_detuning_vs_gate_time}.

\emph{Back-action metric.}
Let $a,b,c$ denote the annihilation operators of the three modes $(\omega_a,\omega_b,\omega_c)$ and assume a sum-frequency three-wave interaction in the RWA,
\begin{equation}
  H_{\mathrm{int}}^{(\mathrm{RWA})}
  = g\big(c\,b\,a^\dagger + c^\dagger b^\dagger a\big),
  \qquad (\omega_c+\omega_b\approx\omega_a).
\end{equation}
The Heisenberg equation for the bus mode gives
\[
  \dot a(t)=i\,[H_{\mathrm{int}}^{(\mathrm{RWA})},a]
  = -\,i\,g\,c(t)b(t).
\]
We compare this instantaneous nonlinear drive to the \MS\ forcing amplitude on the bus, which in the quadrature form of Eq.~\eqref{eq:ms_gate_hamiltonian} is $\eta\,\Omega_r\,\sqrt{2}$.
Projecting onto the $\ket{ee}_y$ spin branch with $P_{ee,y}=\ket{ee}_y\!\bra{ee}_y$, define the conditional correlator
\begin{equation}
  \langle c b\rangle_{ee_y}(t)
  = \frac{\mathrm{Tr}\!\left[(P_{ee,y}\!\otimes c b)\,\rho(t)\right]}
         {\mathrm{Tr}\!\left[(P_{ee,y}\!\otimes \mathbb{I})\,\rho(t)\right]}.
\end{equation}
This motivates the back-action ratio (main text Eq.~\eqref{eq:backaction_def})
\begin{equation}
  \mathcal{R}_{ee_y}(t)
  = \frac{|g|}{\eta\,\Omega_r\,\sqrt{2}}\;
    \big|\langle c b\rangle_{ee_y}(t)\big|,
  \label{eq:backaction_def_app}
\end{equation}
which we use as a diagnostic in Fig.~\ref{fig:ms_gate_populations_and_energies_3_modes}.

\textbf{Figure~\ref{fig:ms_gate_two_level_osc_v_gate_time}.}
We simulate a three-mode (sum-frequency) \co{} interaction during an \MS{} gate, scanning the effective two-level (TL) oscillation period and the gate time.
The mode frequencies satisfy $\omega_a:\omega_b:\omega_c=4:3:1$, with the highest-frequency mode $a$ used as the bus.
We normalize $\omega_a=1$, so the bus period is $T_\text{bus}=2\pi/\omega_a$.

\emph{Scan parameters.}
The TL period is scanned as $T_\text{TL}\in[500,\,5000]\times T_\text{bus}$.
Operationally, $T_\text{TL}$ sets the strength of the nonlinear coupling $g$ by requiring that the isolated two-level exchange $\ket{1_a,0_b,0_c}\leftrightarrow\ket{0_a,1_b,1_c}$ has phase-oscillation period $T_\text{TL}$; on resonance this gives $g=2\pi/T_\text{TL}$ up to factors of $\sqrt{2}$ arising from the ladder operator expansion (since the TL Rabi frequency equals $g$ in our units).
Scanning $T_\text{TL}$ therefore rescales $g$ while keeping the mode frequencies fixed.
The \MS{} gate time is scanned as $T_\text{gate}\in[50,\,500]\times T_\text{bus}$.
For each $T_\text{gate}$ we set the gate detuning $\delta_\text{gate}=2\pi/T_\text{gate}$ (loop number $k=1$); the drive amplitude $\Omega_r$ follows the standard relation used elsewhere in the appendix for a maximally entangling gate.

\emph{Hamiltonian and initial state.}
The motional Hamiltonian includes the RWA three-wave term
\[
H_\text{int}^{(\mathrm{RWA})}
= g\big(c\,b\,a^\dagger + c^\dagger b^\dagger a\big),
\qquad (\omega_b+\omega_c\approx\omega_a),
\]
added to the linear normal-mode Hamiltonian and the \MS{} drive in quadrature form acting on the bus [Eq.~\eqref{eq:ms_gate_hamiltonian}].
All three modes are initialized in their ground states and the two-qubit register in $\ket{gg}$.

\emph{Numerics and truncation.}
We evolve pure states (no ensemble averaging needed here, since all modes start in $\ket{0}$).
Fock cutoffs are $(N_c,N_b,N_a)=(5,5,11)$ for the $(c,b,a)$ modes, respectively.
Convergence was spot-checked by increasing individual cutoffs on a subset of grid points and verifying that fidelities changed by less than $10^{-3}$.

\emph{Outputs.}
For each grid point we compute the Bell-state fidelity at $t=T_\text{gate}$ with target $(\ket{gg}-i\ket{ee})/\sqrt{2}$.
The figure reports the fidelity over the $(T_\text{TL},T_\text{gate})$ scan.

\textbf{Figure~\ref{fig:ms_gate_temp_v_resonance_scan}.}
We study how temperature broadens the region over which the \MS{}-gate fidelity is degraded by a three-mode (sum-frequency) \co{} interaction with frequencies ordered as $\omega_c<\omega_b<\omega_a$ and $\omega_a$ the bus.
We normalize $\omega_a=1$, so $T_\text{bus}=2\pi/\omega_a$.

\emph{Detuning parameterization.}
We scan detuning by varying only the lowest-frequency spectator while holding the bus and intermediate spectator fixed:
\[
\boldsymbol{\omega}=(\omega_c,\omega_b,\omega_a)
=\big(\omega_\mathrm{split}-\Delta_\mathrm{mot},\ \omega_c-\omega_\mathrm{split},\ \omega_a \big),
\]
We report $\Delta_\mathrm{mot}$ in units of $\omega_a$.

\emph{Gate settings.}
The gate time is fixed to $T_\text{gate}=200\,T_\text{bus}$ with $\delta_\text{gate}=2\pi/T_\text{gate}$ (loop number $k=1$).
The TL oscillation period is fixed to $T_\text{TL}=5000\,T_\text{bus}$, which sets the nonlinear coupling strength $g\propto 1/T_\text{TL}$.
The spins are initialized in $\ket{gg}$.

\emph{Initialization and ensemble method.}
The lowest-frequency mode $c$ is prepared in a thermal state with mean occupation $\bar n_\text{spec}\in\{0.1,\,1,\,10\}$.
Modes $b$ (intermediate spectator) and $a$ (bus) start in their ground states.
Thermal dynamics are simulated with the weighted sum over pure Fock states (ensemble method) of Eq.~(\ref{eq:expectation_value_sims}).

\emph{Numerics and truncation.}
We use Fock cutoffs $(N_c,N_b,N_a)=(N_c(\bar n_\text{spec}),\,3,\,8)$.
For each ensemble member with spectator occupancy $n_c$, the $c$-mode basis is truncated adaptively as $N_c=n_c+3$.
The largest pure Fock state in the ensemble was determined by the thermal weight cutoff of $10^{-4}$.

\emph{Output.}
For each $(\Delta_\mathrm{mot},\bar n_\text{spec})$ we compute the Bell-state fidelity at $t=T_\text{gate}$ for the target $(\ket{gg}-i\ket{ee})/\sqrt{2}$.
Increasing $\bar n_\text{spec}$ broadens the detuning window over which fidelity is noticeably reduced.

\textbf{Figure~\ref{fig:ms_gate_fidelity_vs_num_loops}.}
This figure uses the weighted sum over pure states method of Eq.~(\ref{eq:expectation_value_sims}) to compute the gate fidelity versus the number of phase-space loops completed by the bus mode, while keeping the total gate time fixed. 
An on-resonance three-mode (sum-frequency) \co{} interaction is used with the highest-frequency mode acting as the bus.

\[
[\omega_c, \omega_b, \omega_a] = [0.25, 0.75, 1],\qquad T_\text{bus} = \frac{2\pi}{\omega_a}.
\]

The lowest-frequency spectator is initialized thermally, while the other two modes start in their ground states:
\[
\bar n_\text{spec} \in \{0.1,\,1,\,10\}.
\]

We vary the loop number $k$ by adjusting the MS detuning and Rabi frequency so that the bus trajectory encloses $k$ loops during the same gate time:
\[
\delta_\text{MS} = \frac{2\pi k}{T_\text{gate}},\qquad T_\text{gate} \text{ fixed},
\]
and choose the drive amplitude to produce the standard closed-loop MS trajectory at that $k$. 
The bus-mode displacement scales as
\[
d_\text{max} = \frac{1}{\sqrt{2k}}.
\]
Larger $k$ therefore corresponds to smaller excursions and faster phase-space traversal, which typically improves robustness to the nonlinear coupling.

\textbf{Figure~\ref{fig:mode_coupling_resonances_linear}} 
We identify near-resonant three-mode couplings in long linear chains by:
(i) computing equilibrium ion positions in the specified axial potential (anharmonic nearly–equally spaced or harmonic reference),
(ii) constructing and diagonalizing the Hessian to obtain normal-mode frequencies $\{\omega_j\}$ and eigenvectors, and
(iii) evaluating the cubic coupling tensor $T_{nmp}$ in mode coordinates.

\emph{Triad selection.}
For each ordered triplet $(n,m,p)$ we form the sum-frequency detuning
\[
\Delta_{nmp}=\omega_p-\omega_m-\omega_n.
\]
We keep triads with $|\Delta_{nmp}|<0.01\,\omega_0$, where $\omega_0$ is the system's characteristic frequency (we take $\omega_0\equiv\omega_z$), and with tensor magnitude exceeding $|T_{nmp}|>10^{-2}$ in our natural units. 
For each retained triad we build the RWA interaction Hamiltonian $\mathcal{H}_{nmp}^{\mathrm{RWA}}\propto T_{nmp}\,(a_n a_m a_p^\dagger + \mathrm{h.c.})$, extract the effective TL subspace $\{|1_p,0_n,0_m\rangle,\ |0_p,1_n,1_m\rangle\}$, and compute the corresponding two-level oscillation period $\tau_\mathrm{TL}$ and resonance quality as in Sec.~\ref{subsec:TLS}. 
Triads that pass this TL criterion are labeled “coupled” (red).

\emph{Parameters.}
Unless stated otherwise, the $N=25$ chain uses $\omega_x=2\pi\times3.1~\mathrm{MHz}$ and $\omega_y=2\pi\times3.0~\mathrm{MHz}$. 
For the anharmonic nearly–equally spaced case we set $l_0=4.4~\mu\mathrm{m}$ (and also consider a tighter chain at $l_0=2.7~\mu\mathrm{m}$). 
These parameters are based on the experiment in Ref.~\cite{Cetina2022}.
The harmonic references use an axial frequency chosen to match the lowest axial mode of the respective anharmonic case.

\textbf{Figure~\ref{fig:linear_chain_scan_axial_confinement}} 
We sweep the radial-to-axial ratio $\beta=\omega_y/\omega_z$ from 12 to 10.25 in $N_{\mathrm{scan}}=500$ steps, holding $\omega_y=2\pi\times3.0~\mathrm{MHz}$ and $\omega_x=2\pi\times5.0~\mathrm{MHz}$. 
For each $\beta$ we set $\omega_z=\omega_y/\beta$, compute equilibrium positions and normal modes, evaluate the cubic coupling tensor in mode coordinates, and apply the same triad-selection pipeline as Fig.~\ref{fig:mode_coupling_resonances_linear}:
(i) near-resonance test $|\Delta_{nmp}|<0.01\,\omega_0$ with $\Delta_{nmp}=\omega_p-\omega_m-\omega_n$,
(ii) tensor threshold $|T_{nmp}|>10^{-2}$ (natural units), and
(iii) two-level reduction to extract $T_\mathrm{TL}$ under the RWA. 
The $y$-branch spectrum and the distribution of $T_\mathrm{TL}$ over the scan are reported in panels (a) and (b), respectively.

\textbf{Figures~\ref{fig:equilibrium_positions_planar} and \ref{fig:mode_coupling_resonances_planar}.}
We compare two experimental 2D ion crystals: a Penning-trap array of $^9\mathrm{Be}^+$ ions \cite{Gilmore2021} and a monolithic rf-trap array of $^{40}\mathrm{Ca}^+$ ions \cite{Kiesenhofer2023}. 
Although their trapping physics differ, we make a direct comparison by matching the in-plane anisotropy parameters $(\beta,\delta)$ so that the equilibrium geometry is the same up to the species and trap dependent length scale~\cite{Johnson2025}. 
For the rf case we use $\omega_z=2\pi\times343~\mathrm{kHz}$, $\omega_y=2\pi\times680~\mathrm{kHz}$, and $\omega_x=2\pi\times2196~\mathrm{kHz}$ with $N=91$ ions.
For the Penning case we use $\omega_z=2\pi\times1.58~\mathrm{MHz}$ and $B=4.4588~\mathrm{T}$, and choose the rotation frequency $f_\mathrm{rot}$ so that $(\beta,\delta)$ matches the anisotropy of the rf crystal. 
This ensures that the tensor matrix elements, which depend only on the equilibrium positions, are identical between the two cases up to an overall scaling factor.

\emph{Coupling identification.}
For each configuration we compute equilibrium positions, diagonalize the Hessian to obtain mode eigenpairs, evaluate the cubic tensor in mode coordinates, and select triads $(n,m,p)$ by:
(i) near-resonance test $|\Delta_{nmp}|<\Delta_\mathrm{th}$ with $\Delta_{nmp}=\omega_p-\omega_m-\omega_n$,
(ii) tensor threshold $|T_{nmp}|>T_\mathrm{min}$ in natural units, and
(iii) resonance criteria given in Sec.~\ref{subsec:TLS} applied to the RWA TL subspace for each triad.
We use $(\Delta_\mathrm{th},T_\mathrm{min},S_\mathrm{min})=(10^{-3},10^{-3},10^{-1})$ for rf and Penning crystals.
The first two criteria are used to filter triads before applying the more expensive TL reduction.
Spectra are plotted in MHz; coupled triads are highlighted in blue and orange with marker styles distinguishing radial-radial vs.\ radial-axial interactions.

\section{Equally Spaced Ion Chain}
\label{app:equally_spaced_ion_chain}
In experiments an anharmonic trapping potential is engineered to ensure the ions are close to equally spaced.
This section demonstrates how equilibrium positions of the ions can be found using the potential suggested in ref.~\cite{Lin2009}.
Let the ions be organized in a chain along the $z$ axis, and let the axial trapping potential be given by:  

\begin{equation}
    U_{\text{trap}} = \sum_i^N \left( \frac{1}{2} a_2 z_i^2 + \frac{1}{4} a_4 z_i^4 \right)
\label{eq:anharmonic_axial_potential}
\end{equation}

where $a_2$ and $a_4$ are the coefficients of the quadratic and quartic terms, respectively, and $z_i$ is the axial position of the $i$th ion.
Assume that $a_4$ is positive, however, $a_2$ can be positive or negative.
Meanwhile, the trapping potential in the other two directions is harmonic: 

\begin{equation}
    U_{\text{radial}} = \sum_i^N \left( \frac{1}{2} m \omega_x x_i^2 + \frac{1}{2} m \omega_y y_i^2 \right) 
\label{eq:radial_potential}
\end{equation}

where $\omega_x$ and $\omega_y$ are the radial trapping frequencies in the $x$ and $y$ directions, respectively.
Since the ions are arranged in a linear chain, the problem is one-dimensional.
Therefore the equilibrium positions of the ions can be found by minimizing the total potential energy with respect to the positions of the ions, $z_i$.
Although there are two parameters in the potential, $a_2$ and $a_4$, only one parameter is needed to minimize the variance in the spacing of neighboring ions.
The parameter is given as: 

\begin{equation}
    b = \left( \frac{|a_2|}{q^2 k_e} \right)^{2/3} \left( \frac{a_2}{a_4} \right)
\label{eq:b_parameter}
\end{equation}

where $q$ is the charge of the ions, and $k_e$ is Coulomb's constant.   
The system can be normalized with the introduction of a characteristic length, $l_0$, and a characteristic energy, $E_0$: 

\begin{equation}
    l_0 = \left( \frac{k_e q^2}{|a_2|} \right)^{1/3}, \quad E_0 = \frac{k_e q^2}{l_0}. 
\label{eq:characteristic_length_energy}
\end{equation}

Then the total potential energy constrained to the axial direction is:  

\begin{equation}
    U_{\text{total}} = \sum_i^N \left( \frac{1}{2} \operatorname{sgn}(b) z_i^2 + \frac{1}{4} |b| z_i^4 \right) + \frac{1}{2} \sum_{i=1}^{N} \sum_{j\neq i}^{N} \frac{1}{|z_i - z_j|}    
\label{eq:total_axial_potential_dimensionless}
\end{equation}

where energy has been normalized by $E_0$, and length has been normalized by $l_0$, and $\operatorname{sgn} (b)$ is the sign of $b$.    
The value of $b$ is determined by minimizing the variation in the spacing of the qubit ions in the chain.
Let the ions be labled $1, 2, 3, \ldots, N$, and let the equilibrium positions of the ions be $z_1, z_2, z_3, \ldots, z_N$. 
Let the qubit ions be all ions after $N_{\text{aux}}$ ions, and before $N - N_{\text{aux}}$ ions.   
Let $d_i = z_{i+1} - z_i$ be the spacing between the $i$th and $i+1$th ions.
The variance in the spacing of the qubit ions is given by:

\begin{equation}
    s_z = \frac{1}{N - 2 N_{\text{aux}}} \sum_{i=N_{\text{aux}}}^{N - N_{\text{aux}} - 1} \left( \frac{d_i}{\bar{d}} - 1 \right)^2  
\label{eq:spacing_variance}
\end{equation}

where $N_{\text{qubit}} = N - 2 N_{\text{aux}}$ is the number of qubit ions, and $\bar{d}$ is the average spacing between the qubit ions.   
Once $s_z$ is minimized with respect to $b$, it is convenient to rescale the dimensionless equilibrium positions of the ions by $\bar d$, such that the average spacing between the qubit ions is unity.    
This can be done by setting $l_0 = L_\text{target} / \bar d$, where $L_\text{target}$ is the desired average spacing between the qubit ions. 
The dimensionless trapping frequencies can be found by normalizing $\omega_x$ and $\omega_y$ by the characteristic frequency, $\omega_0 = \sqrt{E_0 / m l_0^2}$ = $\sqrt{|a_2| / m}$.

Since the trapping potential is anharmonic, the Tressian tensor will have non-zero contributions from the trapping potential in the axial direction.    
The third order derivatives of the trapping potential in the axial direction are:

\begin{equation}
    T_{iii}^{zzz}= 6 a_4 z_i. 
\label{eq:tressian_zzz}
\end{equation}

\bibliographystyle{apsrev4-2}
\bibliography{bib}

@article{Verlet1967,
  title = {Computer “Experiments” on Classical Fluids. I. Thermodynamical Properties of Lennard-Jones Molecules},
  volume = {159},
  ISSN = {0031-899X},
  url = {http://dx.doi.org/10.1103/PhysRev.159.98},
  DOI = {10.1103/physrev.159.98},
  number = {1},
  journal = {Physical Review},
  publisher = {American Physical Society (APS)},
  author = {Verlet,  Loup},
  year = {1967},
  month = jul,
  pages = {98–103}
}

@article{Jozsa1994,
  author  = {Richard Jozsa},
  title   = {Fidelity for Mixed Quantum States},
  journal = {Journal of Modern Optics},
  year    = {1994},
  volume  = {41},
  number  = {12},
  pages   = {2315--2323},
  doi     = {10.1080/09500349414552171},
}

@article{Cirac1995,
  title = {Quantum Computations with Cold Trapped Ions},
  volume = {74},
  ISSN = {1079-7114},
  url = {http://dx.doi.org/10.1103/PhysRevLett.74.4091},
  DOI = {10.1103/physrevlett.74.4091},
  number = {20},
  journal = {Physical Review Letters},
  publisher = {American Physical Society (APS)},
  author = {Cirac,  J. I. and Zoller,  P.},
  year = {1995},
  month = may,
  pages = {4091–4094}
}

@article{James1998,
  title = {Quantum dynamics of cold trapped ions with application to quantum computation},
  volume = {66},
  ISSN = {1432-0649},
  url = {http://dx.doi.org/10.1007/s003400050373},
  DOI = {10.1007/s003400050373},
  number = {2},
  journal = {Applied Physics B: Lasers and Optics},
  publisher = {Springer Science and Business Media LLC},
  author = {James,  D.F.V.},
  year = {1998},
  month = feb,
  pages = {181–190}
}

@article{Srensen2000,
  title = {Entanglement and quantum computation with ions in thermal motion},
  volume = {62},
  ISSN = {1094-1622},
  url = {http://dx.doi.org/10.1103/PhysRevA.62.022311},
  DOI = {10.1103/physreva.62.022311},
  number = {2},
  journal = {Physical Review A},
  publisher = {American Physical Society (APS)},
  author = {Sørensen,  Anders and Mølmer,  Klaus},
  year = {2000},
  month = jul 
}

@article{Kielpinski2002,
  title = {Architecture for a large-scale ion-trap quantum computer},
  volume = {417},
  ISSN = {1476-4687},
  url = {http://dx.doi.org/10.1038/nature00784},
  DOI = {10.1038/nature00784},
  number = {6890},
  journal = {Nature},
  publisher = {Springer Science and Business Media LLC},
  author = {Kielpinski,  D. and Monroe,  C. and Wineland,  D. J.},
  year = {2002},
  month = jun,
  pages = {709–711}
}

@article{Marquet2003,
  title = {Phonon-phonon interactions due to non-linear effects in a linear ion trap},
  volume = {76},
  ISSN = {1432-0649},
  url = {http://dx.doi.org/10.1007/s00340-003-1097-7},
  DOI = {10.1007/s00340-003-1097-7},
  number = {3},
  journal = {Applied Physics B: Lasers and Optics},
  publisher = {Springer Science and Business Media LLC},
  author = {Marquet,  C. and Schmidt-Kaler,  F. and James,  D.F.V.},
  year = {2003},
  month = mar,
  pages = {199–208}
}

@article{Porras2006,
  title = {Quantum Manipulation of Trapped Ions in Two Dimensional Coulomb Crystals},
  volume = {96},
  ISSN = {1079-7114},
  url = {http://dx.doi.org/10.1103/PhysRevLett.96.250501},
  DOI = {10.1103/physrevlett.96.250501},
  number = {25},
  journal = {Physical Review Letters},
  publisher = {American Physical Society (APS)},
  author = {Porras,  D. and Cirac,  J. I.},
  year = {2006},
  month = jun 
}

@article{Zhu2006,
  title = {Trapped Ion Quantum Computation with Transverse Phonon Modes},
  volume = {97},
  ISSN = {1079-7114},
  url = {http://dx.doi.org/10.1103/PhysRevLett.97.050505},
  DOI = {10.1103/physrevlett.97.050505},
  number = {5},
  journal = {Physical Review Letters},
  publisher = {American Physical Society (APS)},
  author = {Zhu,  Shi-Liang and Monroe,  C. and Duan,  L.-M.},
  year = {2006},
  month = aug 
}

@article{Dodin2008,
  title = {Manley–Rowe relations for an arbitrary discrete system},
  volume = {372},
  ISSN = {0375-9601},
  url = {http://dx.doi.org/10.1016/j.physleta.2008.08.011},
  DOI = {10.1016/j.physleta.2008.08.011},
  number = {39},
  journal = {Physics Letters A},
  publisher = {Elsevier BV},
  author = {Dodin,  I.Y. and Zhmoginov,  A.I. and Fisch,  N.J.},
  year = {2008},
  month = sep,
  pages = {6094–6096}
}

@article{Fishman2008,
  title = {Structural phase transitions in low-dimensional ion crystals},
  volume = {77},
  ISSN = {1550-235X},
  url = {http://dx.doi.org/10.1103/PhysRevB.77.064111},
  DOI = {10.1103/physrevb.77.064111},
  number = {6},
  journal = {Physical Review B},
  publisher = {American Physical Society (APS)},
  author = {Fishman,  Shmuel and De Chiara,  Gabriele and Calarco,  Tommaso and Morigi,  Giovanna},
  year = {2008},
  month = feb 
}

@article{Roos2008,
  title = {Nonlinear coupling of continuous variables at the single quantum level},
  volume = {77},
  ISSN = {1094-1622},
  url = {http://dx.doi.org/10.1103/PhysRevA.77.040302},
  DOI = {10.1103/physreva.77.040302},
  number = {4},
  journal = {Physical Review A},
  publisher = {American Physical Society (APS)},
  author = {Roos,  C. F. and Monz,  T. and Kim,  K. and Riebe,  M. and H\"{a}ffner,  H. and James,  D. F. V. and Blatt,  R.},
  year = {2008},
  month = apr 
}

@article{Lin2009,
  title = {Large-scale quantum computation in an anharmonic linear ion trap},
  volume = {86},
  ISSN = {1286-4854},
  url = {http://dx.doi.org/10.1209/0295-5075/86/60004},
  DOI = {10.1209/0295-5075/86/60004},
  number = {6},
  journal = {EPL (Europhysics Letters)},
  publisher = {IOP Publishing},
  author = {Lin,  G.-D. and Zhu,  S.-L. and Islam,  R. and Kim,  K. and Chang,  M.-S. and Korenblit,  S. and Monroe,  C. and Duan,  L.-M.},
  year = {2009},
  month = jun,
  pages = {60004}
}

@article{Nie2009,
  title = {Theory of cross phase modulation for the vibrational modes of trapped ions},
  volume = {373},
  ISSN = {0375-9601},
  url = {http://dx.doi.org/10.1016/j.physleta.2008.11.045},
  DOI = {10.1016/j.physleta.2008.11.045},
  number = {4},
  journal = {Physics Letters A},
  publisher = {Elsevier BV},
  author = {Nie,  X. Rebecca and Roos,  Christian F. and James,  Daniel F.V.},
  year = {2009},
  month = jan,
  pages = {422–425}
}

@article{Patacchini2009,
  title = {Explicit time-reversible orbit integration in Particle In Cell codes with static homogeneous magnetic field},
  volume = {228},
  ISSN = {0021-9991},
  url = {http://dx.doi.org/10.1016/j.jcp.2008.12.021},
  DOI = {10.1016/j.jcp.2008.12.021},
  number = {7},
  journal = {Journal of Computational Physics},
  publisher = {Elsevier BV},
  author = {Patacchini,  L. and Hutchinson,  I.H.},
  year = {2009},
  month = apr,
  pages = {2604–2615}
}

@article{Home2011,
  title = {Normal modes of trapped ions in the presence of anharmonic trap potentials},
  volume = {13},
  ISSN = {1367-2630},
  url = {http://dx.doi.org/10.1088/1367-2630/13/7/073026},
  DOI = {10.1088/1367-2630/13/7/073026},
  number = {7},
  journal = {New Journal of Physics},
  publisher = {IOP Publishing},
  author = {Home,  J P and Hanneke,  D and Jost,  J D and Leibfried,  D and Wineland,  D J},
  year = {2011},
  month = jul,
  pages = {073026}
}

@article{Britton2012,
  title = {Engineered two-dimensional Ising interactions in a trapped-ion quantum simulator with hundreds of spins},
  volume = {484},
  ISSN = {1476-4687},
  url = {http://dx.doi.org/10.1038/nature10981},
  DOI = {10.1038/nature10981},
  number = {7395},
  journal = {Nature},
  publisher = {Springer Science and Business Media LLC},
  author = {Britton,  Joseph W. and Sawyer,  Brian C. and Keith,  Adam C. and Wang,  C.-C. Joseph and Freericks,  James K. and Uys,  Hermann and Biercuk,  Michael J. and Bollinger,  John J.},
  year = {2012},
  month = apr,
  pages = {489–492}
}

@article{Fowler2012,
  title = {Surface codes: Towards practical large-scale quantum computation},
  volume = {86},
  ISSN = {1094-1622},
  url = {http://dx.doi.org/10.1103/PhysRevA.86.032324},
  DOI = {10.1103/physreva.86.032324},
  number = {3},
  journal = {Physical Review A},
  publisher = {American Physical Society (APS)},
  author = {Fowler,  Austin G. and Mariantoni,  Matteo and Martinis,  John M. and Cleland,  Andrew N.},
  year = {2012},
  month = sep 
}

@article{Hayes2012,
  title = {Coherent Error Suppression in Multiqubit Entangling Gates},
  volume = {109},
  ISSN = {1079-7114},
  url = {http://dx.doi.org/10.1103/PhysRevLett.109.020503},
  DOI = {10.1103/physrevlett.109.020503},
  number = {2},
  journal = {Physical Review Letters},
  publisher = {American Physical Society (APS)},
  author = {Hayes,  D. and Clark,  S. M. and Debnath,  S. and Hucul,  D. and Inlek,  I. V. and Lee,  K. W. and Quraishi,  Q. and Monroe,  C.},
  year = {2012},
  month = jul 
}

@article{Kaufmann2012,
  title = {Precise Experimental Investigation of Eigenmodes in a Planar Ion Crystal},
  volume = {109},
  ISSN = {1079-7114},
  url = {http://dx.doi.org/10.1103/PhysRevLett.109.263003},
  DOI = {10.1103/physrevlett.109.263003},
  number = {26},
  journal = {Physical Review Letters},
  publisher = {American Physical Society (APS)},
  author = {Kaufmann,  H. and Ulm,  S. and Jacob,  G. and Poschinger,  U. and Landa,  H. and Retzker,  A. and Plenio,  M. B. and Schmidt-Kaler,  F.},
  year = {2012},
  month = dec 
}

@article{Johansson2013,
  title = {QuTiP 2: A Python framework for the dynamics of open quantum systems},
  volume = {184},
  ISSN = {0010-4655},
  url = {http://dx.doi.org/10.1016/j.cpc.2012.11.019},
  DOI = {10.1016/j.cpc.2012.11.019},
  number = {4},
  journal = {Computer Physics Communications},
  publisher = {Elsevier BV},
  author = {Johansson,  J.R. and Nation,  P.D. and Nori,  Franco},
  year = {2013},
  month = apr,
  pages = {1234–1240}
}

@article{Wang2013,
  title = {Phonon-mediated quantum spin simulator employing a planar ionic crystal in a Penning trap},
  volume = {87},
  ISSN = {1094-1622},
  url = {http://dx.doi.org/10.1103/PhysRevA.87.013422},
  DOI = {10.1103/physreva.87.013422},
  number = {1},
  journal = {Physical Review A},
  publisher = {American Physical Society (APS)},
  author = {Wang,  C.-C. Joseph and Keith,  Adam C. and Freericks,  J. K.},
  year = {2013},
  month = jan 
}

@article{Wu2013,
  title = {A complicated Duffing oscillator in the surface-electrode ion trap},
  volume = {114},
  ISSN = {1432-0649},
  url = {http://dx.doi.org/10.1007/s00340-013-5541-z},
  DOI = {10.1007/s00340-013-5541-z},
  number = {1–2},
  journal = {Applied Physics B},
  publisher = {Springer Science and Business Media LLC},
  author = {Wu,  Hao-Yu and Xie,  Yi and Wan,  Wei and Chen,  Liang and Zhou,  Fei and Feng,  Mang},
  year = {2013},
  month = jun,
  pages = {81–88}
}

@article{McAneny2014,
  title = {Intrinsic anharmonic effects on the phonon frequencies and effective spin-spin interactions in a quantum simulator made from trapped ions in a linear Paul trap},
  volume = {90},
  ISSN = {1094-1622},
  url = {http://dx.doi.org/10.1103/PhysRevA.90.053405},
  DOI = {10.1103/physreva.90.053405},
  number = {5},
  journal = {Physical Review A},
  publisher = {American Physical Society (APS)},
  author = {McAneny,  M. and Freericks,  J. K.},
  year = {2014},
  month = nov 
}

@article{Pihajoki2014,
  title = {Explicit methods in extended phase space for inseparable Hamiltonian problems},
  volume = {121},
  ISSN = {1572-9478},
  url = {http://dx.doi.org/10.1007/s10569-014-9597-9},
  DOI = {10.1007/s10569-014-9597-9},
  number = {3},
  journal = {Celestial Mechanics and Dynamical Astronomy},
  publisher = {Springer Science and Business Media LLC},
  author = {Pihajoki,  Pauli},
  year = {2014},
  month = nov,
  pages = {211–231}
}

@article{Lemmer2015,
  title = {Two-Dimensional Spectroscopy for the Study of Ion Coulomb Crystals},
  volume = {114},
  ISSN = {1079-7114},
  url = {http://dx.doi.org/10.1103/PhysRevLett.114.073001},
  DOI = {10.1103/physrevlett.114.073001},
  number = {7},
  journal = {Physical Review Letters},
  publisher = {American Physical Society (APS)},
  author = {Lemmer,  A. and Cormick,  C. and Schmiegelow,  C. T. and Schmidt-Kaler,  F. and Plenio,  M. B.},
  year = {2015},
  month = feb 
}

@article{Wang2015,
  title = {Quantum Computation under Micromotion in a Planar Ion Crystal},
  volume = {5},
  ISSN = {2045-2322},
  url = {http://dx.doi.org/10.1038/srep08555},
  DOI = {10.1038/srep08555},
  number = {1},
  journal = {Scientific Reports},
  publisher = {Springer Science and Business Media LLC},
  author = {Wang,  S.-T. and Shen,  C. and Duan,  L.-M.},
  year = {2015},
  month = feb 
}

@article{Bohnet2016,
  title = {Quantum spin dynamics and entanglement generation with hundreds of trapped ions},
  volume = {352},
  ISSN = {1095-9203},
  url = {http://dx.doi.org/10.1126/science.aad9958},
  DOI = {10.1126/science.aad9958},
  number = {6291},
  journal = {Science},
  publisher = {American Association for the Advancement of Science (AAAS)},
  author = {Bohnet,  Justin G. and Sawyer,  Brian C. and Britton,  Joseph W. and Wall,  Michael L. and Rey,  Ana Maria and Foss-Feig,  Michael and Bollinger,  John J.},
  year = {2016},
  month = jun,
  pages = {1297–1301}
}

@article{Lechner2016,
  title = {Electromagnetically-induced-transparency ground-state cooling of long ion strings},
  volume = {93},
  ISSN = {2469-9934},
  url = {http://dx.doi.org/10.1103/PhysRevA.93.053401},
  DOI = {10.1103/physreva.93.053401},
  number = {5},
  journal = {Physical Review A},
  publisher = {American Physical Society (APS)},
  author = {Lechner,  Regina and Maier,  Christine and Hempel,  Cornelius and Jurcevic,  Petar and Lanyon,  Ben P. and Monz,  Thomas and Brownnutt,  Michael and Blatt,  Rainer and Roos,  Christian F.},
  year = {2016},
  month = may 
}

@article{Richerme2016,
  title = {Two-dimensional ion crystals in radio-frequency traps for quantum simulation},
  volume = {94},
  ISSN = {2469-9934},
  url = {http://dx.doi.org/10.1103/PhysRevA.94.032320},
  DOI = {10.1103/physreva.94.032320},
  number = {3},
  journal = {Physical Review A},
  publisher = {American Physical Society (APS)},
  author = {Richerme,  Philip},
  year = {2016},
  month = sep 
}

@article{Torrisi2016,
  title = {Perpendicular laser cooling with a rotating-wall potential in a Penning trap},
  volume = {93},
  ISSN = {2469-9934},
  url = {http://dx.doi.org/10.1103/PhysRevA.93.043421},
  DOI = {10.1103/physreva.93.043421},
  number = {4},
  journal = {Physical Review A},
  publisher = {American Physical Society (APS)},
  author = {Torrisi,  Steven B. and Britton,  Joseph W. and Bohnet,  Justin G. and Bollinger,  John J.},
  year = {2016},
  month = apr 
}

@article{Ding2017co,
  title = {Quantum Parametric Oscillator with Trapped Ions},
  volume = {119},
  ISSN = {1079-7114},
  url = {http://dx.doi.org/10.1103/PhysRevLett.119.150404},
  DOI = {10.1103/physrevlett.119.150404},
  number = {15},
  journal = {Physical Review Letters},
  publisher = {American Physical Society (APS)},
  author = {Ding,  Shiqian and Maslennikov,  Gleb and Habl\"{u}tzel,  Roland and Loh,  Huanqian and Matsukevich,  Dzmitry},
  year = {2017},
  month = oct 
}

@article{Ding2017,
  title = {Cross-Kerr Nonlinearity for Phonon Counting},
  volume = {119},
  ISSN = {1079-7114},
  url = {http://dx.doi.org/10.1103/PhysRevLett.119.193602},
  DOI = {10.1103/physrevlett.119.193602},
  number = {19},
  journal = {Physical Review Letters},
  publisher = {American Physical Society (APS)},
  author = {Ding,  Shiqian and Maslennikov,  Gleb and Habl\"{u}tzel,  Roland and Matsukevich,  Dzmitry},
  year = {2017},
  month = nov 
}

@article{Meurer2017,
  title = {SymPy: symbolic computing in Python},
  volume = {3},
  ISSN = {2376-5992},
  url = {http://dx.doi.org/10.7717/peerj-cs.103},
  DOI = {10.7717/peerj-cs.103},
  journal = {PeerJ Computer Science},
  publisher = {PeerJ},
  author = {Meurer, Aaron and Smith, Christopher P. and Paprocki, Mateusz and {\v{C}}ert\'ik, Ond\v{r}ej and Kirpichev, Sergey B. and Rocklin, Matthew and Kumar, AMiT and Ivanov, Sergiu and Moore, Jason K. and Singh, Sartaj and Rathnayake, Thilina and Vig, Sean and Granger, Brian E. and Muller, Richard P. and Bonazzi, Francesco and Gupta, Harsh and Vats, Shivam and Johansson, Fredrik and Pedregosa, Fabian and Curry, Matthew J. and Terrel, Andy R. and Rou\v{c}ka, \v{S}t\v{e}p\'an and Saboo, Ashutosh and Fernando, Isuru and Kulal, Sumith and Cimrman, Robert and Scopatz, Anthony},
  year = {2017},
  month = jan,
  pages = {e103}
}

@article{Nimmrichter2017,
  title = {Quantum and classical dynamics of a three-mode absorption refrigerator},
  volume = {1},
  ISSN = {2521-327X},
  url = {http://dx.doi.org/10.22331/q-2017-12-11-37},
  DOI = {10.22331/q-2017-12-11-37},
  journal = {Quantum},
  publisher = {Verein zur Forderung des Open Access Publizierens in den Quantenwissenschaften},
  author = {Nimmrichter,  Stefan and Dai,  Jibo and Roulet,  Alexandre and Scarani,  Valerio},
  year = {2017},
  month = dec,
  pages = {37}
}

@article{OGorman2017,
  title = {Quantum computation with realistic magic-state factories},
  volume = {95},
  ISSN = {2469-9934},
  url = {http://dx.doi.org/10.1103/PhysRevA.95.032338},
  DOI = {10.1103/physreva.95.032338},
  number = {3},
  journal = {Physical Review A},
  publisher = {American Physical Society (APS)},
  author = {O’Gorman,  Joe and Campbell,  Earl T.},
  year = {2017},
  month = mar 
}

@article{Zhang2017,
  title = {Observation of a many-body dynamical phase transition with a 53-qubit quantum simulator},
  volume = {551},
  ISSN = {1476-4687},
  url = {http://dx.doi.org/10.1038/nature24654},
  DOI = {10.1038/nature24654},
  number = {7682},
  journal = {Nature},
  publisher = {Springer Science and Business Media LLC},
  author = {Zhang,  J. and Pagano,  G. and Hess,  P. W. and Kyprianidis,  A. and Becker,  P. and Kaplan,  H. and Gorshkov,  A. V. and Gong,  Z.-X. and Monroe,  C.},
  year = {2017},
  month = nov,
  pages = {601–604}
}

@article{Ding2018,
  title = {Quantum Simulation with a Trilinear Hamiltonian},
  volume = {121},
  ISSN = {1079-7114},
  url = {http://dx.doi.org/10.1103/PhysRevLett.121.130502},
  DOI = {10.1103/physrevlett.121.130502},
  number = {13},
  journal = {Physical Review Letters},
  publisher = {American Physical Society (APS)},
  author = {Ding,  Shiqian and Maslennikov,  Gleb and Habl\"{u}tzel,  Roland and Matsukevich,  Dzmitry},
  year = {2018},
  month = sep 
}

@article{Leung2018,
  title = {Robust 2-Qubit Gates in a Linear Ion Crystal Using a Frequency-Modulated Driving Force},
  volume = {120},
  ISSN = {1079-7114},
  url = {http://dx.doi.org/10.1103/PhysRevLett.120.020501},
  DOI = {10.1103/physrevlett.120.020501},
  number = {2},
  journal = {Physical Review Letters},
  publisher = {American Physical Society (APS)},
  author = {Leung,  Pak Hong and Landsman,  Kevin A. and Figgatt,  Caroline and Linke,  Norbert M. and Monroe,  Christopher and Brown,  Kenneth R.},
  year = {2018},
  month = jan 
}

@article{Pagano2018,
  title = {Cryogenic trapped-ion system for large scale quantum simulation},
  volume = {4},
  ISSN = {2058-9565},
  url = {http://dx.doi.org/10.1088/2058-9565/aae0fe},
  DOI = {10.1088/2058-9565/aae0fe},
  number = {1},
  journal = {Quantum Science and Technology},
  publisher = {IOP Publishing},
  author = {Pagano,  G and Hess,  P W and Kaplan,  H B and Tan,  W L and Richerme,  P and Becker,  P and Kyprianidis,  A and Zhang,  J and Birckelbaw,  E and Hernandez,  M R and Wu,  Y and Monroe,  C},
  year = {2018},
  month = oct,
  pages = {014004}
}

@article{Bruzewicz2019,
  title = {Trapped-ion quantum computing: Progress and challenges},
  volume = {6},
  ISSN = {1931-9401},
  url = {http://dx.doi.org/10.1063/1.5088164},
  DOI = {10.1063/1.5088164},
  number = {2},
  journal = {Applied Physics Reviews},
  publisher = {AIP Publishing},
  author = {Bruzewicz,  Colin D. and Chiaverini,  John and McConnell,  Robert and Sage,  Jeremy M.},
  year = {2019},
  month = may 
}

@article{Jordan2019,
  title = {Near Ground-State Cooling of Two-Dimensional Trapped-Ion Crystals with More than 100 Ions},
  volume = {122},
  ISSN = {1079-7114},
  url = {http://dx.doi.org/10.1103/PhysRevLett.122.053603},
  DOI = {10.1103/physrevlett.122.053603},
  number = {5},
  journal = {Physical Review Letters},
  publisher = {American Physical Society (APS)},
  author = {Jordan,  Elena and Gilmore,  Kevin A. and Shankar,  Athreya and Safavi-Naini,  Arghavan and Bohnet,  Justin G. and Holland,  Murray J. and Bollinger,  John J.},
  year = {2019},
  month = feb 
}

@article{Landsman2019,
  title = {Two-qubit entangling gates within arbitrarily long chains of trapped ions},
  volume = {100},
  ISSN = {2469-9934},
  url = {http://dx.doi.org/10.1103/PhysRevA.100.022332},
  DOI = {10.1103/physreva.100.022332},
  number = {2},
  journal = {Physical Review A},
  publisher = {American Physical Society (APS)},
  author = {Landsman,  K. A. and Wu,  Y. and Leung,  P. H. and Zhu,  D. and Linke,  N. M. and Brown,  K. R. and Duan,  L. and Monroe,  C.},
  year = {2019},
  month = aug 
}

@article{Maslennikov2019,
  title = {Quantum absorption refrigerator with trapped ions},
  volume = {10},
  ISSN = {2041-1723},
  url = {http://dx.doi.org/10.1038/s41467-018-08090-0},
  DOI = {10.1038/s41467-018-08090-0},
  number = {1},
  journal = {Nature Communications},
  publisher = {Springer Science and Business Media LLC},
  author = {Maslennikov,  Gleb and Ding,  Shiqian and Habl\"{u}tzel,  Roland and Gan,  Jaren and Roulet,  Alexandre and Nimmrichter,  Stefan and Dai,  Jibo and Scarani,  Valerio and Matsukevich,  Dzmitry},
  year = {2019},
  month = jan 
}

@article{Shankar2019,
  title = {Modeling near ground-state cooling of two-dimensional ion crystals in a Penning trap using electromagnetically induced transparency},
  volume = {99},
  ISSN = {2469-9934},
  url = {http://dx.doi.org/10.1103/PhysRevA.99.023409},
  DOI = {10.1103/physreva.99.023409},
  number = {2},
  journal = {Physical Review A},
  publisher = {American Physical Society (APS)},
  author = {Shankar,  Athreya and Jordan,  Elena and Gilmore,  Kevin A. and Safavi-Naini,  Arghavan and Bollinger,  John J. and Holland,  Murray J.},
  year = {2019},
  month = feb 
}

@article{Tang2019,
  title = {First principles simulation of ultracold ion crystals in a Penning trap with Doppler cooling and a rotating wall potential},
  volume = {26},
  ISSN = {1089-7674},
  url = {http://dx.doi.org/10.1063/1.5099256},
  DOI = {10.1063/1.5099256},
  number = {7},
  journal = {Physics of Plasmas},
  publisher = {AIP Publishing},
  author = {Tang,  Chen and Meiser,  Dominic and Bollinger,  John J. and Parker,  Scott E.},
  year = {2019},
  month = jul 
}

@article{Dubin2020,
  title = {Normal modes,  rotational inertia,  and thermal fluctuations of trapped ion crystals},
  volume = {27},
  ISSN = {1089-7674},
  url = {http://dx.doi.org/10.1063/5.0021732},
  DOI = {10.1063/5.0021732},
  number = {10},
  journal = {Physics of Plasmas},
  publisher = {AIP Publishing},
  author = {Dubin,  Daniel H. E.},
  year = {2020},
  month = oct 
}

@article{Jain2020,
  title = {Scalable Arrays of Micro-Penning Traps for Quantum Computing and Simulation},
  volume = {10},
  ISSN = {2160-3308},
  url = {http://dx.doi.org/10.1103/PhysRevX.10.031027},
  DOI = {10.1103/physrevx.10.031027},
  number = {3},
  journal = {Physical Review X},
  publisher = {American Physical Society (APS)},
  author = {Jain,  S. and Alonso,  J. and Grau,  M. and Home,  J. P.},
  year = {2020},
  month = aug 
}

@article{Shankar2020,
  title = {Broadening of the drumhead-mode spectrum due to in-plane thermal fluctuations of two-dimensional trapped ion crystals in a Penning trap},
  volume = {102},
  ISSN = {2469-9934},
  url = {http://dx.doi.org/10.1103/PhysRevA.102.053106},
  DOI = {10.1103/physreva.102.053106},
  number = {5},
  journal = {Physical Review A},
  publisher = {American Physical Society (APS)},
  author = {Shankar,  Athreya and Tang,  Chen and Affolter,  Matthew and Gilmore,  Kevin and Dubin,  Daniel H. E. and Parker,  Scott and Holland,  Murray J. and Bollinger,  John J.},
  year = {2020},
  month = nov 
}

@article{Wang2020,
  title = {Coherently Manipulated 2D Ion Crystal in a Monolithic Paul Trap},
  volume = {3},
  ISSN = {2511-9044},
  url = {http://dx.doi.org/10.1002/qute.202000068},
  DOI = {10.1002/qute.202000068},
  number = {11},
  journal = {Advanced Quantum Technologies},
  publisher = {Wiley},
  author = {Wang,  Ye and Qiao,  Mu and Cai,  Zhengyang and Zhang,  Kuan and Jin,  Naijun and Wang,  Pengfei and Chen,  Wentao and Luan,  Chunyang and Du,  Botao and Wang,  Haiyan and Song,  Yipu and Yum,  Dahyun and Kim,  Kihwan},
  year = {2020},
  month = oct 
}

@article{Zhang2020,
  title = {Submicrosecond entangling gate between trapped ions via Rydberg interaction},
  volume = {580},
  ISSN = {1476-4687},
  url = {http://dx.doi.org/10.1038/s41586-020-2152-9},
  DOI = {10.1038/s41586-020-2152-9},
  number = {7803},
  journal = {Nature},
  publisher = {Springer Science and Business Media LLC},
  author = {Zhang,  Chi and Pokorny,  Fabian and Li,  Weibin and Higgins,  Gerard and P\"{o}schl,  Andreas and Lesanovsky,  Igor and Hennrich,  Markus},
  year = {2020},
  month = apr,
  pages = {345–349}
}

@article{Chen2021,
  title = {Quantum computation and simulation with vibrational modes of trapped ions},
  volume = {30},
  ISSN = {1674-1056},
  url = {http://dx.doi.org/10.1088/1674-1056/ac01e3},
  DOI = {10.1088/1674-1056/ac01e3},
  number = {6},
  journal = {Chinese Physics B},
  publisher = {IOP Publishing},
  author = {Chen,  Wentao and Gan,  Jaren and Zhang,  Jing-Ning and Matuskevich,  Dzmitry and Kim,  Kihwan},
  year = {2021},
  month = jun,
  pages = {060311}
}

@article{Gilmore2021,
  title = {Quantum-enhanced sensing of displacements and electric fields with two-dimensional trapped-ion crystals},
  volume = {373},
  ISSN = {1095-9203},
  url = {http://dx.doi.org/10.1126/science.abi5226},
  DOI = {10.1126/science.abi5226},
  number = {6555},
  journal = {Science},
  publisher = {American Association for the Advancement of Science (AAAS)},
  author = {Gilmore,  Kevin A. and Affolter,  Matthew and Lewis-Swan,  Robert J. and Barberena,  Diego and Jordan,  Elena and Rey,  Ana Maria and Bollinger,  John J.},
  year = {2021},
  month = aug,
  pages = {673–678}
}

@article{Kalincev2021,
  title = {Motional heating of spatially extended ion crystals},
  volume = {6},
  ISSN = {2058-9565},
  url = {http://dx.doi.org/10.1088/2058-9565/abee99},
  DOI = {10.1088/2058-9565/abee99},
  number = {3},
  journal = {Quantum Science and Technology},
  publisher = {IOP Publishing},
  author = {Kalincev,  D and Dreissen,  L S and Kulosa,  A P and Yeh,  C-H and F\"{u}rst,  H A and Mehlst\"{a}ubler,  T E},
  year = {2021},
  month = may,
  pages = {034003}
}

@article{Kang2021,
  title = {Batch Optimization of Frequency-Modulated Pulses for Robust Two-Qubit Gates in Ion Chains},
  volume = {16},
  ISSN = {2331-7019},
  url = {http://dx.doi.org/10.1103/PhysRevApplied.16.024039},
  DOI = {10.1103/physrevapplied.16.024039},
  number = {2},
  journal = {Physical Review Applied},
  publisher = {American Physical Society (APS)},
  author = {Kang,  Mingyu and Liang,  Qiyao and Zhang,  Bichen and Huang,  Shilin and Wang,  Ye and Fang,  Chao and Kim,  Jungsang and Brown,  Kenneth R.},
  year = {2021},
  month = aug 
}

@article{Kiethe2021,
  title = {Finite-temperature spectrum at the symmetry-breaking linear to zigzag transition},
  volume = {103},
  ISSN = {2469-9969},
  url = {http://dx.doi.org/10.1103/PhysRevB.103.104106},
  DOI = {10.1103/physrevb.103.104106},
  number = {10},
  journal = {Physical Review B},
  publisher = {American Physical Society (APS)},
  author = {Kiethe,  Jan and Timm,  Lars and Landa,  Haggai and Kalincev,  Dimitri and Morigi,  Giovanna and Mehlst\"{a}ubler,  Tanja E.},
  year = {2021},
  month = mar 
}

@article{Monroe2021,
  title = {Programmable quantum simulations of spin systems with trapped ions},
  volume = {93},
  ISSN = {1539-0756},
  url = {http://dx.doi.org/10.1103/RevModPhys.93.025001},
  DOI = {10.1103/revmodphys.93.025001},
  number = {2},
  journal = {Reviews of Modern Physics},
  publisher = {American Physical Society (APS)},
  author = {Monroe,  C. and Campbell,  W. C. and Duan,  L.-M. and Gong,  Z.-X. and Gorshkov,  A. V. and Hess,  P. W. and Islam,  R. and Kim,  K. and Linke,  N. M. and Pagano,  G. and Richerme,  P. and Senko,  C. and Yao,  N. Y.},
  year = {2021},
  month = apr 
}

@article{Qiao2021,
  title = {Double-Electromagnetically-Induced-Transparency Ground-State Cooling of Stationary Two-Dimensional Ion Crystals},
  volume = {126},
  ISSN = {1079-7114},
  url = {http://dx.doi.org/10.1103/PhysRevLett.126.023604},
  DOI = {10.1103/physrevlett.126.023604},
  number = {2},
  journal = {Physical Review Letters},
  publisher = {American Physical Society (APS)},
  author = {Qiao,  Mu and Wang,  Ye and Cai,  Zhengyang and Du,  Botao and Wang,  Pengfei and Luan,  Chunyang and Chen,  Wentao and Noh,  Heung-Ryoul and Kim,  Kihwan},
  year = {2021},
  month = jan 
}

@article{Tang2021,
  title = {Equilibration of the planar modes of ultracold two-dimensional ion crystals in a Penning trap},
  volume = {104},
  ISSN = {2469-9934},
  url = {http://dx.doi.org/10.1103/PhysRevA.104.023325},
  DOI = {10.1103/physreva.104.023325},
  number = {2},
  journal = {Physical Review A},
  publisher = {American Physical Society (APS)},
  author = {Tang,  Chen and Shankar,  Athreya and Meiser,  Dominic and Dubin,  Daniel H. E. and Bollinger,  John J. and Parker,  Scott E.},
  year = {2021},
  month = aug 
}

@article{Cetina2022,
  title = {Control of Transverse Motion for Quantum Gates on Individually Addressed Atomic Qubits},
  volume = {3},
  ISSN = {2691-3399},
  url = {http://dx.doi.org/10.1103/PRXQuantum.3.010334},
  DOI = {10.1103/prxquantum.3.010334},
  number = {1},
  journal = {PRX Quantum},
  publisher = {American Physical Society (APS)},
  author = {Cetina,  M. and Egan,  L.N. and Noel,  C. and Goldman,  M.L. and Biswas,  D. and Risinger,  A.R. and Zhu,  D. and Monroe,  C.},
  year = {2022},
  month = mar 
}

@article{Kranzl2022,
  title = {Controlling long ion strings for quantum simulation and precision measurements},
  volume = {105},
  ISSN = {2469-9934},
  url = {http://dx.doi.org/10.1103/PhysRevA.105.052426},
  DOI = {10.1103/physreva.105.052426},
  number = {5},
  journal = {Physical Review A},
  publisher = {American Physical Society (APS)},
  author = {Kranzl,  Florian and Joshi,  Manoj K. and Maier,  Christine and Brydges,  Tiff and Franke,  Johannes and Blatt,  Rainer and Roos,  Christian F.},
  year = {2022},
  month = may 
}

@article{Manovitz2022,
  title = {Trapped-Ion Quantum Computer with Robust Entangling Gates and Quantum Coherent Feedback},
  volume = {3},
  ISSN = {2691-3399},
  url = {http://dx.doi.org/10.1103/PRXQuantum.3.010347},
  DOI = {10.1103/prxquantum.3.010347},
  number = {1},
  journal = {PRX Quantum},
  publisher = {American Physical Society (APS)},
  author = {Manovitz,  Tom and Shapira,  Yotam and Gazit,  Lior and Akerman,  Nitzan and Ozeri,  Roee},
  year = {2022},
  month = mar 
}

@article{Murali2022,
  title = {Toward systematic architectural design of near-term trapped ion quantum computers},
  volume = {65},
  ISSN = {1557-7317},
  url = {http://dx.doi.org/10.1145/3511064},
  DOI = {10.1145/3511064},
  number = {3},
  journal = {Communications of the ACM},
  publisher = {Association for Computing Machinery (ACM)},
  author = {Murali,  Prakash and Debroy,  Dripto M. and Brown,  Kenneth R. and Martonosi,  Margaret},
  year = {2022},
  month = feb,
  pages = {101–109}
}

@article{Polloreno2022,
  title = {Individual qubit addressing of rotating ion crystals in a Penning trap},
  volume = {4},
  ISSN = {2643-1564},
  url = {http://dx.doi.org/10.1103/PhysRevResearch.4.033076},
  DOI = {10.1103/physrevresearch.4.033076},
  number = {3},
  journal = {Physical Review Research},
  publisher = {American Physical Society (APS)},
  author = {Polloreno,  Anthony M. and Rey,  Ana Maria and Bollinger,  John J.},
  year = {2022},
  month = jul 
}

@article{Sutherland2022,
  title = {One- and two-qubit gate infidelities due to motional errors in trapped ions and electrons},
  volume = {105},
  ISSN = {2469-9934},
  url = {http://dx.doi.org/10.1103/PhysRevA.105.022437},
  DOI = {10.1103/physreva.105.022437},
  number = {2},
  journal = {Physical Review A},
  publisher = {American Physical Society (APS)},
  author = {Sutherland,  R. Tyler and Yu,  Qian and Beck,  Kristin M. and H\"{a}ffner,  Hartmut},
  year = {2022},
  month = feb 
}

@article{Kiesenhofer2023,
  title = {Controlling Two-Dimensional Coulomb Crystals of More Than 100 Ions in a Monolithic Radio-Frequency Trap},
  volume = {4},
  ISSN = {2691-3399},
  url = {http://dx.doi.org/10.1103/PRXQuantum.4.020317},
  DOI = {10.1103/prxquantum.4.020317},
  number = {2},
  journal = {PRX Quantum},
  publisher = {American Physical Society (APS)},
  author = {Kiesenhofer,  Dominik and Hainzer,  Helene and Zhdanov,  Artem and Holz,  Philip C. and Bock,  Matthias and Ollikainen,  Tuomas and Roos,  Christian F.},
  year = {2023},
  month = apr 
}

@article{Chen2024,
  title = {Benchmarking a trapped-ion quantum computer with 30 qubits},
  volume = {8},
  ISSN = {2521-327X},
  url = {http://dx.doi.org/10.22331/q-2024-11-07-1516},
  DOI = {10.22331/q-2024-11-07-1516},
  journal = {Quantum},
  publisher = {Verein zur Forderung des Open Access Publizierens in den Quantenwissenschaften},
  author = {Chen,  Jwo-Sy and Nielsen,  Erik and Ebert,  Matthew and Inlek,  Volkan and Wright,  Kenneth and Chaplin,  Vandiver and Maksymov,  Andrii and Páez,  Eduardo and Poudel,  Amrit and Maunz,  Peter and Gamble,  John},
  year = {2024},
  month = nov,
  pages = {1516}
}

@article{Guo2024,
  title = {A site-resolved two-dimensional quantum simulator with hundreds of trapped ions},
  volume = {630},
  ISSN = {1476-4687},
  url = {http://dx.doi.org/10.1038/s41586-024-07459-0},
  DOI = {10.1038/s41586-024-07459-0},
  number = {8017},
  journal = {Nature},
  publisher = {Springer Science and Business Media LLC},
  author = {Guo,  S.-A. and Wu,  Y.-K. and Ye,  J. and Zhang,  L. and Lian,  W.-Q. and Yao,  R. and Wang,  Y. and Yan,  R.-Y. and Yi,  Y.-J. and Xu,  Y.-L. and Li,  B.-W. and Hou,  Y.-H. and Xu,  Y.-Z. and Guo,  W.-X. and Zhang,  C. and Qi,  B.-X. and Zhou,  Z.-C. and He,  L. and Duan,  L.-M.},
  year = {2024},
  month = may,
  pages = {613–618}
}

@article{Hawaldar2024,
  title = {Bilayer Crystals of Trapped Ions for Quantum Information Processing},
  volume = {14},
  ISSN = {2160-3308},
  url = {http://dx.doi.org/10.1103/PhysRevX.14.031030},
  DOI = {10.1103/physrevx.14.031030},
  number = {3},
  journal = {Physical Review X},
  publisher = {American Physical Society (APS)},
  author = {Hawaldar,  Samarth and Shahi,  Prakriti and Carter,  Allison L. and Rey,  Ana Maria and Bollinger,  John J. and Shankar,  Athreya},
  year = {2024},
  month = aug 
}

@article{Hou2024_2D,
  title = {Individually addressed entangling gates in a two-dimensional ion crystal},
  volume = {15},
  ISSN = {2041-1723},
  url = {http://dx.doi.org/10.1038/s41467-024-53405-z},
  DOI = {10.1038/s41467-024-53405-z},
  number = {1},
  journal = {Nature Communications},
  publisher = {Springer Science and Business Media LLC},
  author = {Hou,  Y.-H. and Yi,  Y.-J. and Wu,  Y.-K. and Chen,  Y.-Y. and Zhang,  L. and Wang,  Y. and Xu,  Y.-L. and Zhang,  C. and Mei,  Q.-X. and Yang,  H.-X. and Ma,  J.-Y. and Guo,  S.-A. and Ye,  J. and Qi,  B.-X. and Zhou,  Z.-C. and Hou,  P.-Y. and Duan,  L.-M.},
  year = {2024},
  month = nov 
}

@article{Jain2024,
  title = {Penning micro-trap for quantum computing},
  volume = {627},
  ISSN = {1476-4687},
  url = {http://dx.doi.org/10.1038/s41586-024-07111-x},
  DOI = {10.1038/s41586-024-07111-x},
  number = {8004},
  journal = {Nature},
  publisher = {Springer Science and Business Media LLC},
  author = {Jain,  Shreyans and S\"{a}gesser,  Tobias and Hrmo,  Pavel and Torkzaban,  Celeste and Stadler,  Martin and Oswald,  Robin and Axline,  Chris and Bautista-Salvador,  Amado and Ospelkaus,  Christian and Kienzler,  Daniel and Home,  Jonathan},
  year = {2024},
  month = mar,
  pages = {510–514}
}

@article{Johnson2024,
  title = {Rapid cooling of the in-plane motion of two-dimensional ion crystals in a Penning trap to millikelvin temperatures},
  volume = {109},
  ISSN = {2469-9934},
  url = {http://dx.doi.org/10.1103/PhysRevA.109.L021102},
  DOI = {10.1103/physreva.109.l021102},
  number = {2},
  journal = {Physical Review A},
  publisher = {American Physical Society (APS)},
  author = {Johnson,  Wes and Shankar,  Athreya and Zaris,  John and Bollinger,  John J. and Parker,  Scott E.},
  year = {2024},
  month = feb 
}

@misc{Loschnauer2024, 
  doi = {10.48550/ARXIV.2407.07694},
  url = {https://arxiv.org/abs/2407.07694},
  author = {L\"{o}schnauer,  C. M. and Toba,  J. Mosca and Hughes,  A. C. and King,  S. A. and Weber,  M. A. and Srinivas,  R. and Matt,  R. and Nourshargh,  R. and Allcock,  D. T. C. and Ballance,  C. J. and Matthiesen,  C. and Malinowski,  M. and Harty,  T. P.},
  title = {Scalable,  high-fidelity all-electronic control of trapped-ion qubits},
  publisher = {arXiv},
  year = {2024},
  copyright = {arXiv.org perpetual,  non-exclusive license}
}

@article{McMahon2024,
  title = {Individual-Ion Addressing and Readout in a Penning Trap},
  volume = {133},
  ISSN = {1079-7114},
  url = {http://dx.doi.org/10.1103/PhysRevLett.133.173201},
  DOI = {10.1103/physrevlett.133.173201},
  number = {17},
  journal = {Physical Review Letters},
  publisher = {American Physical Society (APS)},
  author = {McMahon,  Brian J. and Brown,  Kenton R. and Herold,  Creston D. and Sawyer,  Brian C.},
  year = {2024},
  month = oct 
}

@article{Ruzic2024,
  title = {Leveraging motional-mode balancing and simply parametrized waveforms to perform frequency-robust entangling gates},
  volume = {22},
  ISSN = {2331-7019},
  url = {http://dx.doi.org/10.1103/PhysRevApplied.22.014007},
  DOI = {10.1103/physrevapplied.22.014007},
  number = {1},
  journal = {Physical Review Applied},
  publisher = {American Physical Society (APS)},
  author = {Ruzic,  Brandon P. and Chow,  Matthew N.H. and Burch,  Ashlyn D. and Lobser,  Daniel S. and Revelle,  Melissa C. and Wilson,  Joshua M. and Yale,  Christopher G. and Clark,  Susan M.},
  year = {2024},
  month = jul 
}

@article{Schwerdt2024,
  title = {Scalable Architecture for Trapped-Ion Quantum Computing Using rf Traps and Dynamic Optical Potentials},
  volume = {14},
  ISSN = {2160-3308},
  url = {http://dx.doi.org/10.1103/PhysRevX.14.041017},
  DOI = {10.1103/physrevx.14.041017},
  number = {4},
  journal = {Physical Review X},
  publisher = {American Physical Society (APS)},
  author = {Schwerdt,  David and Peleg,  Lee and Shapira,  Yotam and Priel,  Nadav and Florshaim,  Yanay and Gross,  Avram and Zalic,  Ayelet and Afek,  Gadi and Akerman,  Nitzan and Stern,  Ady and Kish,  Amit Ben and Ozeri,  Roee},
  year = {2024},
  month = oct 
}

@article{Sun2024,
  title = {Two-dimensional ion crystals in a hybrid optical cavity trap for quantum information processing},
  volume = {109},
  ISSN = {2469-9934},
  url = {http://dx.doi.org/10.1103/PhysRevA.109.032426},
  DOI = {10.1103/physreva.109.032426},
  number = {3},
  journal = {Physical Review A},
  publisher = {American Physical Society (APS)},
  author = {Sun,  Zewen and Teoh,  Yi Hong and Rajabi,  Fereshteh and Islam,  Rajibul},
  year = {2024},
  month = mar 
}

@article{Wolf2024,
  title = {Efficient site-resolved imaging and spin-state detection in dynamic two-dimensional ion crystals},
  volume = {21},
  ISSN = {2331-7019},
  url = {http://dx.doi.org/10.1103/PhysRevApplied.21.054067},
  DOI = {10.1103/physrevapplied.21.054067},
  number = {5},
  journal = {Physical Review Applied},
  publisher = {American Physical Society (APS)},
  author = {Wolf,  Robert N. and Pham,  Joseph H. and Jee,  Julian Y. Z. and Rischka,  Alexander and Biercuk,  Michael J.},
  year = {2024},
  month = may 
}

@article{Johnson2025,
  title = {Adiabatic cooling of planar motion in a Penning-trap ion crystal to sub-millikelvin temperatures},
  volume = {112},
  ISSN = {2469-9934},
  url = {http://dx.doi.org/10.1103/1xtw-m3j2},
  DOI = {10.1103/1xtw-m3j2},
  number = {6},
  journal = {Physical Review A},
  publisher = {American Physical Society (APS)},
  author = {Johnson,  Wes and Bullock,  Bryce and Shankar,  Athreya and Zaris,  John and Bollinger,  John J. and Parker,  Scott E.},
  year = {2025},
  month = dec 
}

@article{Le2025,
  title = {Amplitude-noise-resilient entangling gates for trapped ions},
  volume = {24},
  ISSN = {2331-7019},
  url = {http://dx.doi.org/10.1103/yrj7-tvw8},
  DOI = {10.1103/yrj7-tvw8},
  number = {5},
  journal = {Physical Review Applied},
  publisher = {American Physical Society (APS)},
  author = {Le,  Nguyen H. and Orozco-Ruiz,  Modesto and Kulmiya,  Sahra A. and Urquhart,  James G. and Hile,  Samuel J. and Hensinger,  Winfried K. and Mintert,  Florian},
  year = {2025},
  month = nov 
}

@article{Nagies2025,
  title = {The role of higher-order terms in trapped-ion quantum computing with magnetic gradient induced coupling},
  volume = {10},
  ISSN = {2058-9565},
  url = {http://dx.doi.org/10.1088/2058-9565/adc1fe},
  DOI = {10.1088/2058-9565/adc1fe},
  number = {2},
  journal = {Quantum Science and Technology},
  publisher = {IOP Publishing},
  author = {Nagies,  Sebastian and Geier,  Kevin T and Akram,  Javed and Okamoto,  Junichi and Bantounas,  Dimitrios and Wunderlich,  Christof and Johanning,  Michael and Hauke,  Philipp},
  year = {2025},
  month = mar,
  pages = {025051}
}

@article{Valentini2025,
  title = {Demonstration of Two-Dimensional Connectivity for a Scalable Error-Corrected Ion-Trap Quantum Processor Architecture},
  volume = {15},
  ISSN = {2160-3308},
  url = {http://dx.doi.org/10.1103/b9s1-6r44},
  DOI = {10.1103/b9s1-6r44},
  number = {4},
  journal = {Physical Review X},
  publisher = {American Physical Society (APS)},
  author = {Valentini,  M. and van Mourik,  M. W. and Butt,  F. and Wahl,  J. and Dietl,  M. and Pfeifer,  M. and Anmasser,  F. and Colombe,  Y. and R\"{o}ssler,  C. and Holz,  P. C. and Blatt,  R. and Bermudez,  A. and M\"{u}ller,  M. and Monz,  T. and Schindler,  P.},
  year = {2025},
  month = nov 
}

@article{Ringbauer2025,
  title = {Verifiable measurement-based quantum random sampling with trapped ions},
  volume = {16},
  ISSN = {2041-1723},
  url = {http://dx.doi.org/10.1038/s41467-024-55342-3},
  DOI = {10.1038/s41467-024-55342-3},
  number = {1},
  journal = {Nature Communications},
  publisher = {Springer Science and Business Media LLC},
  author = {Ringbauer,  Martin and Hinsche,  Marcel and Feldker,  Thomas and Faehrmann,  Paul K. and Bermejo-Vega,  Juani and Edmunds,  Claire L. and Postler,  Lukas and Stricker,  Roman and Marciniak,  Christian D. and Meth,  Michael and Pogorelov,  Ivan and Blatt,  Rainer and Schindler,  Philipp and Eisert,  Jens and Monz,  Thomas and Hangleiter,  Dominik},
  year = {2025},
  month = jan 
}

@article{Zaris2025,
  title = {Numerical simulations of three-dimensional ion crystal dynamics in a Penning trap using the fast multipole method},
  volume = {91},
  ISSN = {1469-7807},
  url = {http://dx.doi.org/10.1017/S0022377824000990},
  DOI = {10.1017/s0022377824000990},
  number = {2},
  journal = {Journal of Plasma Physics},
  publisher = {Cambridge University Press (CUP)},
  author = {Zaris,  John and Johnson,  Wes and Shankar,  Athreya and Bollinger,  John J. and Parker,  Scott E.},
  year = {2025},
  month = apr 
}

\end{document}